\documentclass{aa}

\usepackage{graphicx}
\usepackage{txfonts}

\usepackage{lipsum}

\usepackage{placeins}
\usepackage[T1]{fontenc}
\usepackage{ae,aecompl}
\usepackage{enumitem}
\usepackage{float} 
\usepackage[usenames,dvipsnames,table,xcdraw]{xcolor}
\usepackage{xcolor}
\usepackage{xspace}
\usepackage[para]{threeparttable} 
\usepackage{multirow}
\usepackage{multicol}
\usepackage{pdfpages}
\usepackage[version=4]{mhchem}

\usepackage{afterpage}

\usepackage{subcaption}
\usepackage[colorlinks=true,linkcolor=Blue,citecolor=Blue,urlcolor=black,anchorcolor=black,bookmarks=true,breaklinks=true,hypertexnames=false,bookmarksnumbered=true]{hyperref}
\begin{document} 
\newcommand{\kepler}{\textit{Kepler}}
\newcommand{\starshadow}{\texttt{STAR SHADOW}}
\newcommand{\pofour}{\texttt{period04}}
\newcommand{\amigo}{\texttt{AMiGO}}
\newcommand{\cthree}{\texttt{C-3PO}}
\newcommand{\ks}{$\rm K_{\rm s}$}

\newcommand\kms{\ifmmode{\rm km\thinspace s^{-1}}
\else km\thinspace s$^{-1}$\fi}
\newcommand{\perd}{$\rm d^{-1}$}
\newcommand{\solar}{$_\odot$}
\newcommand{\M}{$M$}
\newcommand{\Mc}{$M_c$}
\newcommand{\omegai}{$\omega_0$}
\newcommand{\xc}{$X_{\rm c}$}
\newcommand{\xcx}{$X_{\rm c}/X_{\rm i}$}
\newcommand{\fov}{$f_{\rm ov}$}
\newcommand{\logr}{log$(R)$}
\newcommand{\Z}{$Z$}
\newcommand{\teff}{$T_{\rm eff}$}
\newcommand{\logg}{log$(g)$}
\newcommand{\lum}{$L$}
\newcommand{\logl}{log$(L)$}
\newcommand{\logteff}{log$(T_{\rm eff})$}

\newcommand{\chisqr}{$\chi^2$}
\newcommand{\mone}{$^{-1}$}

\newcommand{\gssp}{\texttt{GSSP}}
\newcommand{\gdor}{$\gamma$~Dor}
\newcommand{\dsct}{$\delta$~Sct}
\newcommand{\bcep}{$\beta$~Cep}

\newcommand{\gaia}{\textit{Gaia}}

\newcommand{\tento}[1]{$10^{#1}$}
\newcommand{\timestento}[2]{$#1 \times 10^{#2}$}

   \title{Populations of tidal and pulsating variables in eclipsing binaries}
\author{
   Alex~Kemp\inst{1}\thanks{
    \email{alex.kemp@kuleuven.be}}
    \and
    Jasmine~Vrancken\inst{1}
    \and
    Joey~S.~G.~Mombarg\inst{2}
    \and
    Luc~IJspeert\inst{1}
    \and
    Mykyta Kliapets\inst{1}
    \and
    Andrew~Tkachenko\inst{1}
    \and
    Conny~Aerts\inst{1,3,4}
  }

   \institute{Institute of Astronomy (IvS), KU Leuven, Celestijnenlaan 200D, 3001, Leuven, Belgium
   \and
   {Universit\'e Paris-Saclay, Universit\'e de Paris, Sorbonne Paris Cit\'e, CEA, CNRS, AIM, 91191 Gif-sur-Yvette, France}
   \and
   Department of Astrophysics, IMAPP, Radboud University Nijmegen, PO Box 9010, 6500 GL Nijmegen, The Netherlands
   \and
   Max Planck Institute for Astronomy, Koenigstuhl 17, 69117 Heidelberg, Germany}
   
   \date{}

 
  \abstract
   {In the modern era of large-scale photometric time-domain surveys, relatively rare but information-rich eclipsing binary systems can be leveraged at a population level across the Hertzsprung-Russel diagram. This high-precision photometry is also excellent for assessing and exploiting the asteroseismic properties of these stars, resulting in a powerful synergy that has great potential for shedding light on how stellar interiors and tides affect stellar evolution and mass transfer.}
   {In this work, we seek to characterise a large sample of 14377 main sequence eclipsing binaries in terms of their stellar, asteroseismic, and orbital properties.}
   {We conduct manual vetting on a 4000-target subset of our full 14377-target sample to identify targets with pressure or gravity modes. We infer stellar properties including the mass, convective core mass, radius, and central H fraction for the primary using \gaia\ Data Release 3 effective temperature and luminosity estimates and a grid of asteroseismically calibrated stellar models. We use surface brightness ratio and radius ratio estimates from previous eclipse analysis to study the effect of binarity on our results.}
   {Our manual vetting identifies 751 candidate g-mode pulsators, 131 p-mode pulsators, and a further 48 hybrid pulsators. The inferred stellar properties of the hybrid and p-mode pulsators are highly correlated, while the orbital properties of the hybrid pulsators align best with the g-mode pulsators. The g-mode pulsators themselves show a distribution that peaks around the classical \gdor\ instability region but extends continuously towards higher masses, with no detectable divide between the classical \gdor\ and SPB instability regions. There is evidence at the population level for a heightened level of tidal efficiency in stars showing g-mode or hybrid variability. Correcting the primary mass inference for binarity based on eclipse measurements of the surface brightness and radius ratios results in a relatively small shift towards lower masses.}
   {This work provides a working initial characterisation of this sample from which more detailed analyses folding in asteroseismic information can be built. It also provides a foundational understanding of the limitations and capabilities of this kind of rapid, scalable analysis that will be highly relevant in planning the exploitation of future large-scale binary surveys.}


   \maketitle
%

\section{Introduction}
\label{sec:intro}
In the current age of space-based precision photometry, photometric observations provide the most complete time-domain coverage of binary star systems. Kepler \citep{borucki2010}, TESS 
\citep{ricker2015}, and the upcoming PLATO mission \citep{rauer2025} are designed to provide high-precision photometry for large numbers of targets, resulting in the detection and characterisation of tens of thousands of eclipsing binaries (EBs, \citealt{prsa2011,slawson2011,kirk2016,prsa2022}). These EBs allow invaluable constraints to be placed on the binary components and their orbital properties, information that is highly complementary with surface properties obtained through traditional spectroscopic follow-up or, where unavailable or impractical, estimates based on \gaia's BP/RP low resolution spectra \citep{gaia2023}.

A natural synergy between photometric observations of EBs is that with asteroseismology, as the same high-quality time series that allows measurement of eclipse properties also captures flux variations due to pulsations \citep{gaia2023pulsations}. At the level of individual stars, this synergy has been proved to be powerful; the combination of EB constraints \citep{hambleton2013,schmid2016,guo2019,sekaran2021,guo2021,aerts2024,kemp2024eclipse,kemp2025_k415} and asteroseismic forward modelling \citep{aerts2018} can provide an invaluable window on fundamental stellar mixing physics \citep{pedersen2021,michielsen2021,michielsen2023,michielsen2024foam}. These and similar studies rely heavily on the sensitivity of gravity-mode (g-mode) pulsators to the near-core properties of stars with convective cores and radiative envelopes (main sequence stars with masses greater than 1.2 M\solar). Pressure-mode (p-mode) pulsators \citep{bowman2016,murphy2019,murphy2024} offer complementary information about the outer envelope, making stars in which both g- and p-modes are detected (hybrid pulsators) particularly valuable \citep{kurtz2014,mombarg2020,audenaert2022,kemp2024eclipse,kemp2025_k415}.

At the population level, model-independent measurements of the stellar rotation in the near-core region have revealed that main sequence g-mode pulsators are (near-) rigid rotators \citep{aerts2021}. Further, using nothing but the dominant g-mode oscillation frequency, an estimate of the stellar mass and near-core rotation rate can be made even from sparse \gaia\ photometry \citep{hey2024,aerts2025}.

Recently, large numbers of eclipsing binaries have been identified and have had their eclipse properties measured using TESS photometry \citep{ijspeert2024_14k} (IJ24). Previous analysis of the asteroseismic potential of this sample was limited to an automated detection procedure based on the number of significant frequencies detected that were not associated with eclipse harmonics. We aim to compile a robust, manually vetted sample of pulsating EBs that is well-characterised in terms of their orbital and stellar properties. Of particular interest is the degree of similarity between the distributions of the subpopulations of pulsating vs non-pulsating EBs, and understanding how binarity can induce bias in the inference of stellar properties. Such a well characterised sample is ideal for applying novel asteroseismic inference methodologies such as the dominant-mode-based estimator for the near-core rotation rate in \cite{aerts2025}. Further, it provides a starting point for more traditional asteroseismic analyses for targets of high interest and data quality. 

In this work, we perform manual vetting of the IJ24 sample in order to identify candidate EB pulsators, and use asteroseismically calibrated stellar models from \cite{mombarg2024gaia} (M24) to estimate stellar properties for the sample based on their \gaia\ effective temperature and luminosity estimates. We explore if, and if so how, binarity can result in systematic bias to these results, and confront our mass inferences with 2MASS \citep{skrutskie2006} \ks -magnitude measurements in order to better understand the efficacy of our methodology.

\section{Methodology}
\label{sec:meth}

\subsection{Sample definition}

Beginning with the 69085-strong eclipsing binary sample of IJ24, we first recover IJ24's high-quality sample of over 14377\footnote{The precise number of stars in our sample differs slightly from IJ24 as we filter both duplicate TESS and \gaia\ entries, while IJ24 only filters duplicate TESS entries} stars. This is achieved by demanding that
the automated \starshadow\ \citep{ijspeert2024} eclipse analyses of IJ24 have completed successfully, filtering for duplicate TESS and \gaia\ entries, and removing all binaries with $|e\cos(\omega)|>0.02$ that are consistent with 0 eccentricity to within $3\sigma$. System lumjinosities are calculated using \gaia\ DR3 parallax measurements, while other \gaia\ properties (most notably the effective temperatures) are drawn from \texttt{esp-hs} where available, as it is most suitable for hot stars (7000~K~<~\teff ~<~50000~K, \citealt{gaia2023}), and \texttt{gsp-phot} otherwise.

Using the frequency analysis of IJ24, we filtered for stars which have at least one candidate independent frequency of at least 4~mmag. This threshold was designed to enrich our manually vetted sample with pulsators \citep{gaia2023pulsations}, as well as improve compatibility with existing empirical asteroseismic methods in the literature \citep{hey2024,aerts2025} relying on dominant g-mode oscillations of at least 4~mmag in amplitude. The 4~mmag limit was originally derived from \gaia ~ DR3 time series \citep{gaia2023pulsations}, and so isn't strictly applicable to TESS photometry. For this reason, we only use this limit for the pre-selection phase resulting in our 4000-strong subsample that we subject to manual vetting; it does not form part of the manual vetting criteria. Most of the scientific discussion in this work will focus on this 4000-target subsample, but note that the stellar property inference was conducted on the entire 14377-target sample of IJ24.

\begin{table*}[h!]
\caption{Sample numbers, including overlaps with the automated detection method of IJ24.}
\begin{tabular}{l|lll}
 & Manual inspection & Overlap with IJ24 g-modes & Overlap with IJ24 p-modes \\ \hline
Total & 14377 & 2844 & 669 \\
Not inspected & 10377 & 1142 & 391 \\
Inspected & 4000 & 1702 & 278 \\ \hline
g-mode & 751 & 424 & 32 \\
p-mode & 131 & 34 & 48 \\
Hybrid & 48 & 31 & 16 \\
Non-pulsator & 3070 & 1213 & 182 \\ \hline
Tidal & 1500 & 864 & 164 \\
Tidal variation & 2213 & 716 & 100 \\
Period problem & 237 & 104 & 11 \\
No secondary eclipse & 50 & 18 & 3
\end{tabular}
\tablefoot{Hybrid pulsators are counted as a unique class in this work, while in IJ24 they are the intersection of g-mode and p-mode pulsators. All subsamples referenced in subsequent figures are from the manual classification process.}
\label{tab:luc_compar}
\end{table*}

IJ24 employed an automated detection algorithm demanding multiple independent frequencies in predefined g-mode ($<5$~\perd) or p-mode ($>5$~\perd) regimes. This method identifies 2844 candidate g-mode pulsators and 669 candidate p-mode pulsators, of which 270 are hybrids. Our 4000-binary subsample is more strict in terms of amplitude but less strict in terms of the number of candidate independent frequencies. Detailed information with the overlap with the automated method of IJ24 can be found in Table \ref{tab:luc_compar}. Note that our g-mode and p-mode numbers do not include members of the hybrid classification.

Our 4000-binary subsample of IJ24 was manually inspected in the time and Fourier domains for candidate independent g-mode (751), p-mode (131), or hybrid (48) pulsators. This inspection was done on one year of TESS QLP \citep{huang2020I,huang2020II,kunimoto2021} or SPOC \citep{kunimoto2021} data per target, where the data used was selected based on the following priority ranking from highest to lowest:
\begin{enumerate}
    \item extended mission SPOC data,
    \item nominal mission SPOC data,
    \item extended-mission QLP data,
    \item nominal mission QLP data.
\end{enumerate}
This ranking intended to prioritise the higher quality SPOC time series while available, while also maximally making use of the higher cadence and higher quality TESS extended mission photometry where available.

An additional round of manual classification of these 4000 targets was made for whether tidal variability (including but not limited to oscillatory behaviour) was present in the light curve, whether there was an issue with the period calculation (typically either half or double the true orbital period), or whether there was no secondary detected.

\subsection{Stellar parameter estimation}

Using conditional normalising flows (CNFs, \citealt{winkler2019,hon2024}) trained on the M24 stellar evolution models, we estimate masses (\M), convective core masses (\Mc), radii ($R$), and central H fractions (\xcx) for the 12508 targets with \gaia\ effective temperatures and luminosities. This estimation is based on the previously described \gaia\ \logteff\ and \logl\ estimates, and is performed for the three metallicities ($Z=0.014, Z=0.008, Z=0.0045$) available, marginalising over the initial (rigid) rotations (\omegai) and values for the exponential overshoot (\fov) varied in the M24 grids.

To arrive at maximum likelihood estimates for each parameter and its uncertainty, we first use M24's CNFs to generate 2.5 million samples spanning stellar masses from 1.3~M\solar\ to 9~M\solar, initial rigid rotation frequencies \omegai=$(\Omega_{\rm surf}/\Omega_{\rm crit})_{\rm ZAMS}$ from 0.05 to 0.55, and values of the exponential diffusive mixing coefficient \fov\ 
 from 0.005 to 0.025. \fov\ is used to calculate diffusivity coefficient in the overshoot region
 
\begin{equation*}
D_{\rm ov} = D_0 \cdot {\rm exp}\Big(\frac{-2\cdot z}{(f_{\rm ov}\cdot H_{\rm p}}\Big),
\end{equation*}

where $z$ is the height above the convective boundary, $H_{\rm p}$ is the pressure scale height at the convective boundary, and $D_0$ is the diffusivity coefficient at the edge of the convective boundary \citep{freytag1996,herwig2000,mombarg2024}. Following M24, we uniformly resample the evolution models in \xc\ to avoid bias towards the zero-age main sequence (ZAMS) and terminal-age main sequence (TAMS) caused by the numerics of the evolution models. For all targets, a kernel density estimate (KDE) is then performed to obtain the maximum likelihood estimate and its uncertainty by uniformly distributing 800 points in each stellar parameter (\M, \Mc, \xcx, \logr, \omegai, and \fov).

\section{Results}
\label{sec:res}

In section \ref{sec:res:fid}, we present the results of applying our methodology using `out-of-box' \gaia\ properties. This is equivalent to treating the \gaia\ \logteff\ and \logl\ measurements as the properties of the primary. We also implicitly assume that the primary is the pulsator, which will not always be the case. We assess the effect of binary-induced inaccuracy in these parameters on the inferred stellar properties later in this section. 

\subsection{Basic stellar and orbital properties}
\label{sec:res:fid}

\begin{figure*}
\centering
\begin{subfigure}{0.99\columnwidth}
\centering
\includegraphics[width=0.95\columnwidth]{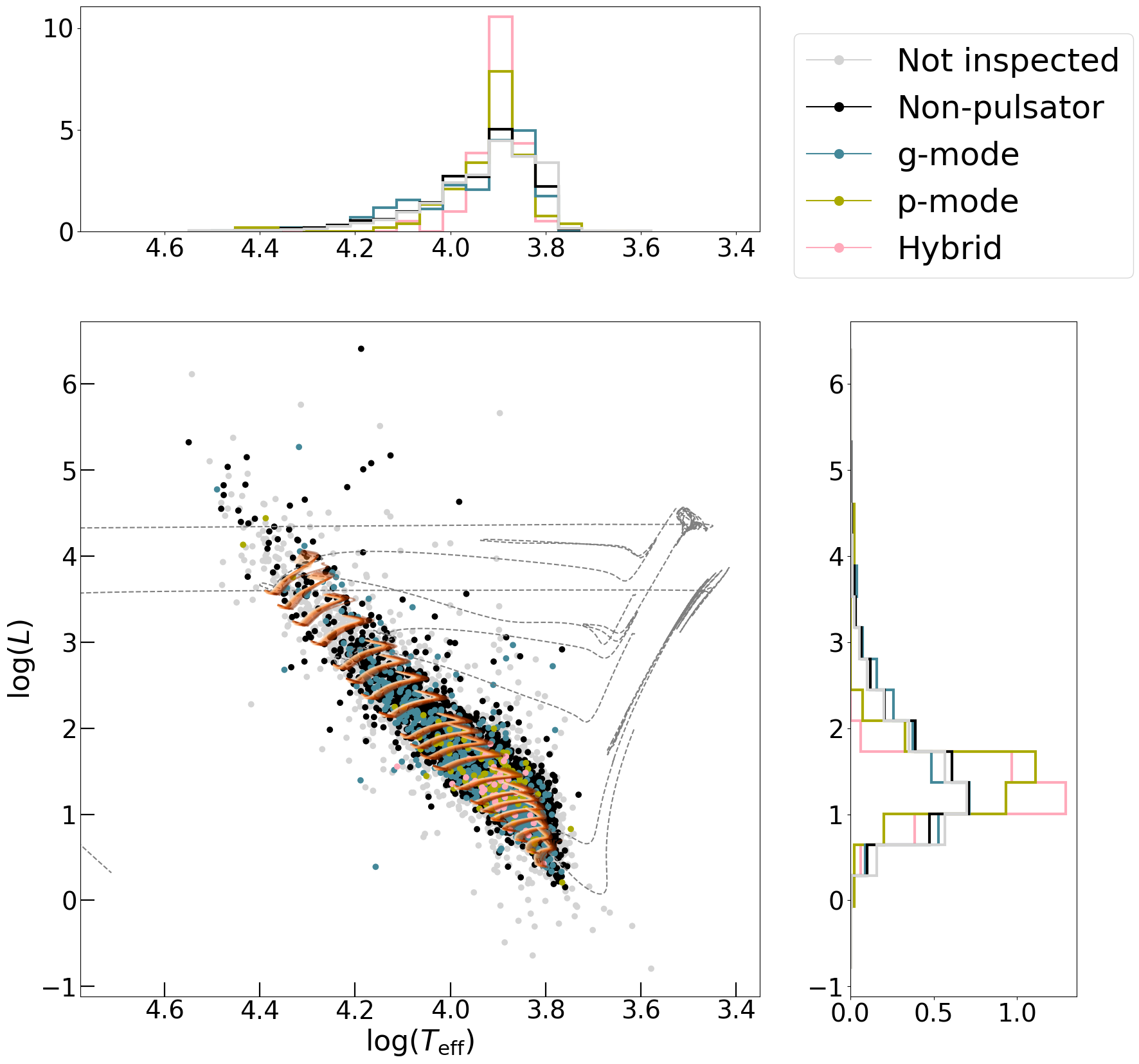}
\end{subfigure}%
\begin{subfigure}{0.99\columnwidth}
\centering
\includegraphics[width=0.95\columnwidth]{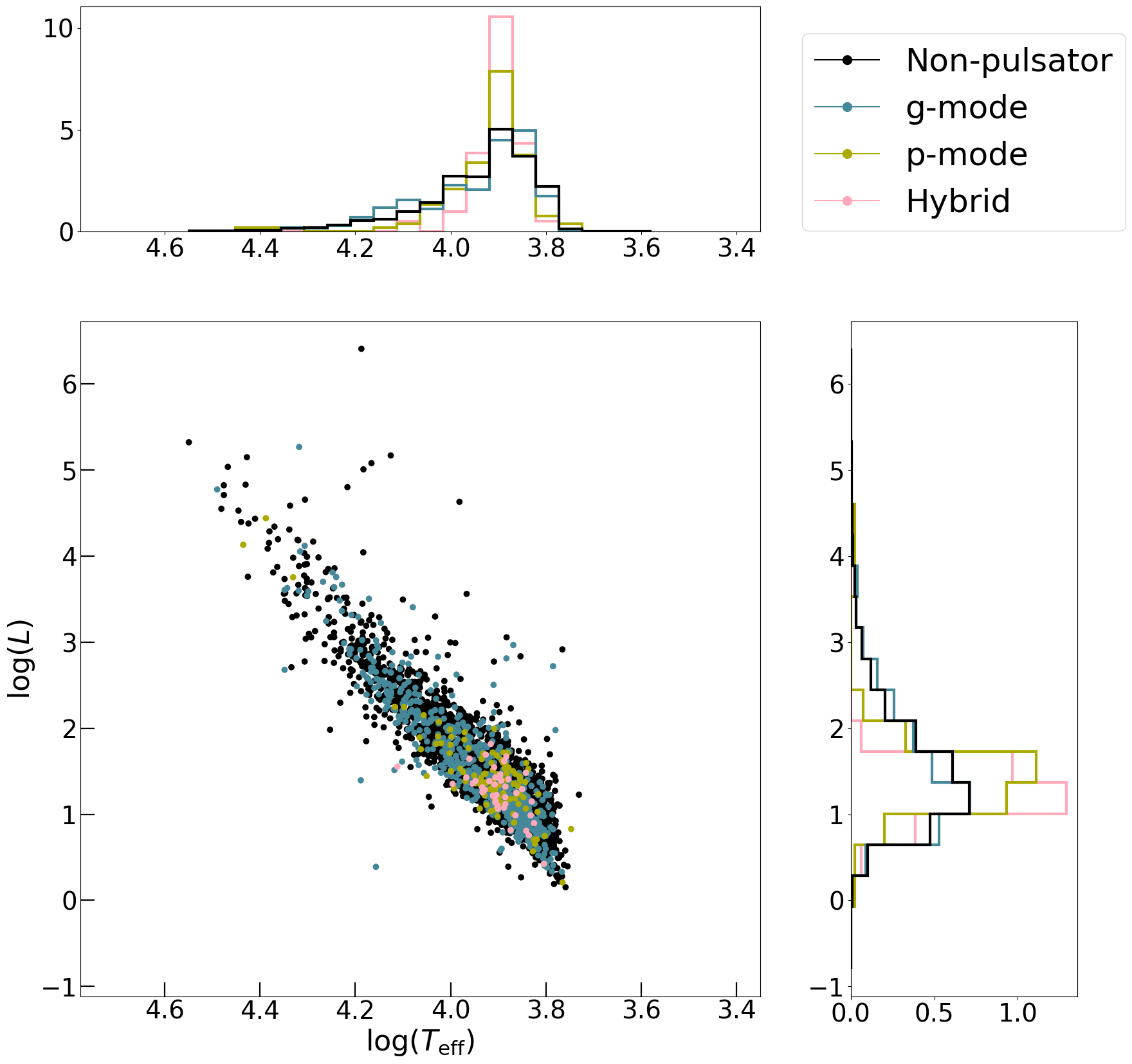}
\end{subfigure}

\begin{subfigure}{0.99\columnwidth}
\centering
\includegraphics[width=0.95\columnwidth]{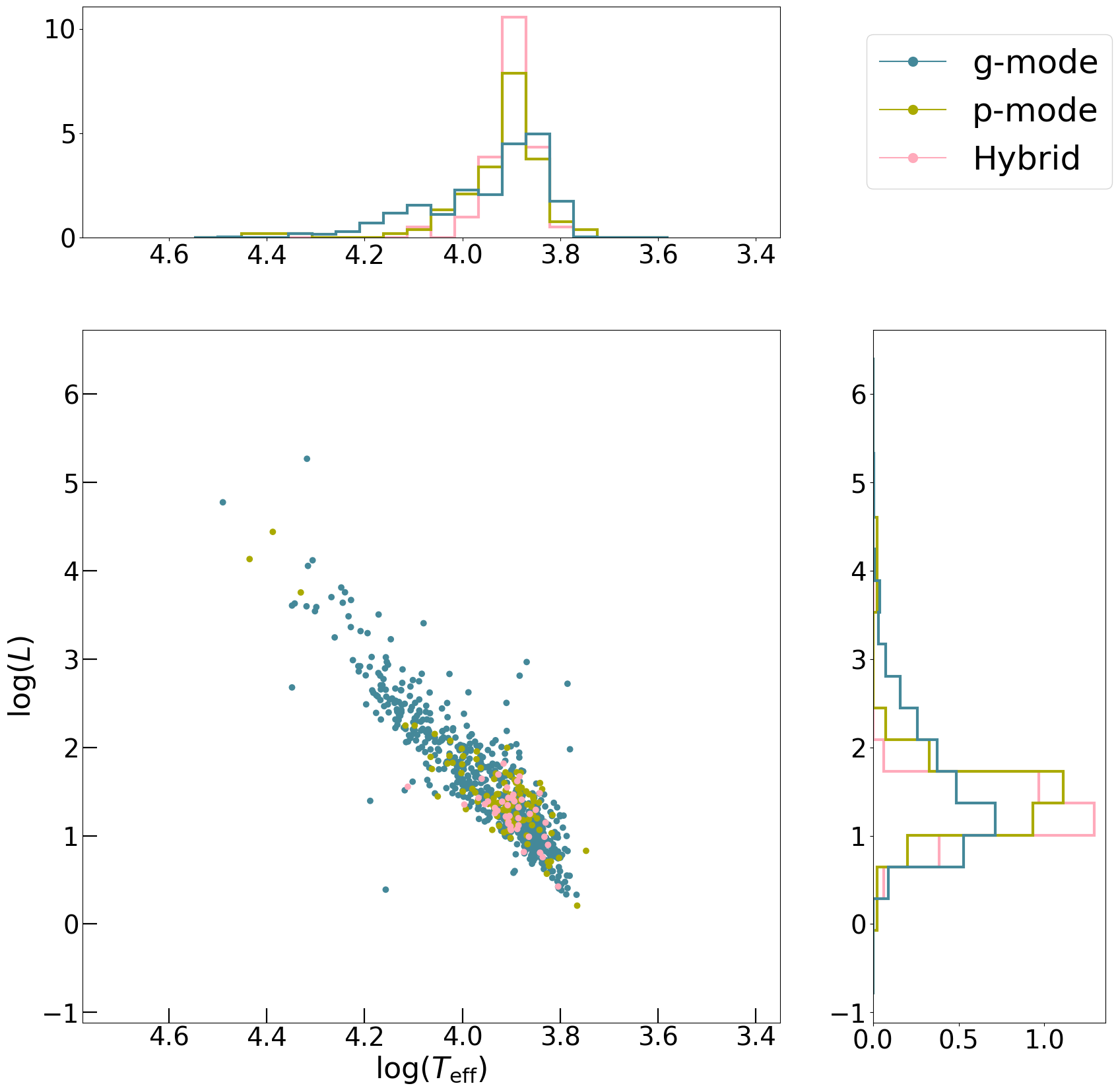}
\end{subfigure}%
\begin{subfigure}{0.99\columnwidth}
\centering
\includegraphics[width=0.95\columnwidth]{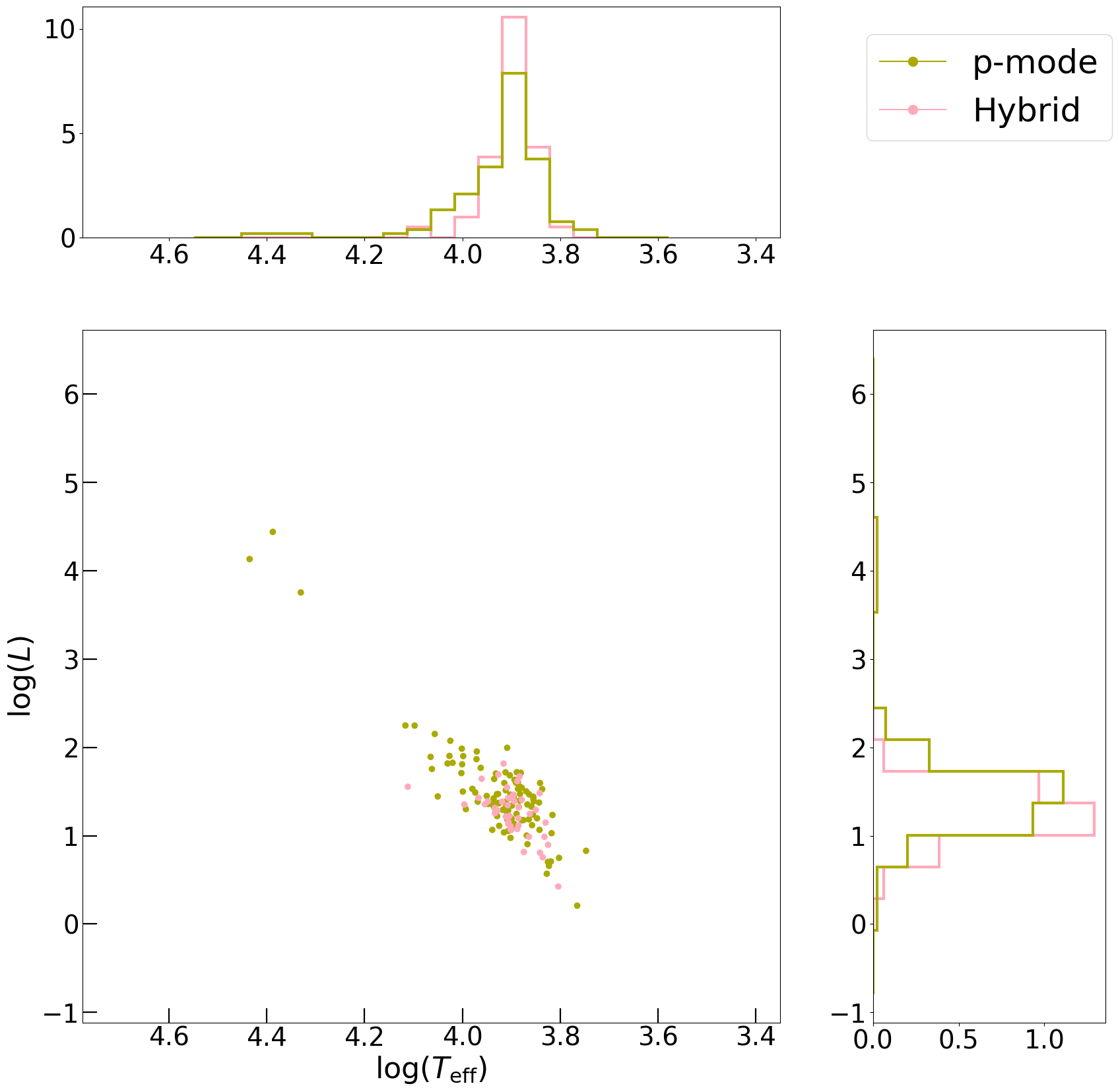}
\end{subfigure}

\caption{HRD of all systems with \gaia\ measurements for the effective temperature and luminosity, with adjoining histograms showing the density distributions for luminosity and temperature for each subclass for each panel. The upper left panel also shows solar metallicity (\Z=0.014) evolutionary tracks from 1.2 to 9 M\solar\ from M24 (solid red/yellow) computed using an exponentially-decaying core-boundary mixing prescription (\fov\ of 0.015), along with full 1.3~M\solar, 5~M\solar, and 9~M\solar\ MIST tracks (grey dashed, \citealt{choi2016}). The coloured variation in each of the M24 tracks shows the effect of varying the rotation from 5-55\% (red-yellow) of the initial Keplerian critical rotation frequency.}
\label{fig:hrdiag}
\end{figure*}

The distribution of our stars across the Hertzsprung-Russel diagram (HRD) is shown in Fig. \ref{fig:hrdiag}. The upper left panel also shows solar metallicity (\Z=0.014) evolutionary tracks from 1.2~\M\solar\ to 9~M\solar\ from M24. P-values for 2-sample Kolmogorov-Smirnov (KS) tests \citep{massey1951kolmogorov,hodges1958} applied to all subsample pairings can be found in Table \ref{tab:kstest}. Unless otherwise stated, the p-values discussed in the main text always adopt the null hypothesis that the relevant pairing of subsamples is drawn from the same underlying distribution; hence, a high p-value implies a high degree of similarity between the distributions, while a low value indicates significant discrepancies. The lower panels highlight the distribution of the g-mode, p-mode, and hybrid pulsators.

By design, our population is dominated by main sequence stars (see IJ24 for details). For most of the sample the agreement with the stellar tracks of M24 is good, although there is some scatter around the theoretical stellar tracks and not all can be explained by metallicity effects. Further, the limited mass range of the M24 grid (1.2-9 M\solar) results in domain issues at both the high mass and low mass end of the HRD that, as we will see later, lead to numerical pile-ups around these masses.

The distribution of the g-mode pulsators is relatively continuous across the HRD; while the 2-sample KS test probability with the non-pulsator class is low (0.045 for \logteff), it is also significantly higher than that of either the p-mode or hybrid samples (0.006 and 0.001, respectively). Qualitatively, there is a slight bimodality in \logteff\ for the g-mode stars that is absent for the non-pulsators. This may be the signature of the \gdor\ \citep{guzik2000, dupret2005,uytterhoeven2011,bouabid2013} and slowly-pulsating B-star (SPB, \citealt{waelkens1991,dziembowski1993,gautschy1993,miglio2007}) instability regions. However, we still must conclude that the distribution of g-mode pulsators across the HRD appears more similar to the distribution of non-pulsators than we would expect for a (pure) sample of g-mode pulsators confined to the traditional \gdor, SPB, and \bcep\ \citep{dziembowski1993bcep,stankov2005} instability regions. Pulsators outside these regions have previously been found by combining \gaia\ astrometry with \kepler\ or TESS photometry \citep{gaia2023,balona2023,hey2024,fritzewski2024ubc1,mombarg2024}. It is possbile that this blurring of the traditional instability regions could be due to fundamental limitations to the precision and accuracy of \gaia\ \teff\ estimates in the SPB regime. However, it is at least as plausible that this phenomena is due to the physical limitations of our excitation models, which are missing proper treatments for physics such as stellar rotation and tides \cite{aerts2024}.

In the context of our eclipsing binary sample, a key question is whether tidal interactions \citep{fuller2012,fuller2017,guo2021,aerts2024} are a dominant force for the g-mode pulsator classes becoming blurred. Alternative explanations include non-binary related missing physics in mode excitation models, such as rotation, but also the possibility of a relatively high level of contamination by rotational variables, which, even with manual vetting, can be difficult to distinguish from genuine g-mode pulsators in TESS photometry \citep{audenaert2021,hey2024}.

The hybrid pulsators have broadly similar distributions to the p-mode pulsators in effective temperature (p-value~=~0.29) and luminosity (p-value~=~0.048). They peak at slightly lower values, and do not extend to as high \logteff\ and \logl\ regions as pure p-mode pulsators. There are also a few examples of hot, bright p-mode pulsators in the \bcep\ region. With this in mind, we may conclude that the sample of p-mode and hybrid pulsators appear to be well-behaved in terms of their position on the HRD.

\begin{figure}
\centering
\begin{subfigure}{0.99\columnwidth}
\centering
\includegraphics[width=0.91\columnwidth]{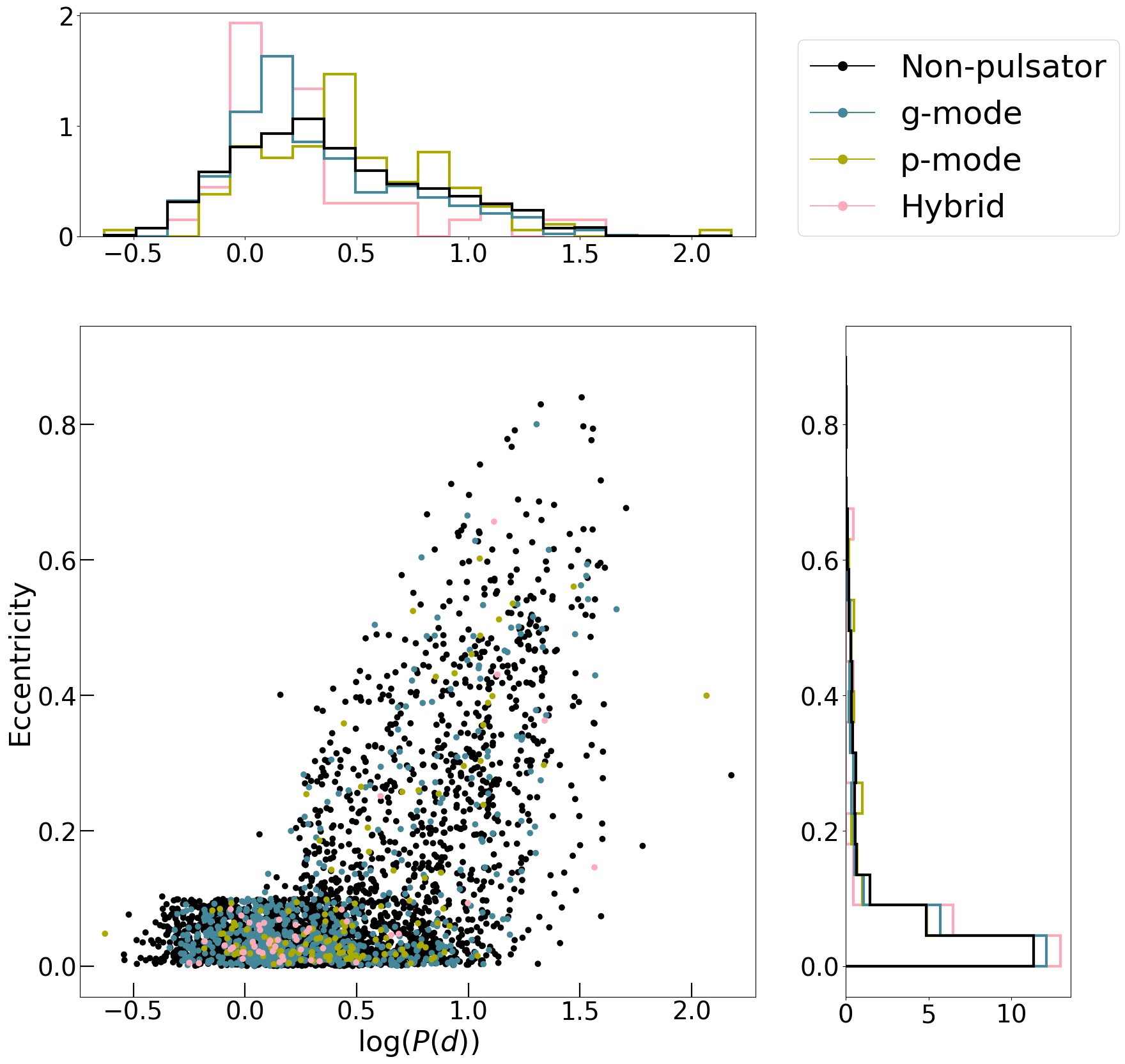}
\end{subfigure}

\begin{subfigure}{0.99\columnwidth}
\centering
\includegraphics[width=0.91\columnwidth]{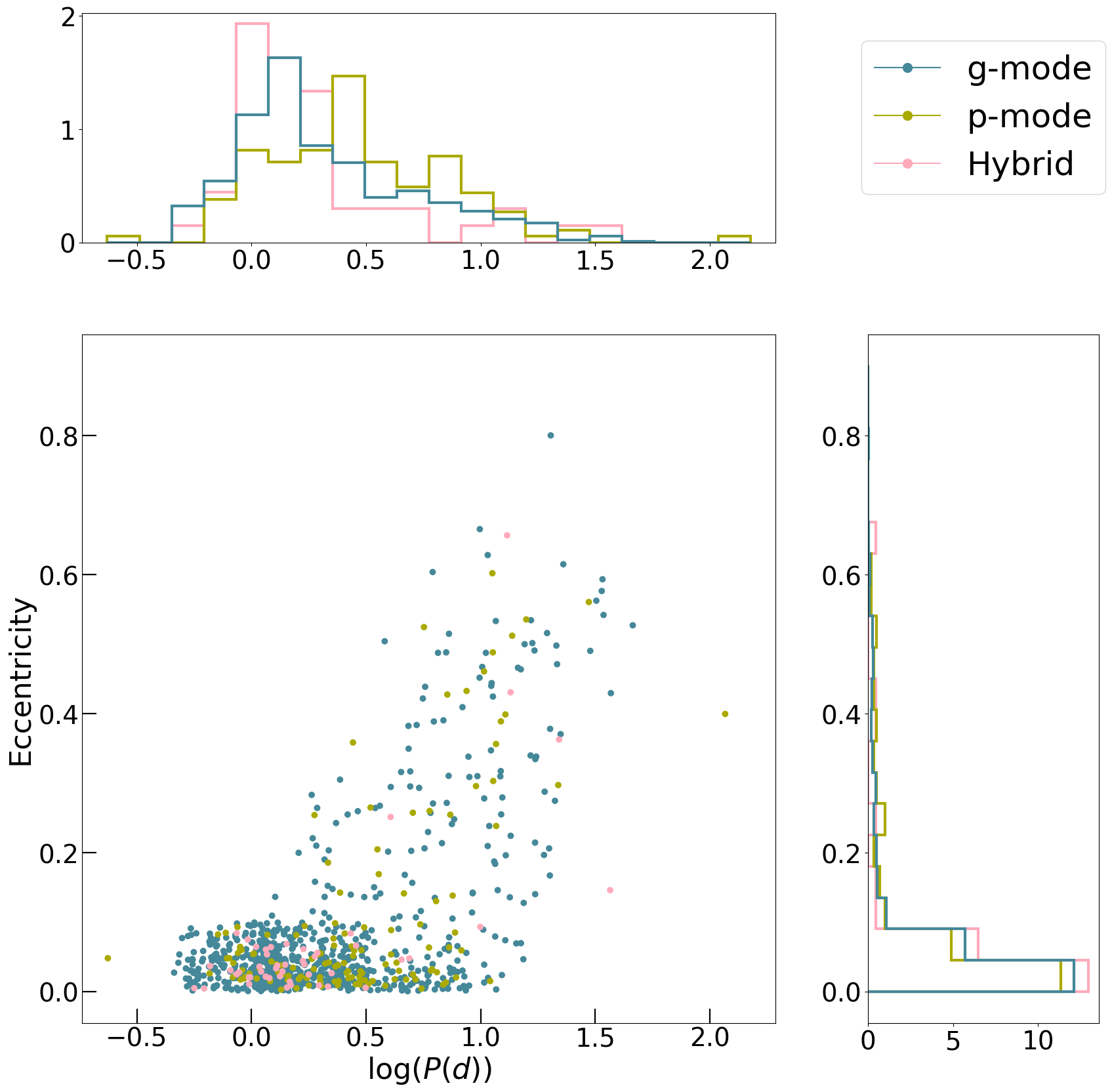}
\end{subfigure}

\begin{subfigure}{0.99\columnwidth}
\centering
\includegraphics[width=0.91\columnwidth]{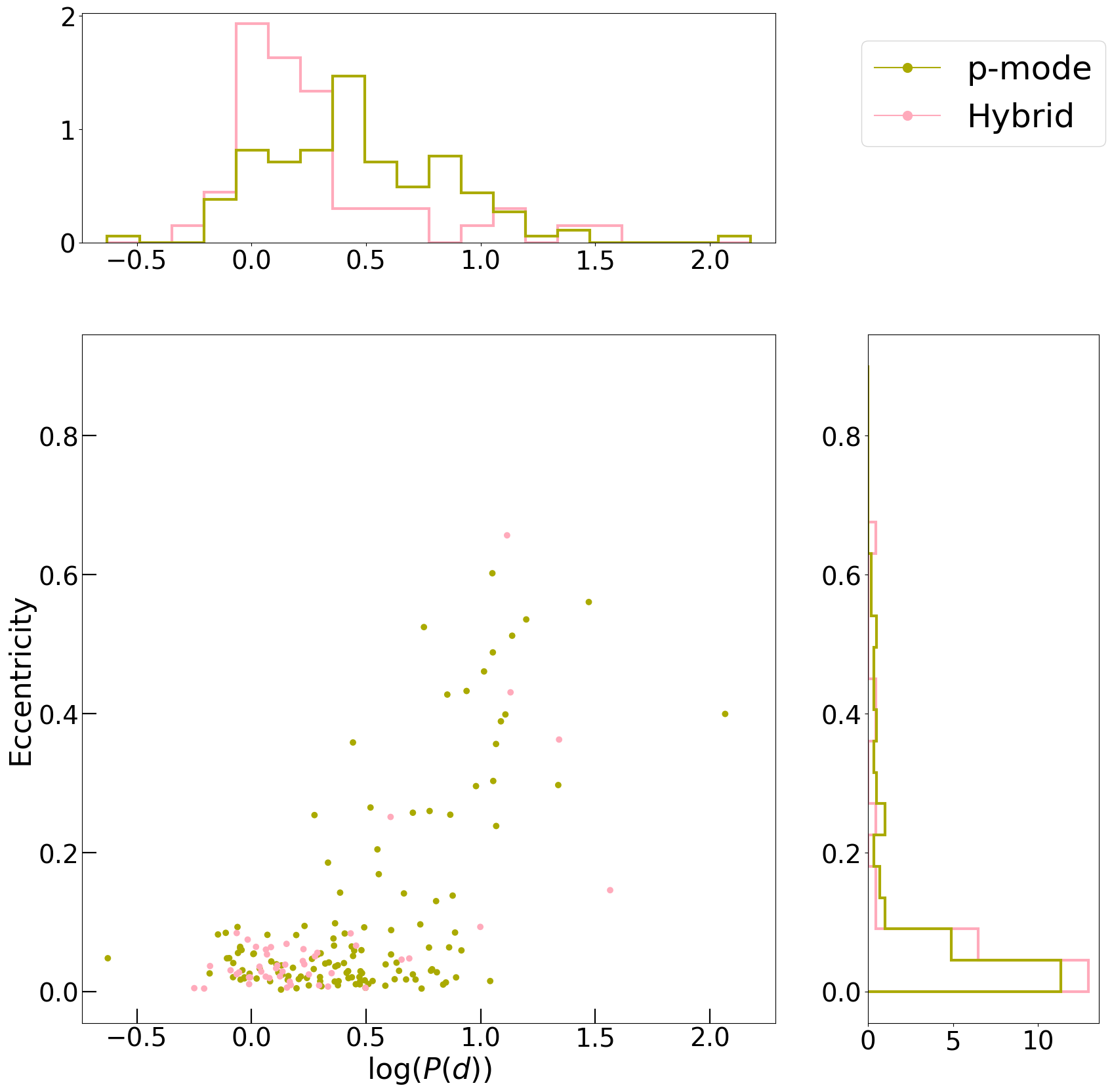}
\end{subfigure}%

\caption{$\log$~P vs eccentricity for all inspected stars.}
\label{fig:period_ecc}
\end{figure}

The period-eccentricity distribution of the 4000-target vetted sample is shown in Fig. \ref{fig:period_ecc}. G-mode pulsators, including hybrids, are preferentially found at short orbital periods compared to both the non-pulsator subclass and the p-mode pulsators. This contrasts with Fig. \ref{fig:hrdiag}, which showed that in terms of stellar surface properties, the distribution of the hybrids and p-mode were far more similar than that of hybrids and g-modes. The p-mode pulsators peak at longer orbital periods and generally have a flatter distribution that is more consistent with the non-pulsator subclass (p-value~=~\timestento{9.34}{-3})) than either the g-mode (p-value~=~\timestento{2.73}{-6}) or hybrid pulsators (p-value~=~\timestento{2.19}{-5})). This preference for short-period, circular binaries is consistent with increased tidal efficiency in the presence of g-mode pulsations, which would lead to more rapid spin-orbit coupling and circularisation \citep{rogers2013,ogilvie2014}.

\begin{figure*}
\centering
\begin{subfigure}{0.99\columnwidth}
\centering
\includegraphics[width=0.95\columnwidth]{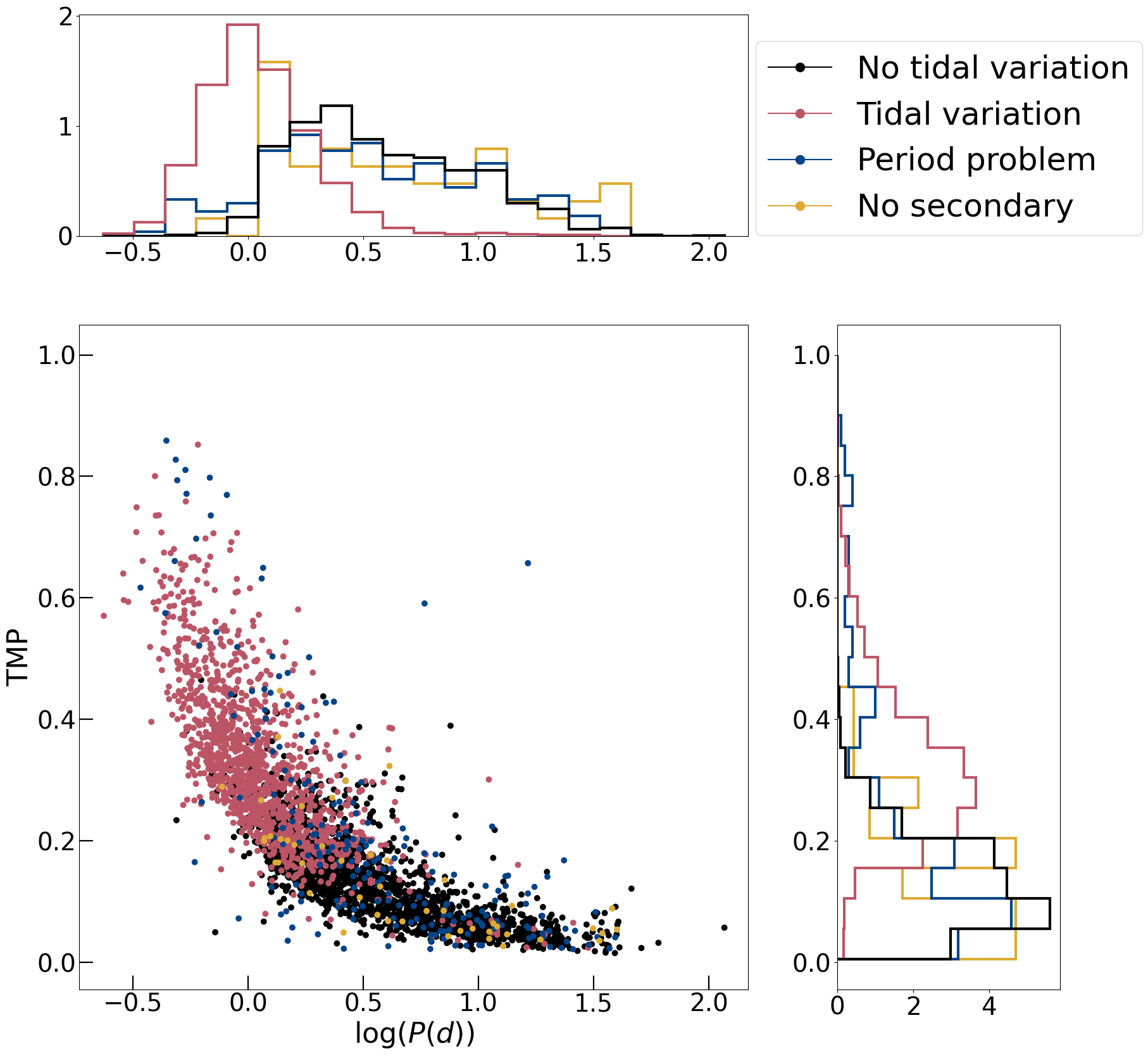}
\end{subfigure}%
\begin{subfigure}{0.99\columnwidth}
\centering
\includegraphics[width=0.95\columnwidth]{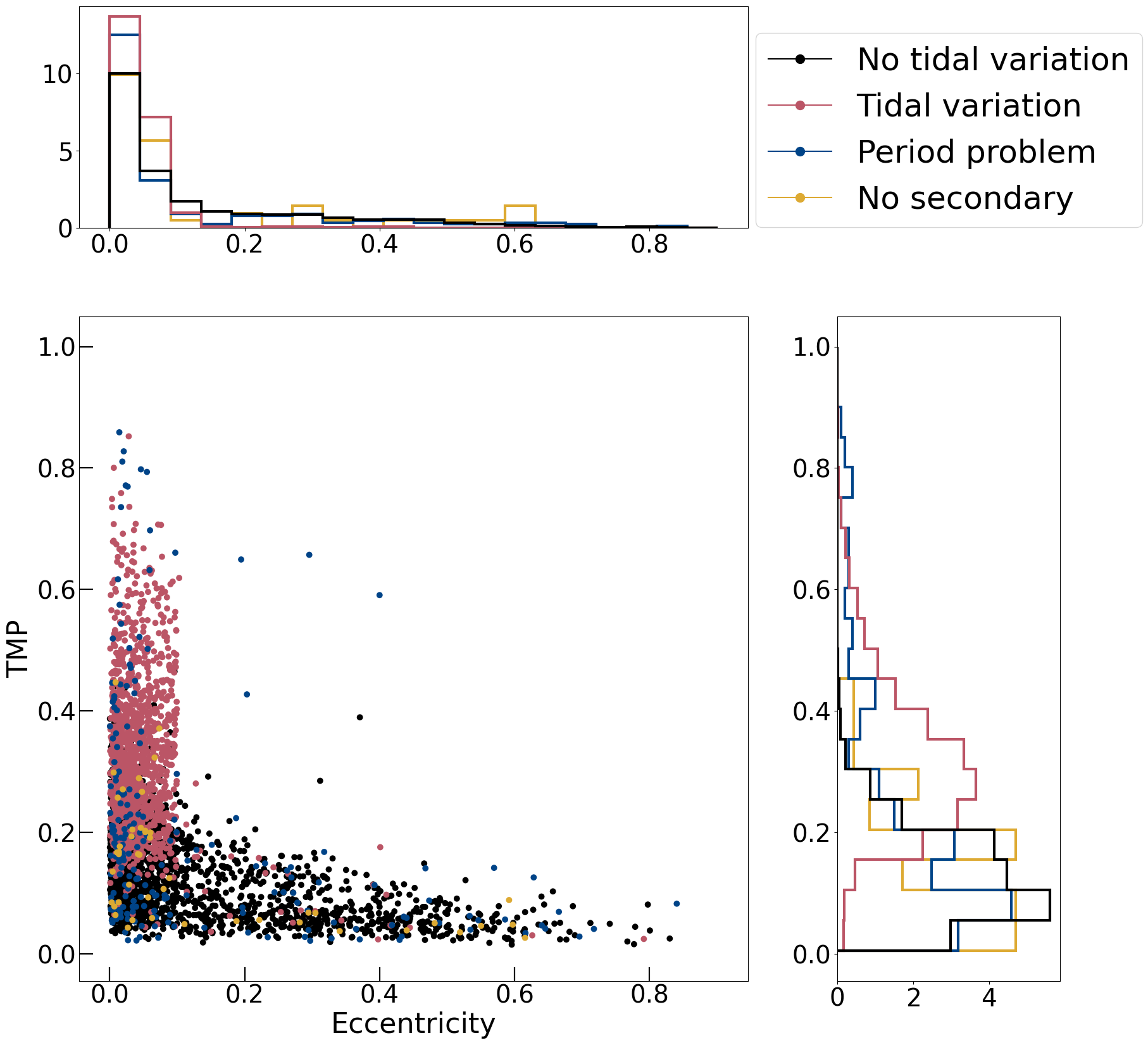}
\end{subfigure}

\begin{subfigure}{0.99\columnwidth}
\centering
\includegraphics[width=0.95\columnwidth]{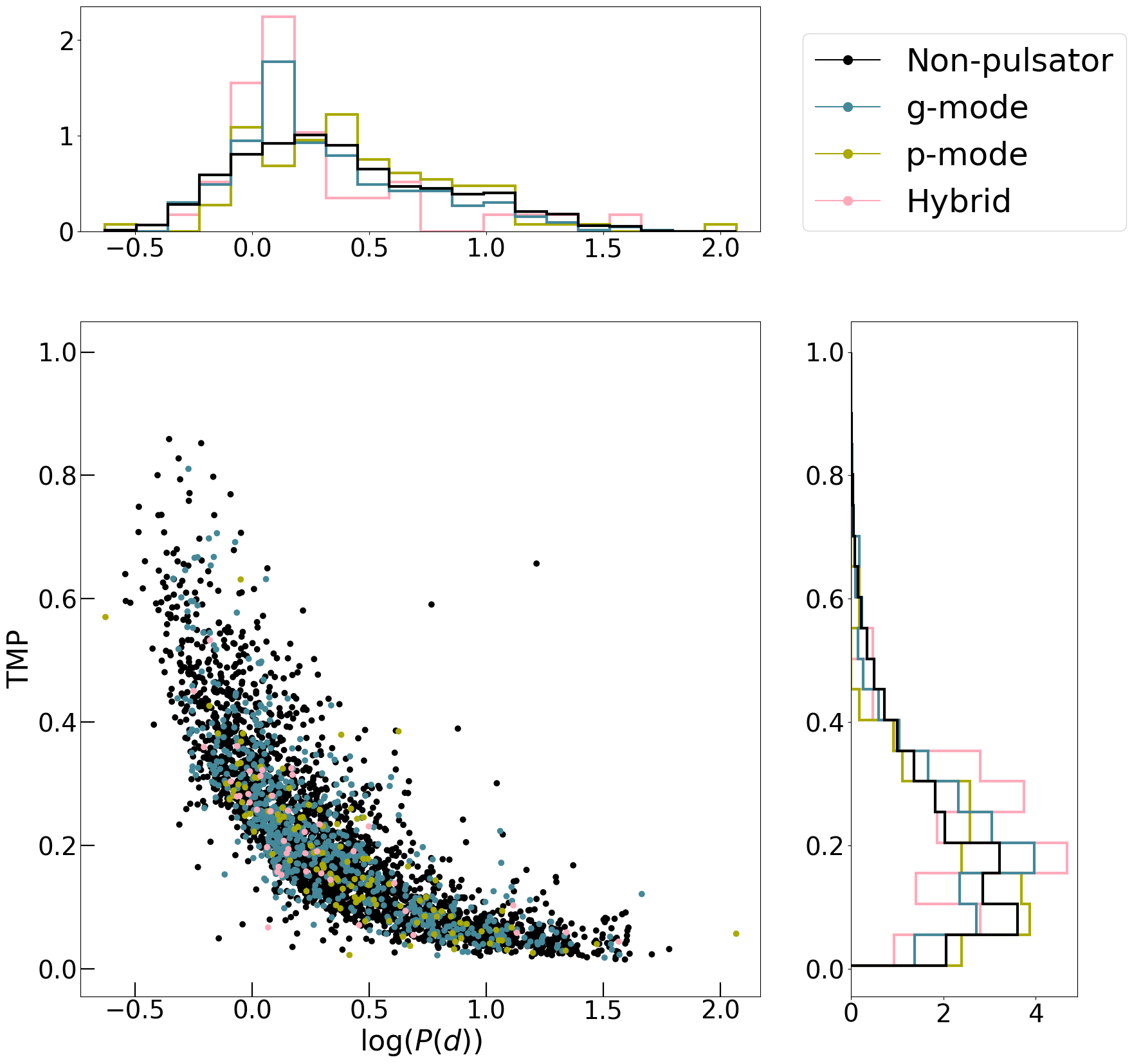}
\end{subfigure}%
\begin{subfigure}{0.99\columnwidth}
\centering
\includegraphics[width=0.95\columnwidth]{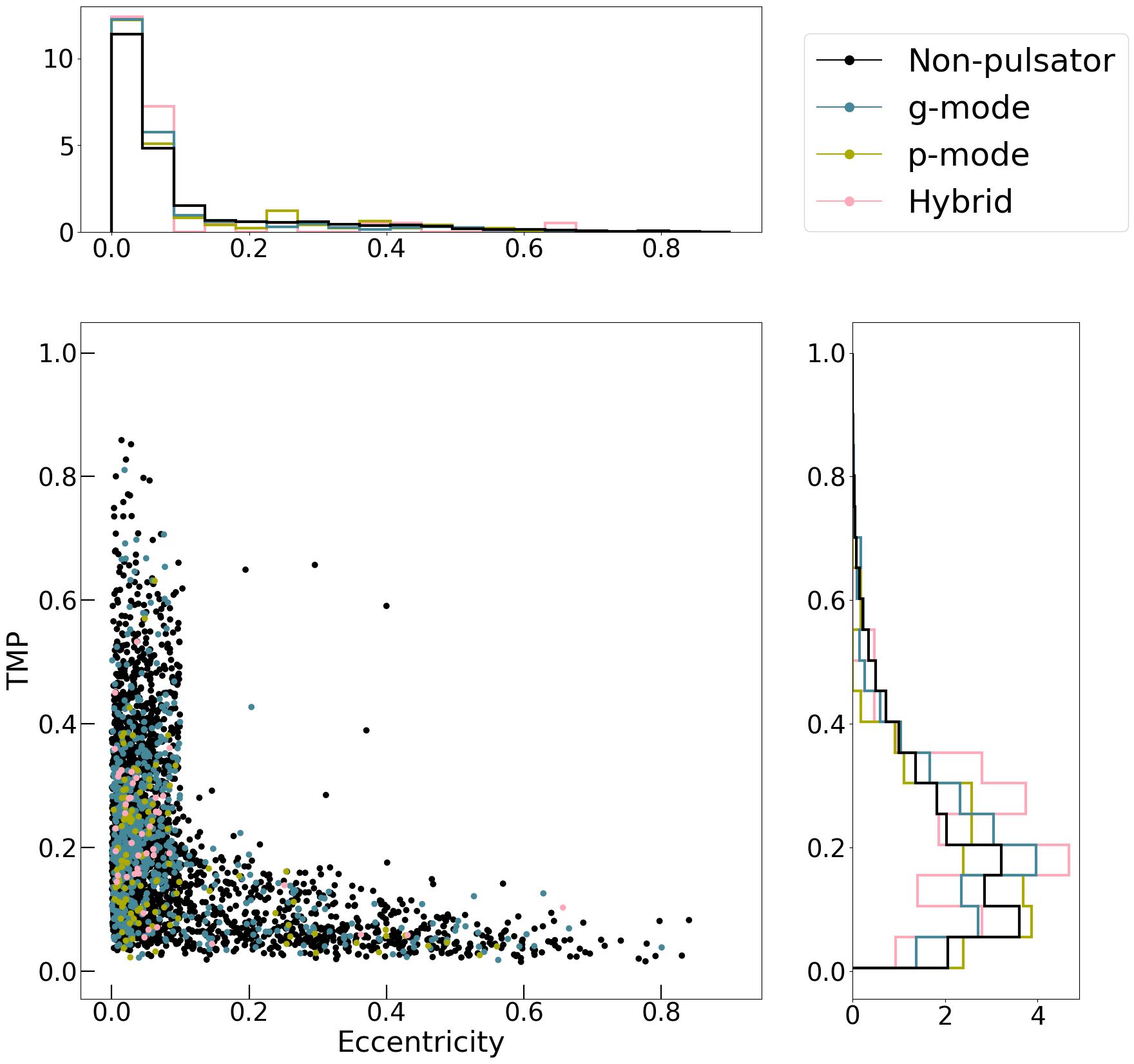}
\end{subfigure}
\caption{Orbital period (left) and eccentricity (right) versus our tidal morphology parameter (TMP, see equation \ref{eq:tmp}). Upper panels show the tidal-variability subclasses, while the bottom panels show the pulsator subclasses introduced in Fig. \ref{fig:hrdiag}}.
\label{fig:morph}
\end{figure*}

In order to directly link pulsator-properties with tidal effects, we tested several simple tidal-morphology parameters (TMPs) based on the amplitudes of the orbital-harmonic series against our manual classification of `tidal variables' (see Table \ref{tab:luc_compar}, which we took as stars showing sort of tidally related variability in their light curves. The most successful of these parameters was calculated by taking the maximum amplitude of the harmonic series and dividing it by the sum of the amplitudes of all orbital harmonics:
\begin{equation}
{\rm TMP}~=~max(A_{\rm orbital\ harmonic})/\Sigma(A_{\rm orbital\ harmonic}).
\label{eq:tmp}
\end{equation}
Fig. \ref{fig:morph} shows the behaviour of this TMP under the lenses of both tidal variability and pulsation subclass, and we note that it is significantly better at separating our manually identified tidal variables from the non-tidal variable class than only using the orbital period.

The top panels of Fig. \ref{fig:morph} illustrate the efficacy of our tidal morphology parameter. At values over 0.3, we have an almost entirely pure sample of targets showing noticeable tidal variability. Approximately 40\% of the targets manually identified as displaying some degree of tidal variability have TMPs below 0.3, with significant overlap with targets not exhibiting tidal variability. Above a value of 0.2 almost all tidal variables are captured at the cost of reducing purity to 80-90\%. Considering the distributions of tidal variability in period and eccentricity space, we note that our manually identified tidal variables are almost exclusively found in close, circular systems. The previously mentioned tidal-morphology cut-off of 0.2 coincides almost exactly with the onset of significant eccentricity in our systems, and more approximately to the longer orbital period systems. 
{As a final note on the performance of the TMP parameter, we also checked its independence from the orbital inclination angle $i$ (Fig. \ref{fig:i_morph}). While the TMP reduces slightly as the viewing angle approaches edge-on, this effect is weak in systems with the moderate-high TMP values that are indicative of tidal variability.}

Comparing the upper and lower panels of Fig. \ref{fig:morph}, there does not appear to be a strong correlation between tidal variability and the g-mode pulsators. This is not consistent with tidal effects dominantly affecting the excitation of g-modes in our sample. If this were the case, we would expect more g-mode pulsators in systems where tidal variability is the strongest. Instead, the g-mode pulsators are distributed relatively evenly between low and moderate values (0-0.4) of our tidal morphology parameter. Comparing the different pulsator classes (see Fig. \ref{fig:puls_morph}), we note that the g-mode pulsators prefer higher values of the tidal-morphology parameter than the p-mode pulsators (p-value~=~0.004), while the hybrids closely match the g-mode distribution (p-value~=~0.73, compared to a p-mode | hybrid p-value of 0.03). While this does constitute evidence for g-mode pulsators being affected by tides more than targets without g-mode pulsations, the effect appears to be small, and insufficient to explain the continuous distribution of g-mode pulsators across the HRD found in this and other works \citep{gaia2023pulsations,mombarg2024gaia,hey2024}.

It is possible that some of the most strongly tidally affected systems possess induced oscillations that escaped detection. Our methodology is not ideal for the detection of tidally induced pulsations, where the pulsations are perfectly synchronous in frequency with the orbital frequency. Because we rely on harmonic models for the eclipse, such pulsation frequencies are highly vulnerable to being removed during eclipse identification or being confused with an artifact of an imperfectly identified orbital harmonic. With this caveat, even small perturbations from the eclipse orbital frequencies can result in a detectable, extractable oscillation frequency under this methodology \citep{kemp2024eclipse,kemp2025_k415}. 

\subsection{Inferred properties}
\label{sec:res:inf}

\begin{figure}
    \centering
    \includegraphics[width=0.99\columnwidth]{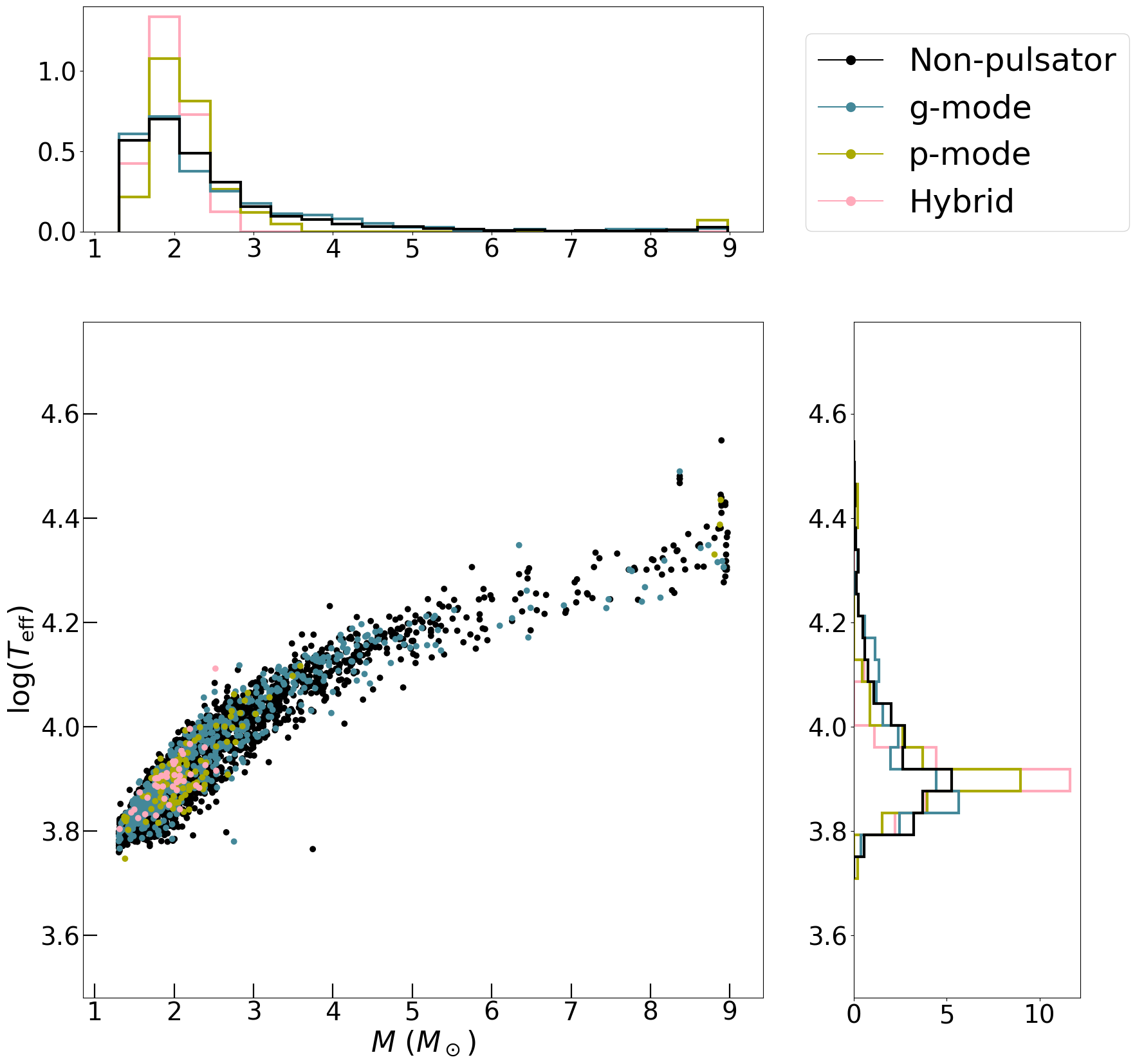}
    \caption{Mass vs temperature. Mass is inferred using M24's \Z=0.014 stellar grid using \gaia\ \logteff\ and \logl\ estimates.}
    \label{fig:mass_temp}
\end{figure}

Fig. \ref{fig:mass_temp} shows the stellar masses inferred from \gaia\ \logteff\ and \logl\ estimates using the stellar grids of M24 (see Section \ref{sec:meth}) plotted against the \gaia\ effective temperatures. As expected, and confirmed via a multivariate linear regression (MLR) based importance analysis (see Fig. \ref{fig:mlr}), we find that the inferred mass is highly correlated with the effective temperature. The anticipated pile-ups caused by domain issues with the M24 grid are clearly visible at low and high masses. Targets with mass estimates less than 1.3~M\solar\ or greater than 8.5~M\solar\ should be treated with particular caution.

The distributions of the inferred pulsator masses appear reasonable in the context of the previously discussed \logteff\ distribution. The hybrid pulsators are clumped around 2~M\solar, with the pure p-mode pulsators peaking at slightly higher masses. The pure g-mode pulsators peak in mass similarly to the hybrid pulsators, with a long high-mass `tail' that extends to the highest masses. We conclude, in the context of the provided \logteff\ measurements, that the behaviour of the inference methods appears sensible; the close agreement between our expectations for the hybrid and p-mode pulsator masses and their inferred masses is particularly reassuring.

\begin{figure*}
\centering
\begin{subfigure}{0.99\columnwidth}
\centering
\includegraphics[width=0.95\columnwidth]{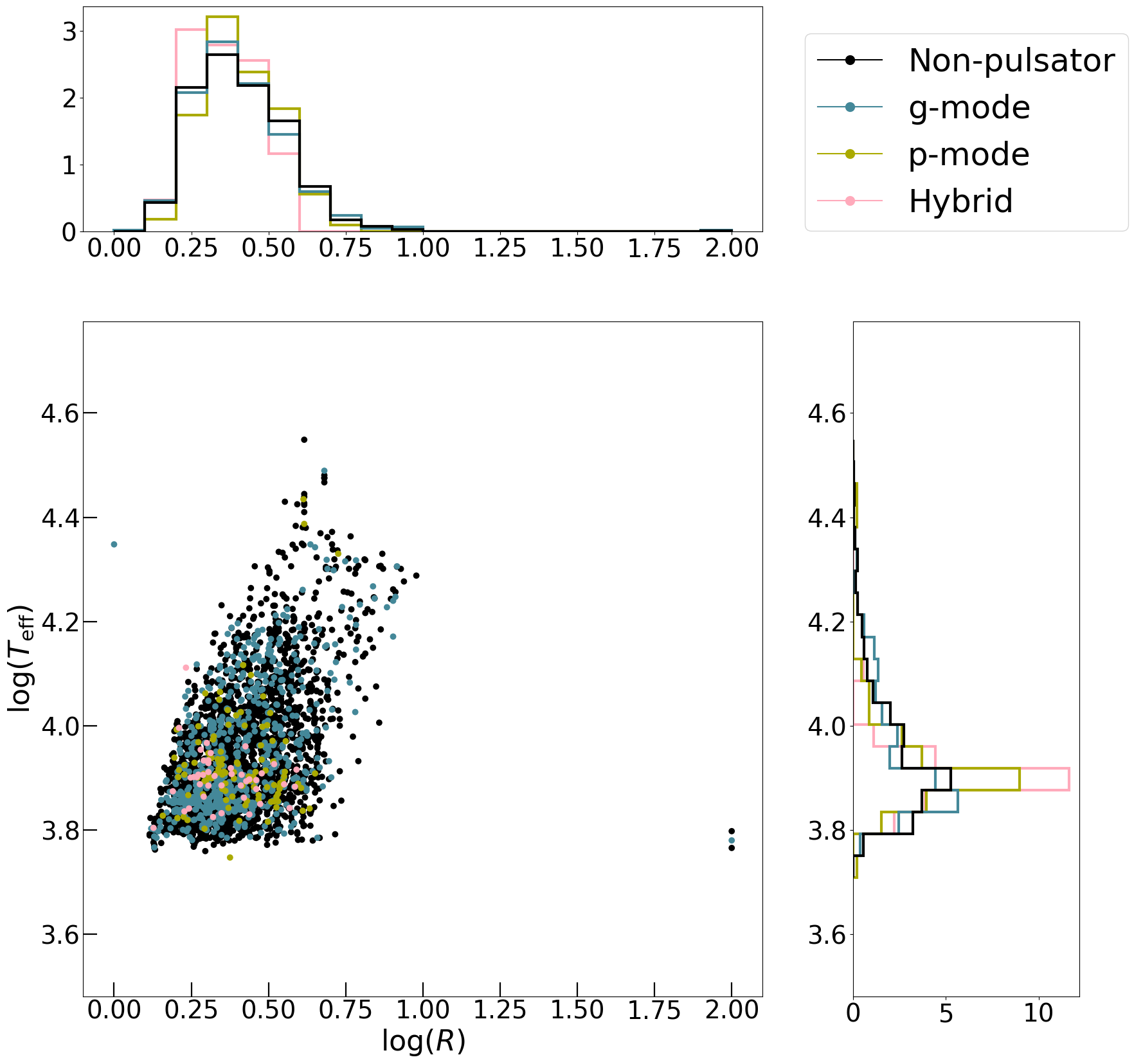}
\end{subfigure}%
\begin{subfigure}{0.99\columnwidth}
\centering
\includegraphics[width=0.95\columnwidth]{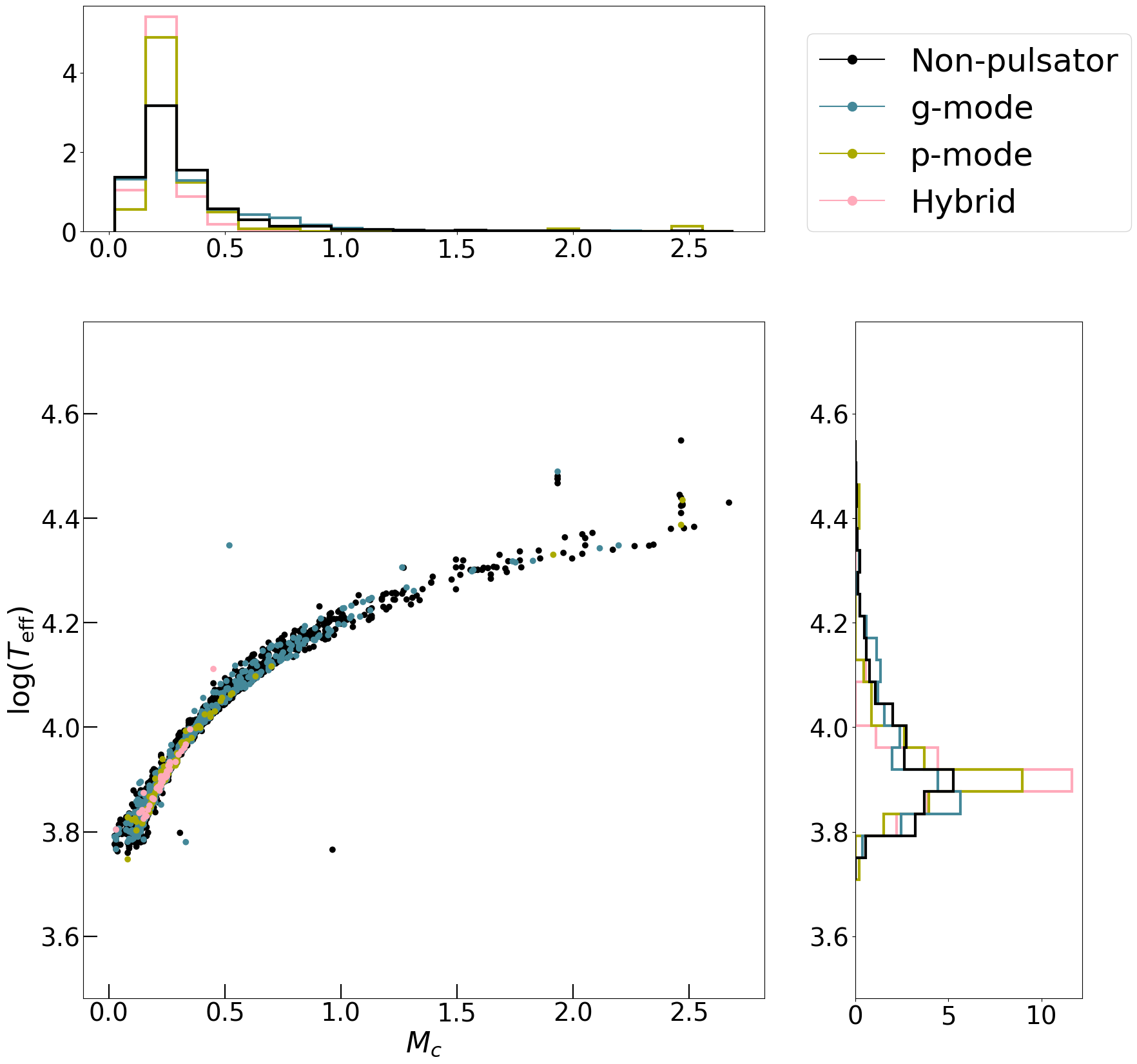}
\end{subfigure}

\begin{subfigure}{0.99\columnwidth}
\centering
\includegraphics[width=0.95\columnwidth]{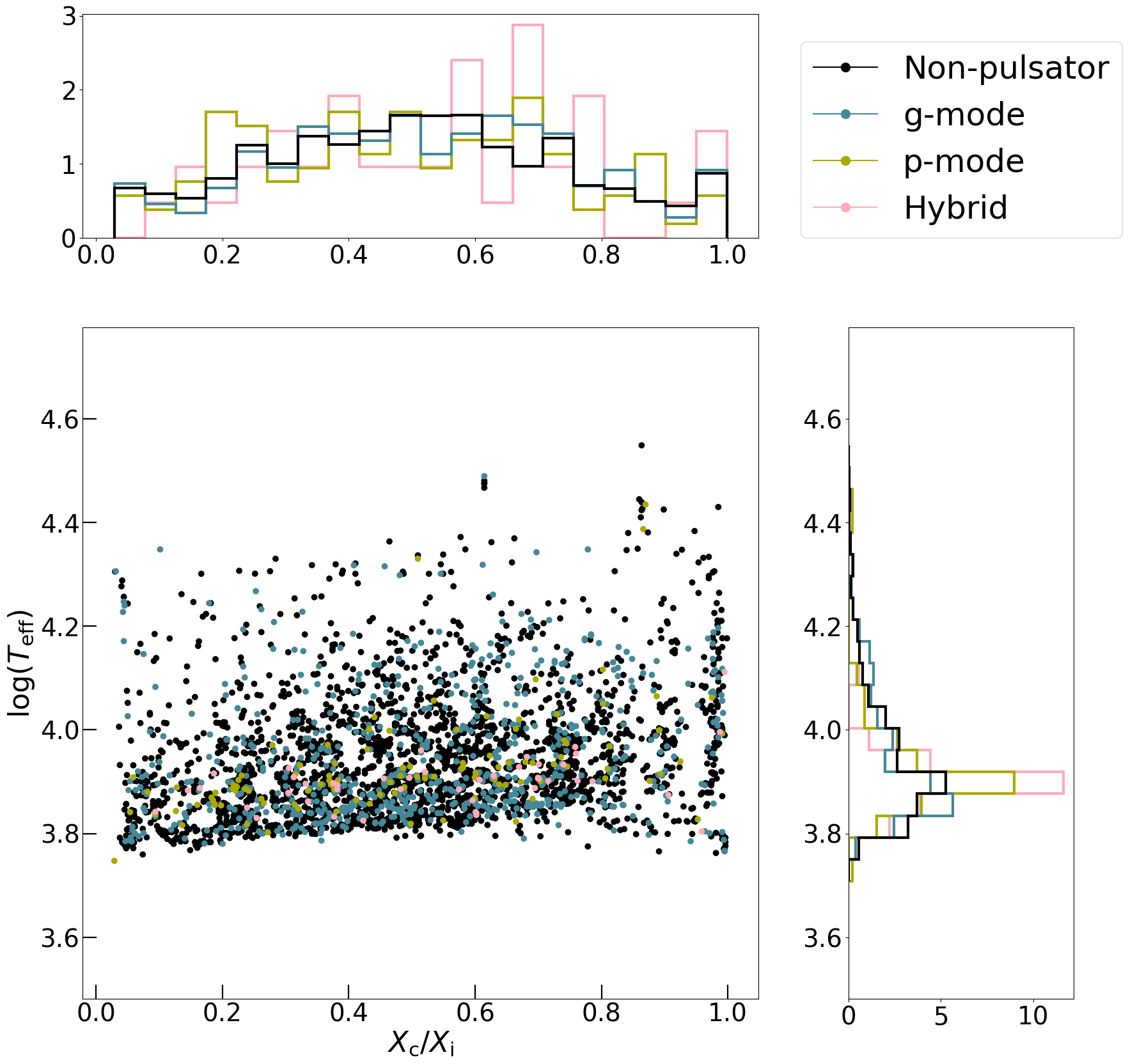}
\end{subfigure}%
\begin{subfigure}{0.99\columnwidth}
\centering
\includegraphics[width=0.95\columnwidth]{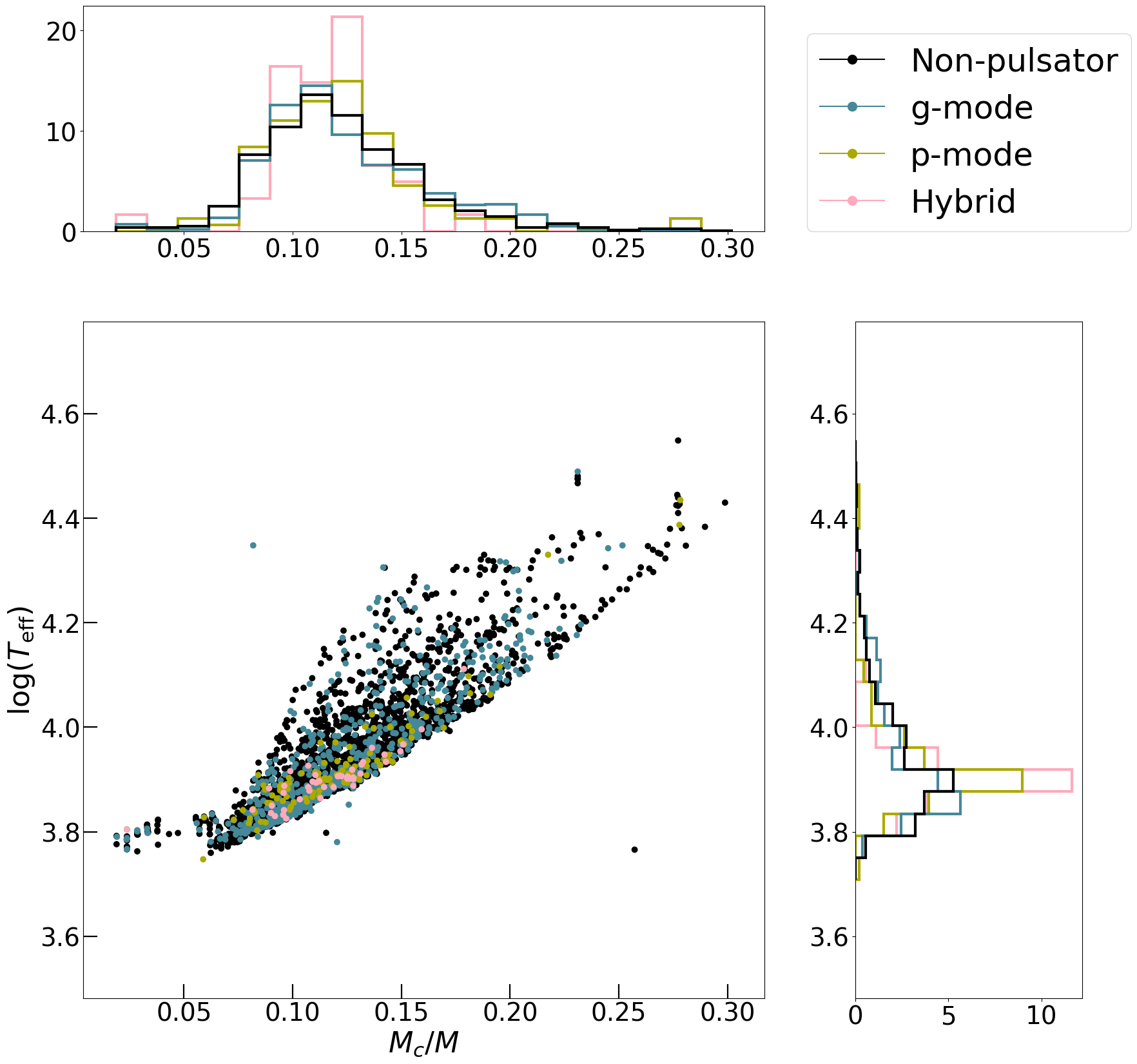}
\end{subfigure}

\caption{Distributions for the radius (upper left), core mass (upper right), central H fraction relative to the initial H fraction \xcx\ (lower left) and core mass fraction (lower left) plotted against \logteff.}
\label{fig:partymix}
\end{figure*}

Fig. \ref{fig:partymix} shows the distributions inferred from \gaia\ \logteff\ and \logl\ estimates using the M24 models for the radius \logr, core mass \Mc, central H fraction relative to the initial central H fraction \xcx, and core mass fraction plotted against \teff. \Mc\ is extremely tightly correlated with \logteff, a behaviour that is reflected in our MLR-based importance analysis (see Fig. \ref{fig:mlr}), and follows a similar distribution to \M. The radius distribution is significantly broader, depending on a combination of the mass, core mass, and central H fraction. There is a weak correlation between \xcx\ and \teff\, with a large degree of scatter in the distribution. The core mass fraction exhibits more scatter than either the mass or the core-mass distribution, but the relation with \teff\ is still clear.

The \Mc, \Mc /\M\ and \logr\ distributions of the different pulsator classes behave as we expect, closely following patterns previously discussed in the context of \logteff\ and \M, but \xcx\ bears a more detailed discussion. Despite the large amount of scatter present in the \xcx\ distribution when considered relative to \teff, its inferred distribution is clearly non-uniform, featuring a broad distribution that peaks between 0.25 and 0.5 depending on the assumed metallicity (see Fig. \ref{fig:xcx_varz}) but extends all the way from ZAMS to TAMS. This behaviour approximately holds for all subclasses.

It is interesting to note that despite taking steps to avoid numerical pile-ups relating to sampling at low or high values of \xc, a pile-up occurs at the ZAMS (\xcx~=~1) nonetheless. Fig. \ref{fig:xcx_varz} illustrates the impact of altering the metallicity on the implied \xcx\ distributions, and shows that as we reduce the metallicity the size of the ZAMS pile-up does reduce, but at the cost of greatly increasing the degree of pileup around the TAMS. Considering that the ZAMS pileup never entirely disappears even in the lowest metallicity case considered, we conclude that metallicity effects cannot fully account for the pile-up. A comprehensive array of figures is provided in the online supplementary material that includes metallicity comparisons for each parameter.

{Returning to the topic of tidal variation, we expect a preference for stronger tidal variability (and therefore higher values of TMP) in more evolved stars (lower values of \xcx) for a given orbital period. Despite the high degree of scatter in the \xcx\ measurements and the aforementioned pile-ups, we recover the expected anti-correlation between \xcx\ and TMP when $\log(P)<1$ and the data is not sparse. Figure \ref{fig:gradient} shows the gradient of the linear fit to the \xcx\ vs TMP data as a function of $\log(P)$, normalised by the mean value of TMP in each $\log(P)$ slice.}

\subsection{Adjusting \gaia\ \logteff\ and \logl\ estimates for binarity using eclipse information}
\label{sec:res:bin}

Neglecting binary evolution effects, the presence of a binary companion will make a target appear cooler and brighter than a single star equivalent to the primary. In Sections \ref{sec:res:fid} and \ref{sec:res:inf}, we considered the distributions obtained from using \gaia\ \logteff\ and \logl\ without modification. In this section, we explore the effect of correcting for the effect of binarity on the \gaia\ \logteff\ and \logl\ distributions.

From the eclipse analysis of IJ24, we have estimates for the surface brightness ratio and radius ratio of each target. In principle, this allows the primary and secondary effective temperatures and luminosities to be disentangled through relying on the Stefan-Boltzmann law and assuming the \gaia\ temperatures are approximately a luminosity-weighted average of the two stars. However, surface brightness ratios and radius ratios are difficult to obtain from eclipse analysis, as they rely on accurate detection and modelling of the eclipse wings. IJ24 measure these properties using two different methods to estimate these parameters: one method relying on physical fits to the eclipse shapes, and the other relying on the ingress and egress timings. However, the results obtained from these two methods can disagree wildly, as shown in Fig. \ref{fig:rrat_sbrat}. We therefore apply a selection filter that discards all targets that do not have radius ratio measurements that agree to better than 0.5 and surface brightness ratios that agree to better than 0.1 (the green points in Fig. \ref{fig:rrat_sbrat}) between the two methods. This cut-off method was selected over an uncertainty-based method to allow better control over which targets were passed through and to prevent a bias towards targets with high formal uncertainties in their radius ratio and surface brightness measurements. The cost to this is a degree of arbitrarity and subjectivity in the cut-offs themselves.

The \logteff\ and \logl\ estimates for the primary and secondary with the \gaia\ measurements for each system are shown in Fig. \ref{fig:teff_l_psb_recovery}. The primary luminosities are restricted to be at least half of the \gaia\ measurement due to our use of the luminosity ratio to define the primary and secondary. The secondary luminosities have significantly more spread while being bounded to half the \gaia\ luminosity. The primary and secondary \logteff\ measurements are messier, as we do not demand that the primary be the hotter star.

\begin{figure}
\centering
\begin{subfigure}{0.99\columnwidth}
\centering
\includegraphics[width=0.95\columnwidth]{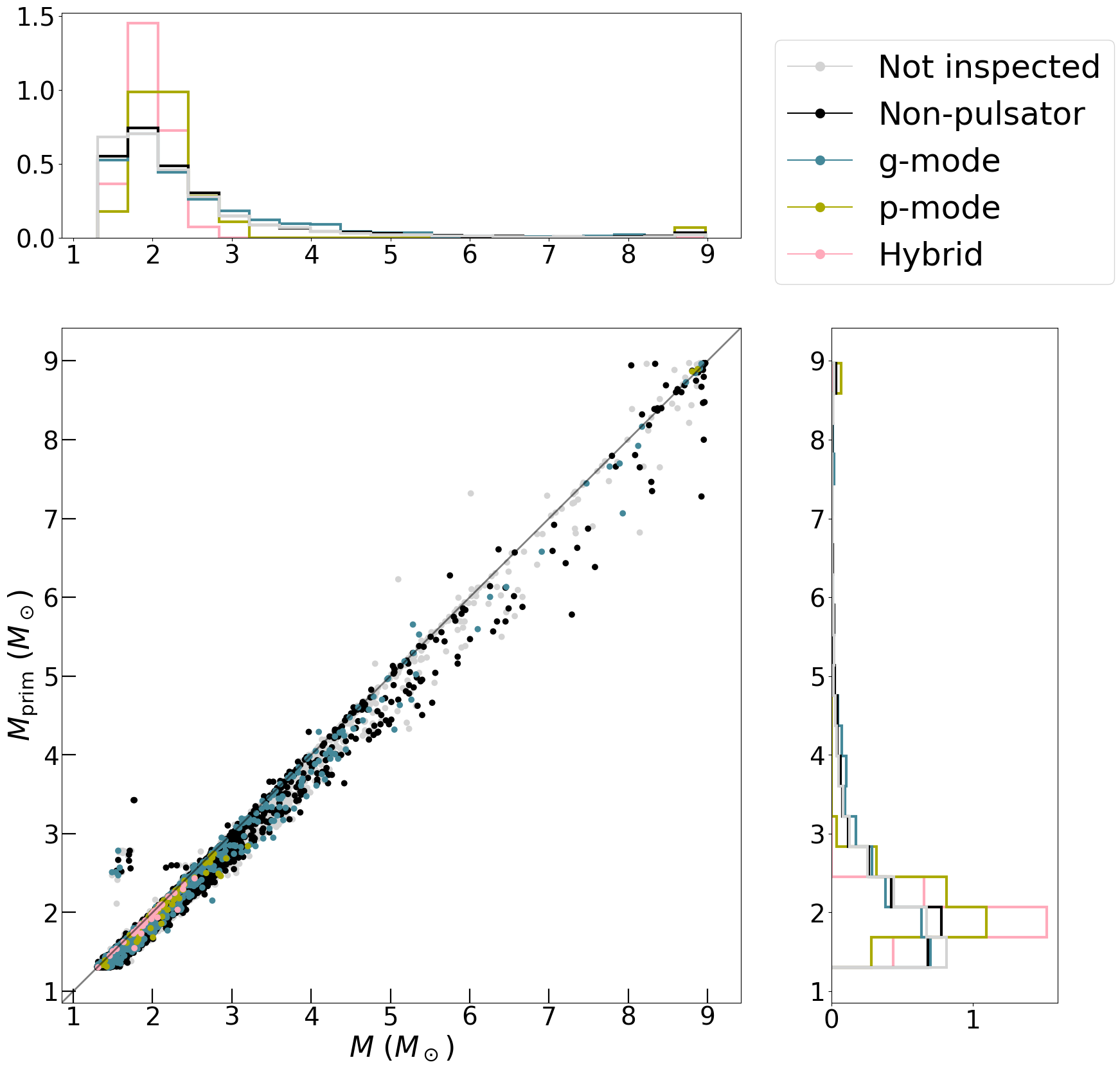}
\end{subfigure}

\begin{subfigure}{0.99\columnwidth}
\centering
\includegraphics[width=0.95\columnwidth]{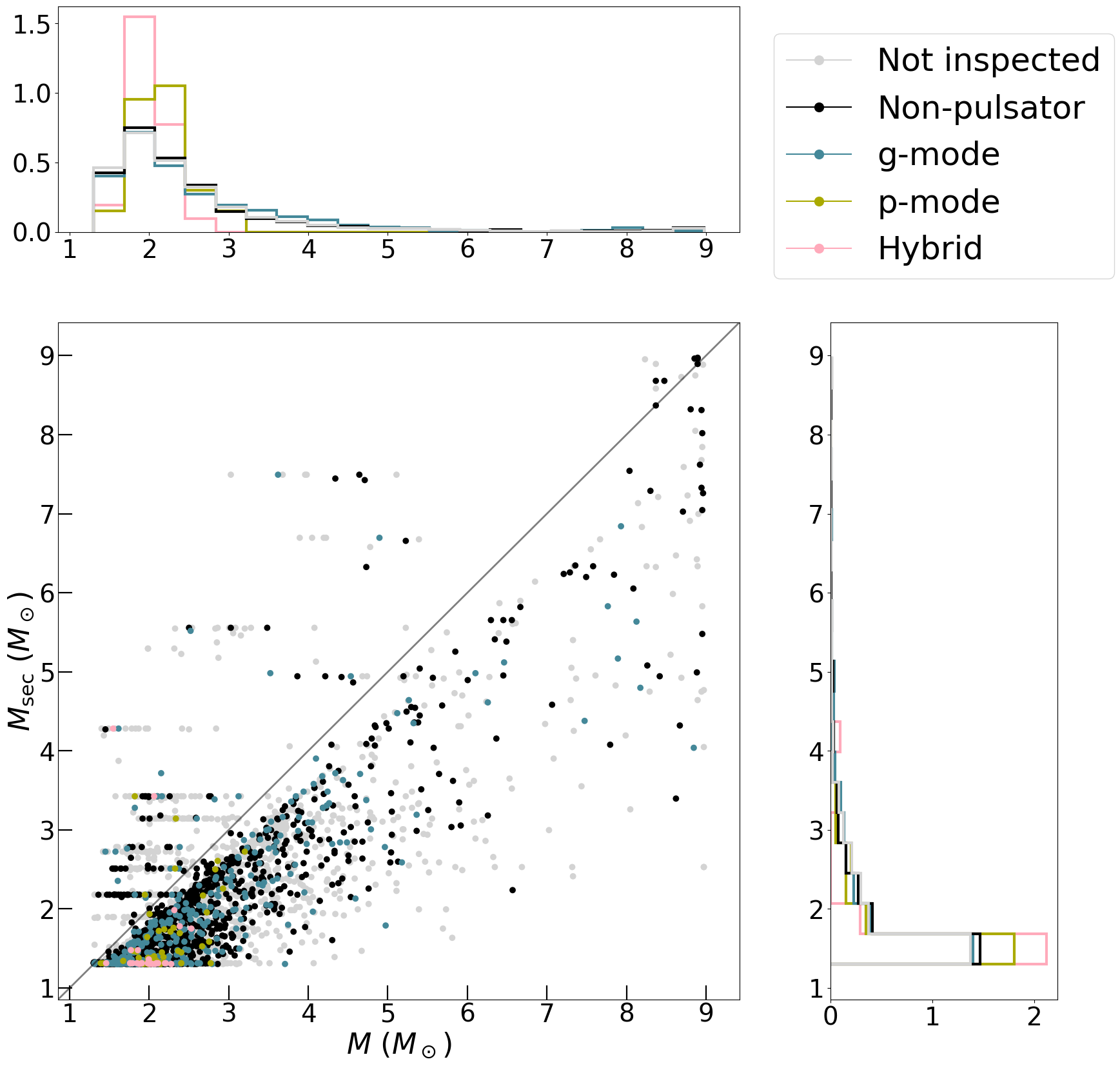}
\end{subfigure}
\caption{Inferred primary (upper panel) and secondary (lower panel) masses using surface brightness ratio and radius ratio measurements from the eclipse geometry (see Fig. \ref{fig:rrat_sbrat}), plotted against the mass inferred from the \gaia\ measurements \logteff\ and \logl\ estimates.}
\label{fig:m_prim_sec}
\end{figure}

Fig. \ref{fig:m_prim_sec} compares the inferred primary and secondary masses using the eclipse-disentangled \gaia\ \logteff\ and \logl\ estimates -- with the masses inferred from the \gaia\ \logteff\ and \logl\ measurements without correcting for binarity. There is a small but significant systematic shift towards lower inferred masses for the primary, and a much larger shift is apparent for the secondary. At first glance, this is in conflict with our prior discussion, which established that \logteff\ appears to have more of an effect on the inferred masses than \logl\ and that the primary \teff\ is statistically higher than the \gaia\ temperature. From this, we would expect an increase in the mass inferences for the primary, while the secondary masses would be lower.

\begin{figure}
\centering
\begin{subfigure}{0.99\columnwidth}
\centering
\includegraphics[width=0.95\columnwidth]{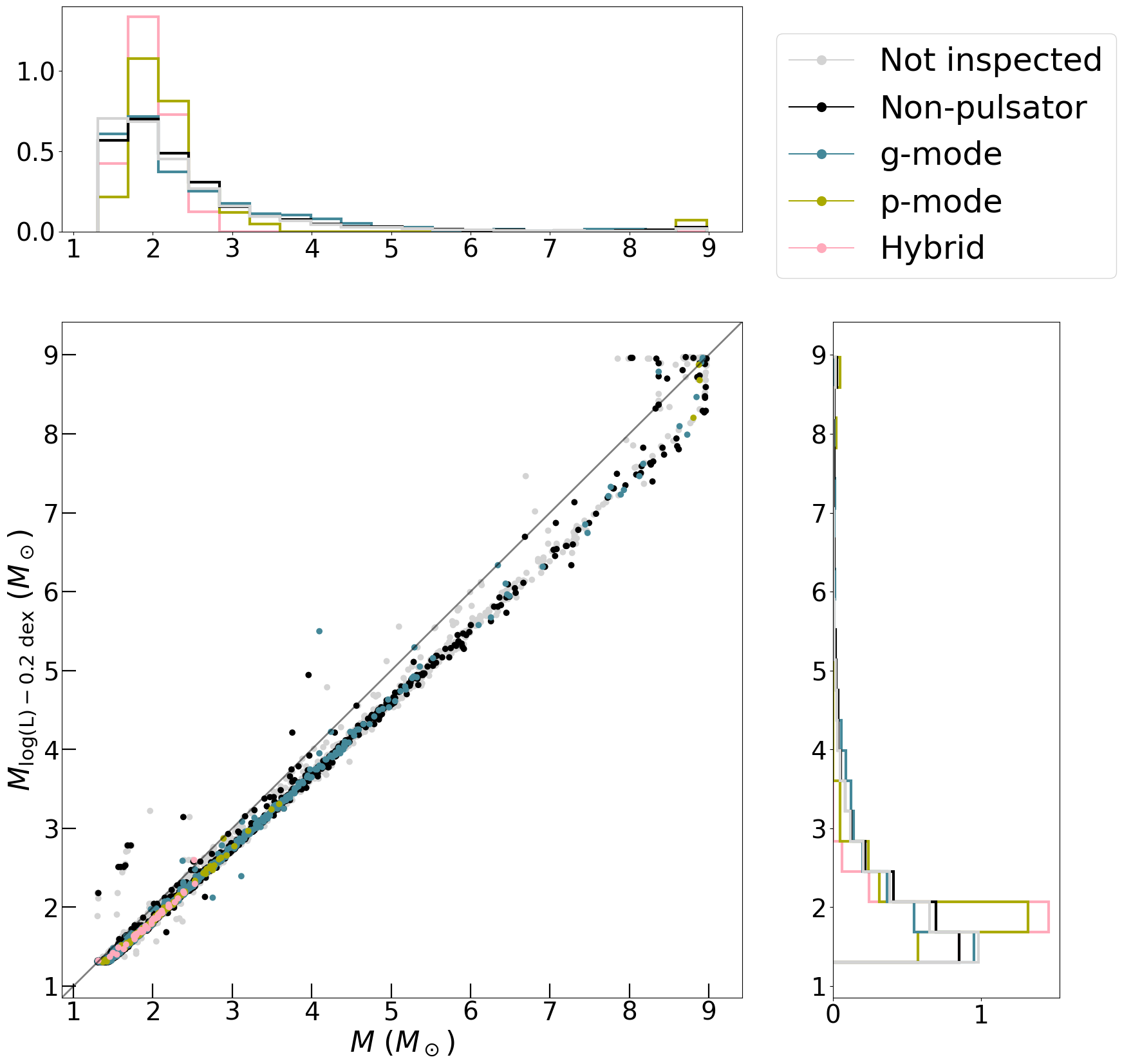}
\end{subfigure}

\begin{subfigure}{0.99\columnwidth}
\centering
\includegraphics[width=0.95\columnwidth]{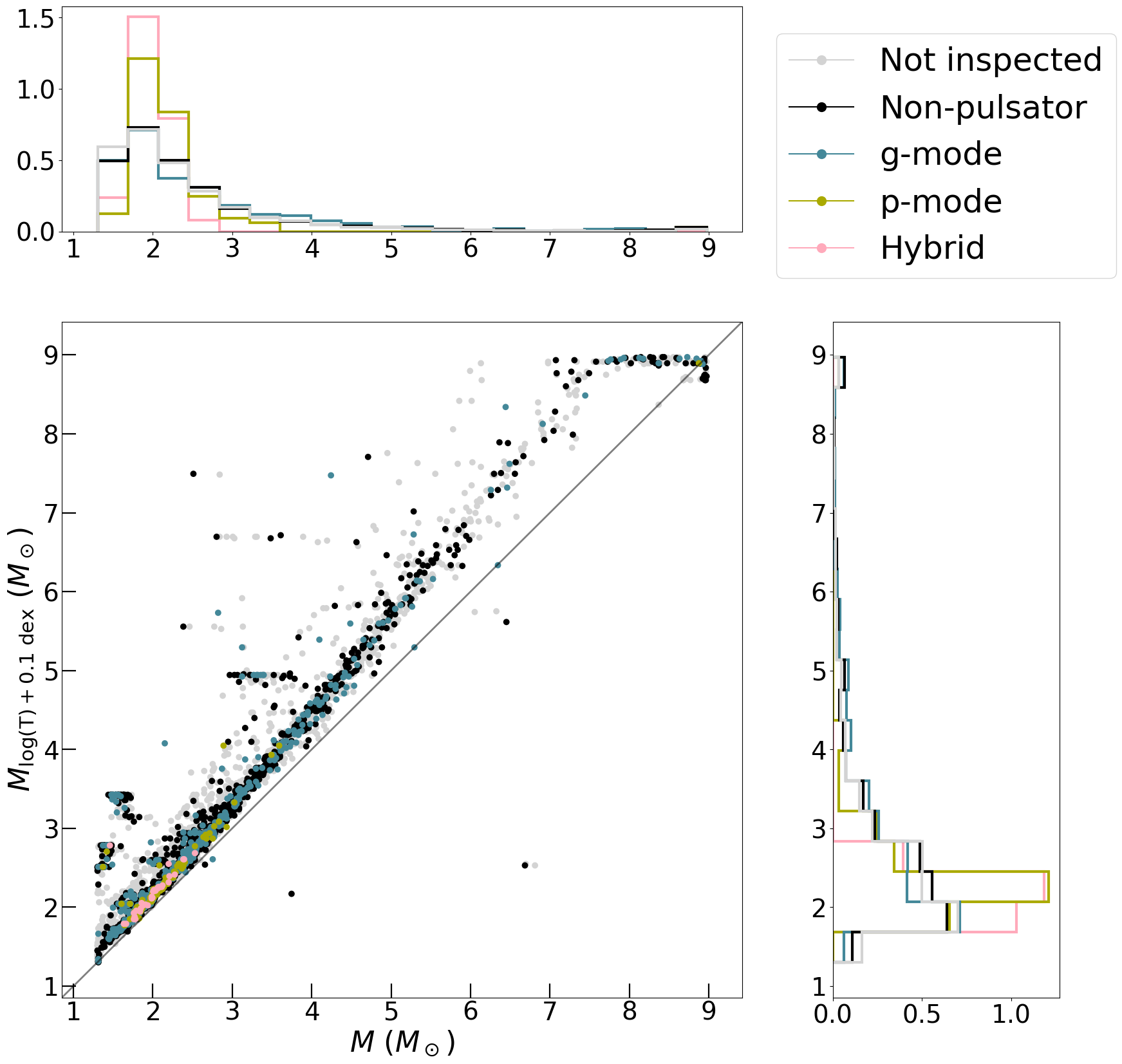}
\end{subfigure}

\caption{Mass inferences for a -0.2~dex shift in \logl\ (upper) and a +0.1~dex shift in \logteff\ (right) plotted against the mass inferred using the \gaia\ measurements \logteff\ and \logl\ estimates.}.
\label{fig:mshift}
\end{figure}

The reason we instead observe a slight decrease in the inferred primary mass is because the shift in \logteff\ caused by binarity is not occurring in isolation; it inevitably comes with a reduction in \logl. Fig. \ref{fig:mshift} shows the independent effects of lowering \logl\ and increasing \logteff, showing that, as we expected, reducing \logl\ leads to lower mass inferences while increasing \logteff\ leads to higher mass inferences, with both of these effects becoming stronger for more massive stars. Thus, the effect of attempting to correct for binarity results in two competing influences on the primary mass estimate, where increasing \logteff\ acts to increase the inferred mass while the \logl\ acts to reduce the inferred mass. For this reason, the primary mass estimates are surprisingly close to the uncorrected \gaia\ \logteff\ and \logl\ inferences. The same competing influences do not hold for the secondary, which becomes both cooler and dimmer relative to the \gaia\ estimates for the system, so there is a much stronger systematic towards lower mass predictions.

We note that much of the scatter in the secondary mass inference is due to a far larger number of the secondaries falling outside the domain of the grid. This is a natural consequence of the secondary masses being lower mass overall, as well as being less constrained in terms of their luminosities and temperatures: a binary with an extreme surface brightness and radius ratio will manifest as very close to the \gaia\ properties for the primary, but very far from these properties for the secondary. {For this reason, the secondary properties should be treated with caution even at a population level.}

{From tidal theory, we expect that close mass ratio binaries should generally be more tidally affected. Figure \ref{fig:q_tmp} shows the implied q vs TMP relation while Fig \ref{fig:q_slice} shows an exemplar TMP distribution for a fixed value of fixed $q$. In these figures, targets showing signs of obvious grid effects in their secondary mass inference have been removed (our conclusions are unchanged if these data are included). Despite the numeric difficulties facing the mass ratio inference, we find that higher TMP values are more common at higher $q$ values, consistent with expectations. However, we do not find evidence for any significant overabundance of high TMP values when $q$ is held constant for stars exhibiting g-mode pulsations compared to non-pulsators.}

Fig. \ref{fig:partymixprim} shows the distributions of the other inferred properties for the primary when correcting for binarity. To briefly summarise, the estimates for the primary star's properties systematically shift to lower radii to a similar degree to the mass, slightly higher core masses and core mass fractions when the \gaia\ estimate is high, and to systematically higher values of \xcx\ (younger stars), with the ZAMS pile-up increasing significantly. All of this is consistent with the reduction in luminosity caused by correcting for binarity having a more significant impact on the inferences than the increase in temperature, as we surmised for the primary's mass inference.

\section{2MASS \ks-magnitudes}

In the previous section, we presented and discussed stellar property inference from \gaia\ \logteff\ and \logl\ measurements for large numbers of EB systems using M24's grids of asteroseismically calibrated stellar models. This was done with consideration to the potential -- and challenges -- of using additional properties derived from the eclipse geometry in an attempt to better understand biases associated with binarity.

A more classical way to infer stellar properties is to use a mass-magnitude relation in conjunction with magnitude measurements that are minimally affected by reddening, such as 2MASS \ks-band magnitude (\ks-mag) measurements. By combining stellar evolution models with synthetic spectra, the \ks-mag can be estimated for any combination of primary and secondary mass. This methodology is currently being pursued in Vrancken et al. (in prep) to provide independent mass estimates for the IJ24 EB sample, and provides a valuable opportunity to test our mass inferences based on \gaia\ \logteff\ and \logl\ estimates.

\begin{figure}
    \centering
    \includegraphics[width=1\linewidth]{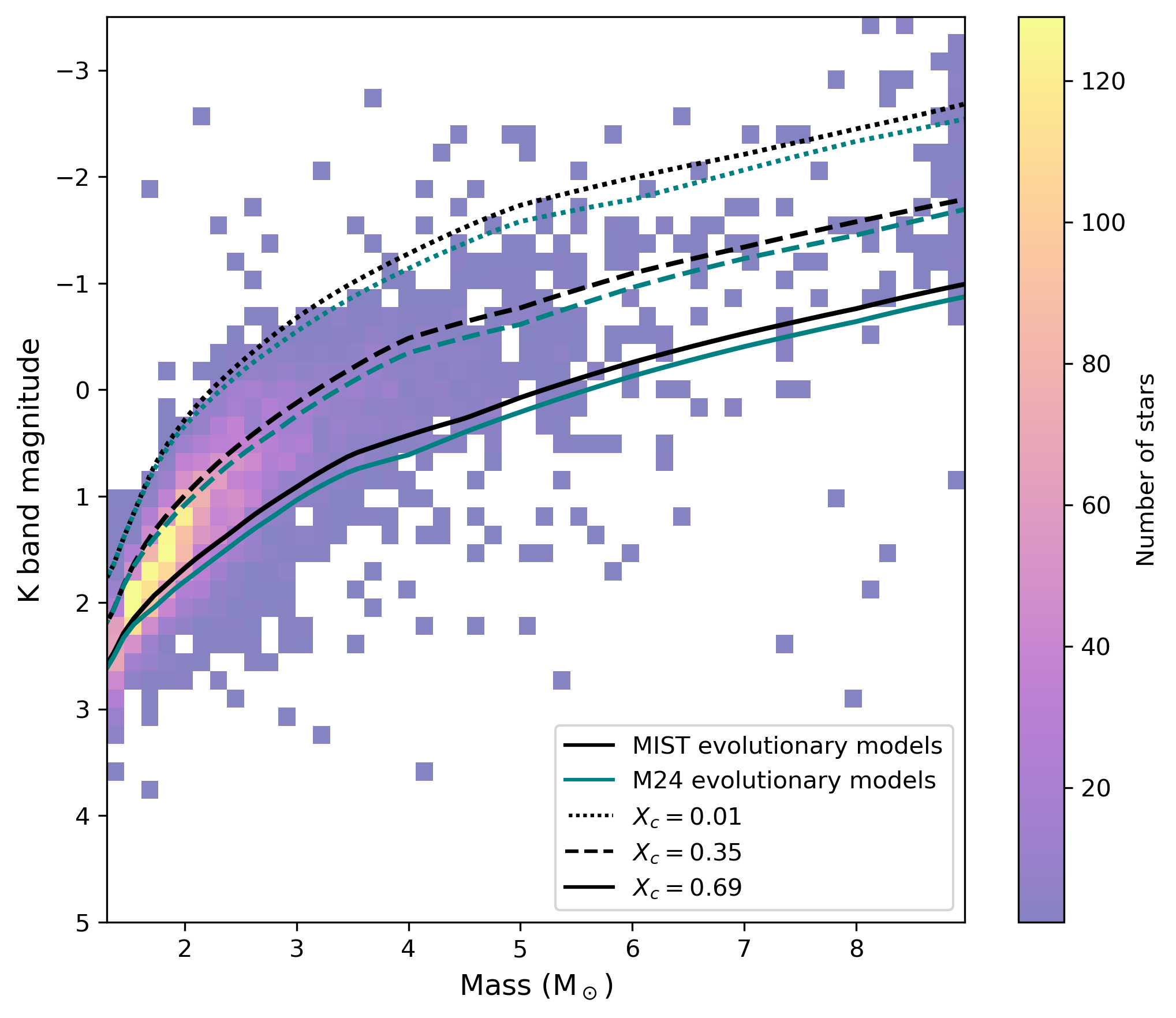}
    \caption{Curves: Single-stellar mass vs \ks-band magnitude (\ks-mag) for three different central H fractions (\xc\ = 0.69, \xc\ = 0.35, and \xc\ = 0.01) using solar metallicity MIST (black) and M24 (blue) grids of non-rotating stellar models.
    2D histogram: 2MASS \ks-mag vs  eclipse-inferred primary mass for the 5243 targets that satisfy the previously discussed radius ratio and surface brightness ratio cuts, have an eclipse-inferred primary mass, and have a 2MASS \ks-mag measurement.}
    \label{fig:jasmasskmag}
\end{figure}

Fig. \ref{fig:jasmasskmag} shows the resulting mass vs \ks-mag relations when fixing the central H fraction to different values and when relying on different grids of stellar evolution models. Pre-computed, publicly available spectral synthesis models are used to calculate the \ks-mag based on \gaia's \logteff\ and \logg\ estimates. The NLTE TLUSTY \citep{hubeny2017} spectral synthesis models were used for O-type stars \citep{lanz2003} and B-type stars \citep{lanz2007}, while the LTE ATLAS models \citep{castelli2003} were used otherwise. We use the resulting mass-magnitude relations presented in Fig. \ref{fig:jasmasskmag} to test our inferred primary masses by considering that, if our primary masses are correct, a plausible secondary that can reproduce the 2MASS \ks-mag measurement should exist.

The 2MASS \ks-mag measurements are shown as a 2-D histogram in Fig. \ref{fig:jasmasskmag}, where our eclipse-inferred primary masses are used for the x-axis. Most of the \ks-mag measurements are bracketed by the \xc ~=~0.69 and \xc ~=~0.01 curves, which is reassuring. Measurements above any given curve represent a case where the calculated \ks-mag for the relevant primary mass is dimmer than the measured 2MASS \ks-mag. For these cases, a secondary can be introduced into the system in order to reproduce the \ks-mag measurement. When the measurement is below the curve, however, the calculated \ks-mag of the primary is already brighter than the measurement for the whole system, and we conclude that the assumed primary mass is incompatible with the 2MASS \ks-mag estimate. 

\begin{figure}
    \centering
    \includegraphics[width=0.93\linewidth]{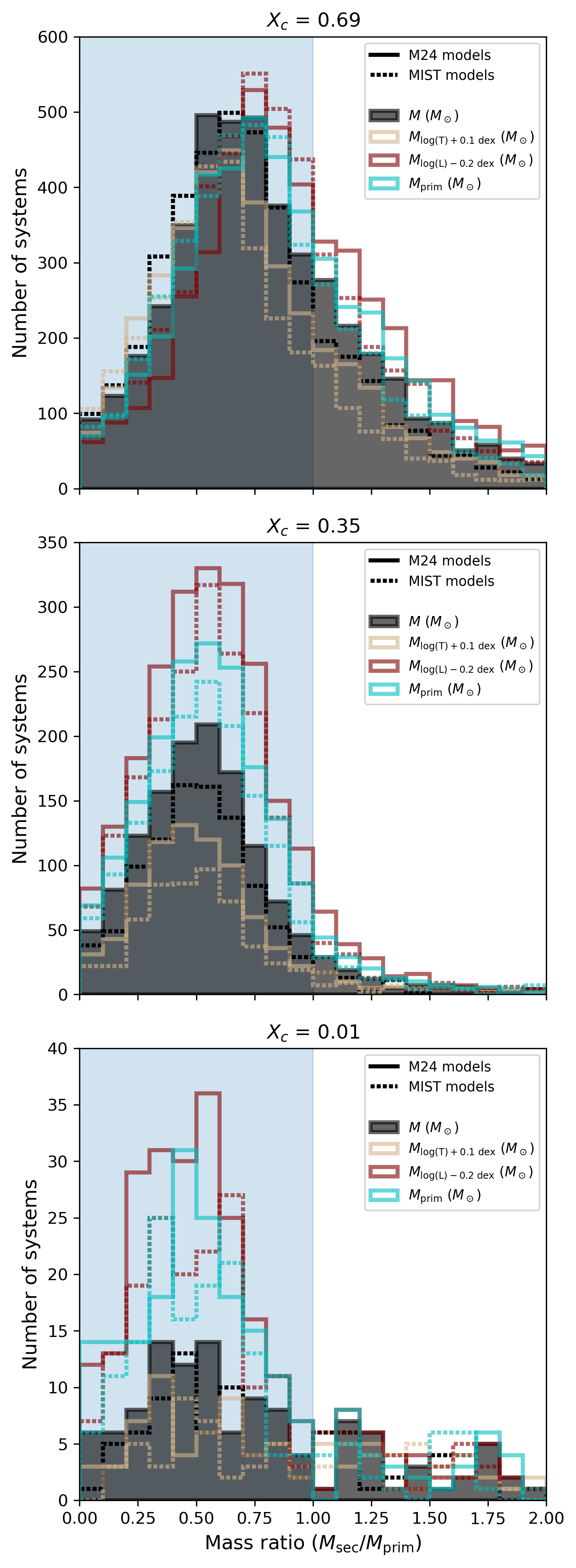}
    \caption{Distributions for the implied mass ratio ($q$) under different evolutionary assumptions (\xc), stellar grids (M24 vs MIST), and primary mass inferences.}
    \label{fig:jasq}
\end{figure}

Table \ref{tab:jastab} summarises the number of incompatible targets for each curve in Fig. \ref{fig:jasmasskmag} under all considered primary mass inference schemes. Fig. \ref{fig:jasq} shows the implied mass ratio distribution for each different \xc\ value, the M24 and MIST \citep{choi2016} stellar tracks, considering four different inferred primary masses obtained: when using the \gaia\ \logteff\ and \logl\ measurements without modification, when increasing the \logteff\ measurement by 0.1~dex, when decreasing the \logl\ measurements by 0.2~dex, and when using eclipse information to isolate the primary. The region of plausibility, where the mass ratio $q$ is between 0 and 1, is highlighted in blue.

The adopted \xc\ value plays a significant role in the shape of the distribution, out-weighing the importance of the selected grid of stellar models. Near the ZAMS (\xc ~=~0.69) the stars are the dimmest, while they become more luminous towards the TAMS (\xc ~=~0.01). Thus, the ZAMS case has the minimum number of targets that are incompatible with the 2MASS \ks-mag measurement (approximately 15\%, see Table \ref{tab:jastab}) but also the most that overshoot the region of plausibility, with significant numbers of stars with implied secondary masses greater than the primary mass ($q>1$).

In practice, the true \xc\ is unknown for any given target, and we can see that the high-$q$ overshoot almost completely disappears as we increase \xc. Therefore, only the few tens of systems that have mass ratios higher than unity for the TAMS assumption are truly incompatible due to the upper bound ($q$~=~1) of the plausibility region. We conclude that the level of incompatibility with the 2MASS \ks-mag measurements is approximately 15\%, and dominated by systems that are over-luminous and over-massive.

The incompatibility fractions under different physics bears a brief discussion. By virtue of having lower calculated luminosities for a given mass, and therefore lower primary mass estimates, the eclipse-inferred primary masses have a slightly lower level of incompatibility (14.3\%) compared to the default \gaia\ mass inference (16.9\%). However, this difference is still smaller than the change in the incompatibility between M24 and MIST (incompatibility fractions of 21.5\% and 17.9\% for \gaia\ and eclipse inferences, respectively), which we already noted was a minor effect compared to the evolutionary age of the star (\xc). 

Overall, we conclude that most of the targets ($\approx 85$\%) have plausible inferred values for $q$, although the fraction of targets with unreconcilable incompatibilities is not negligible. Further, the choice of stellar grid appears to be secondary to knowledge of the evolutionary age of the system when using the \ks-mag measurements to infer secondary masses.

\section{Conclusion}
\label{sec:conc}

In this work, we characterise a large population of eclipsing binaries in terms of their orbital and stellar properties, with attention to their pulsation class. The pulsation classes are determined through manual inspection of high-precision TESS photometry, while the stellar properties are inferred from \gaia\ DR3 estimates for the \logteff\ and \logl\ using M24's asteroseismically calibrated grids of stellar models. Attention is paid to sources of bias stemming from binarity on the \gaia\ \logteff\ and \logl\ estimates and how this propagates into stellar parameter estimates.

The positions of the hybrid and the p-mode pulsators in the HRD are as we expect for these pulsators, peaking strongly around the \dsct\ region, with the hybrid pulsators having somewhat lower masses than the pure p-mode pulsators. The far more numerous g-mode pulsators peak in the \gdor\ region and also populate the SPB region. However, there is no sign of a well-defined break between the two regions in the HRD. Instead, the distribution of g-mode pulsators is continuous towards higher masses. The g-mode pulsators are biased towards shorter orbital periods and towards higher values of our orbital-harmonic-based tidal morphology parameter, which is indicative of the strength of tidal variability in the light curve. However, this effect appears to be small. We conclude that tidal effects are unlikely to be the primary cause of the departures from the classical regions of instability found in this and previous works dealing with large populations of pulsators \citep{gaia2023pulsations,hey2024,mombarg2024}, leaving the physical limitations of mode excitation theories and sample contamination by rotational variables as leading explanations. 

{An alternative explanation that is particularly relevant for our sample is that of secondary contamination. In this scenario, the pulsator is the secondary companion which belongs to a traditional instability region. We disfavour this explanation for two reasons. The first is that such targets are naturally selected against by birth, as the sub-set of intermediate-mass stars with relatively massive companions is by definition smaller than the total set of intermediate mass stars. The second is that there will inevitably be significant dilution in the relative flux variations by light from the primary, making secondary pulsations inherently more difficult to detect -- especially when the mass ratio is not close to unity. Empirically, we see no signs of a preference for pulsator-mass companions in our data for stars that fall outside traditional instability zones. This does not outrule this source of contamination, especially as the inferred properties for the secondary (see Section \ref{sec:res:bin}) should be treated with caution even at a population level. However, the fact remains that there is no evidence for contamination by pulsations on the secondary being responsible for the observed continuity in g-mode pulsators.}

Eclipsing binaries featuring hybrid pulsations more closely match the g-mode distribution when it comes to orbital properties -- and correlation with tidal variability -- than the p-mode distribution, while closely matching the p-mode distribution in terms of surface properties such as \logteff\ and \logl. The correlation with the p-mode properties is intuitive, given that the p-mode distribution is dominated by \dsct\ pulsators and contains only a few (presumed) \bcep\ pulsators. The preference for short-period, circular binaries for both pure g-mode pulsators and hybrid pulsators is consistent with more rapid spin-orbit coupling and circularisation in the presence of g-mode pulsations, presumably driven by increased tidal efficiency and angular momentum transport \citep{aerts2019}.

Stellar mass inferences using unmodified \gaia\ \logteff\ and \logl\ provide a continuous mass distribution for the g-mode pulsators that peaks in the \gdor\ region, while the p-mode and hybrid pulsators return mass inferences consistent with our theoretical expectations for \dsct\ and \dsct/\gdor\ hybrids, respectively.

We use the surface brightness ratio and radius ratio estimates from IJ24's eclipse analysis to disentangle \logteff\ and \logl\ for the primary and secondary, and thereby assess the impact of binarity on our inferred stellar properties. The surface brightness ratio and radius ratio on which this disentangling relies are both difficult to obtain from the eclipse geometry, a fact reflected in the large amount of scatter in the dual-method approach of IJ24 (see Fig. \ref{fig:rrat_sbrat}). Enforcing agreement between the two methods of IJ24 results in a substantial decrease in sample size (from 14377 targets to 7823).

The systematic effect of correcting for binarity on the inferred primary mass is relatively small, showing a slight shift towards lower masses. This is caused by the competing effects of reducing \logl\ while increasing \logteff\ when isolating the primary. The slight bias to lower masses is consistent with the luminosity effect being more important, although this result should be taken with the caveat that the balance between increasing \logteff\ and reducing \logl\ will inevitably be somewhat sensitive to how we propagate the surface brightness and radius ratios into an estimate for \logteff. The other inferred properties behave similarly, with their distributions shifting relative to the basic \gaia\ distributions in ways that are consistent with the dominant effect being a reduction in the luminosity.

Confronting our mass inferences with 2MASS \ks-magnitude measurements by considering whether a plausible value of $q$ can be found, we find that approximately 15\% of our sample has mass inferences that are incompatible with the 2MASS \ks-magnitude measurements. These incompatible systems are dominated by targets where the inferred primary mass results in \ks-magnitudes from the primary that are brighter than the 2MASS \ks-magnitude measurement for the entire system. We also find that the choice of evolutionary track (MIST vs M24) plays only a minor role when compared to the role of the evolutionary age (\xc) in the calculation of $q$.

Our initial characterisation of this large sample of TESS EBs, despite its caveats, provides an excellent starting point from which future work focusing on leveraging the asteroseismic properties of these targets can be built.

\section*{Data availability}
Supplementary material can be found online \href{https://zenodo.org/records/17513039}{https://zenodo.org/records/17513039}. This material includes extra data, figures, and animations. Additional technical data and figures are available upon request from the lead author (alex.kemp@kuleuven.be).

\begin{acknowledgements}
The authors wish to acknowledge useful discussions with Dario Fritzewski, Haotian Wang, and Vincent Vanlaer. The authors also wish to thank the anonymous referee for their constructive report. The research leading to these results has received financial support from the Flemish Government under the long-term structural Methusalem funding program by means of the project SOUL: Stellar evolution in full glory, grant METH/24/012, and from the European Research Council (ERC) under the Horizon Europe programme (Synergy Grant agreement No. 101071505: 4D-STAR). While partially funded by the European Union, views and opinions expressed are however those of the authors only and do not necessarily reflect those of the European Union or the European Research Council. Neither the European Union nor the granting authority can be held responsible for them. The authors also acknowledge the Belgian Federal Science Policy Office (BELSPO) for their financial support in the framework of the PRODEX Programme of the European Space Agency (ESA), facilitating the exploitation of the Gaia data.

\end{acknowledgements}

\bibliographystyle{aa}
\bibliography{bibfile.bib}
\appendix

\onecolumn
\FloatBarrier

\section{Additional tables and figures}

\begin{table}[]
\caption{2-sample KS test probabilities (p-values) for the null hypothesis of each subsample pairing <A>~|~<B> being drawn from the same underlying distribution.}
\label{tab:kstest}
\begin{tabular}{l|llllll}
Parameter & NP|GM & NP|PM & NP|HY & GM|PM & GM|HY & PM|HY \\ \hline
log$(T_{\rm eff})$ & \textbf{0.05173} & 0.00490 & 0.00116 & 0.00433 & 0.00049 & \textbf{0.29272} \\
log$(L)$ & \textbf{0.10388} & 0.01198 & 0.00080 & 0.00215 & 0.00027 & 0.04760 \\
log$(P (d))$ & 0.00000 & 0.00934 & 0.00214 & 0.00000 & \textbf{0.17399} & 0.00002 \\
Eccentricity & 0.00067 & \textbf{0.29576} & \textbf{0.12532} & \textbf{0.16170} & \textbf{0.64109} & \textbf{0.25785} \\
TMP & 0.00000 & \textbf{0.13311} & \textbf{0.08091} & 0.00387 & \textbf{0.72920} & 0.03036 \\
log$(T_{\rm eff})$ prim & \textbf{0.05772} & 0.01837 & 0.00235 & 0.00483 & 0.00097 & \textbf{0.39335} \\
log$(T_{\rm eff})$ sec & \textbf{0.33606} & 0.02243 & 0.00001 & \textbf{0.10567} & 0.00003 & 0.00048 \\
log$(L)$ prim & \textbf{0.08634} & 0.01427 & 0.00261 & 0.00226 & 0.00058 & 0.03672 \\
log$(L)$ sec & \textbf{0.47429} & 0.00327 & 0.00092 & 0.00326 & 0.00230 & \textbf{0.41100} \\ \hline
$Z=0.014$ &  &  &  &  &  &  \\ 
$M$ & \textbf{0.20610} & 0.00356 & 0.00074 & 0.00118 & 0.00031 & \textbf{0.07986} \\
log$(R)$ & \textbf{0.66994} & \textbf{0.44749} & \textbf{0.06242} & \textbf{0.33707} & \textbf{0.07237} & 0.01822 \\
$X_{\rm c}/X_{\rm i}$ & 0.04134 & \textbf{0.69540} & \textbf{0.34505} & \textbf{0.32562} & \textbf{0.87858} & \textbf{0.73671} \\
$M_c$ & \textbf{0.07195} & 0.02412 & 0.00389 & 0.01337 & 0.00155 & \textbf{0.16347} \\
$M_c/M$ & \textbf{0.08573} & \textbf{0.85786} & \textbf{0.15253} & \textbf{0.27424} & \textbf{0.11680} & \textbf{0.52671} \\
$M$ prim & \textbf{0.11206} & 0.03606 & 0.00085 & 0.00218 & 0.00015 & \textbf{0.07886} \\
log$(R)$ prim & \textbf{0.97410} & \textbf{0.18141} & \textbf{0.13820} & \textbf{0.29756} & \textbf{0.13549} & 0.02419 \\
$X_{\rm c}/X_{\rm i}$ prim & \textbf{0.85383} & \textbf{0.06487} & \textbf{0.47922} & 0.02770 & \textbf{0.59221} & 0.04691 \\
$M_c$ prim & 0.04157 & 0.03822 & 0.00664 & 0.01076 & 0.00171 & \textbf{0.39455} \\
$M_c/M$ prim & \textbf{0.14205} & \textbf{0.12917} & \textbf{0.14932} & \textbf{0.09906} & 0.03850 & \textbf{0.61577} \\
$M$ sec & \textbf{0.26239} & 0.02428 & 0.00013 & 0.04791 & 0.00020 & 0.02127 \\
log$(R)$    sec & \textbf{0.90778} & \textbf{0.08663} & 0.04563 & 0.04644 & \textbf{0.06610} & \textbf{0.34993} \\
$X_{\rm c}/X_{\rm i}$ sec & \textbf{0.64186} & \textbf{0.83143} & \textbf{0.42043} & \textbf{0.57386} & \textbf{0.30852} & \textbf{0.91710} \\
$M_c$ sec & \textbf{0.38032} & 0.00846 & 0.00023 & 0.01778 & 0.00040 & 0.04960 \\
$M_c/M$ sec & \textbf{0.68931} & 0.00149 & 0.00055 & 0.01200 & 0.00111 & \textbf{0.06801} \\ \hline
$Z=0.008$ &  &  &  &  &  &  \\
$M$ & \textbf{0.21470} & 0.00537 & 0.00080 & 0.00184 & 0.00030 & \textbf{0.09178} \\
log$(R)$ & \textbf{0.65828} & \textbf{0.41209} & \textbf{0.07461} & \textbf{0.36384} & \textbf{0.07977} & 0.03361 \\
$X_{\rm c}/X_{\rm i}$ & \textbf{0.05629} & \textbf{0.45191} & \textbf{0.53879} & \textbf{0.24212} & \textbf{0.66431} & \textbf{0.44960} \\
$M_c$ & \textbf{0.05692} & 0.01231 & 0.00153 & 0.00715 & 0.00102 & \textbf{0.37903} \\
$M_c/M$ & \textbf{0.08175} & \textbf{0.51362} & \textbf{0.15756} & \textbf{0.15135} & \textbf{0.06057} & \textbf{0.72993} \\
$M$ prim & \textbf{0.13339} & 0.01418 & 0.00989 & 0.00239 & 0.00308 & 0.03504 \\
log$(R)$ prim & \textbf{0.90270} & \textbf{0.52356} & \textbf{0.13728} & \textbf{0.67609} & \textbf{0.14981} & \textbf{0.05732} \\
$X_{\rm c}/X_{\rm i}$ prim & \textbf{0.46715} & \textbf{0.15661} & \textbf{0.57069} & \textbf{0.18486} & \textbf{0.44299} & \textbf{0.36981} \\
$M_c$ prim & \textbf{0.36469} & \textbf{0.07882} & 0.04945 & \textbf{0.05434} & 0.03562 & \textbf{0.59614} \\
$M_c/M$ prim & \textbf{0.51754} & \textbf{0.30238} & \textbf{0.41744} & \textbf{0.73457} & \textbf{0.35243} & \textbf{0.76872} \\
$M$ sec & \textbf{0.59723} & 0.00150 & 0.00239 & 0.00754 & 0.00559 & \textbf{0.09799} \\
log$(R)$    sec & \textbf{0.65349} & 0.04363 & 0.03285 & \textbf{0.10760} & \textbf{0.08144} & \textbf{0.67302} \\
$X_{\rm c}/X_{\rm i}$ sec & \textbf{0.66300} & \textbf{0.17678} & \textbf{0.06644} & \textbf{0.39804} & \textbf{0.11885} & \textbf{0.35559} \\
$M_c$ sec & \textbf{0.68327} & 0.00451 & 0.00125 & 0.00695 & 0.00225 & \textbf{0.22762} \\
$M_c/M$ sec & \textbf{0.91319} & 0.00198 & 0.00057 & 0.00971 & 0.00100 & \textbf{0.22762} \\ \hline
$Z=0.0045$ &  &  &  &  &  &  \\
$M$ & \textbf{0.24138} & 0.00411 & 0.00063 & 0.00125 & 0.00025 & \textbf{0.07986} \\
log$(R)$ & \textbf{0.72216} & \textbf{0.49207} & \textbf{0.09752} & \textbf{0.44455} & \textbf{0.11792} & 0.03435 \\
$X_{\rm c}/X_{\rm i}$ & \textbf{0.13046} & \textbf{0.38515} & \textbf{0.77856} & \textbf{0.21725} & \textbf{0.91677} & \textbf{0.53959} \\
$M_c$ & \textbf{0.05967} & 0.01190 & 0.00221 & 0.00254 & 0.00094 & \textbf{0.43570} \\
$M_c/M$ & 0.04745 & \textbf{0.65752} & \textbf{0.26918} & \textbf{0.14245} & \textbf{0.10643} & \textbf{0.88164} \\
$M$ prim & \textbf{0.07392} & 0.01733 & 0.01802 & 0.00753 & 0.00342 & 0.04298 \\
log$(R)$ prim & \textbf{0.47125} & \textbf{0.05817} & \textbf{0.10399} & \textbf{0.06997} & \textbf{0.11135} & 0.00893 \\
$X_{\rm c}/X_{\rm i}$ prim & \textbf{0.39208} & 0.01266 & \textbf{0.36238} & 0.01476 & \textbf{0.55184} & \textbf{0.16458} \\
$M_c$ prim & 0.03421 & \textbf{0.07934} & 0.02069 & 0.03487 & 0.00980 & \textbf{0.43167} \\
$M_c/M$ prim & \textbf{0.39992} & \textbf{0.28082} & \textbf{0.33818} & \textbf{0.28626} & \textbf{0.21333} & \textbf{0.53428} \\
$M$ sec & \textbf{0.38297} & 0.01476 & 0.00119 & 0.01753 & 0.00121 & \textbf{0.26907} \\
log$(R)$    sec & \textbf{0.93322} & 0.01747 & 0.01908 & 0.01437 & 0.01725 & \textbf{0.63125} \\
$X_{\rm c}/X_{\rm i}$ sec & \textbf{0.97801} & 0.01591 & 0.00262 & 0.00859 & 0.00425 & \textbf{0.41993} \\
$M_c$ sec & \textbf{0.70949} & 0.03723 & 0.00549 & 0.02826 & 0.00441 & \textbf{0.09117} \\
$M_c/M$ sec & \textbf{0.75328} & 0.04444 & 0.00317 & \textbf{0.10155} & 0.00528 & \textbf{0.07185}
\end{tabular}
\tablefoot{The non-pulsator (NP), g-mode (GM), p-mode (PM), and hybrid (HY) subclasses are shown, with probabilities higher than 5~\% highlighted in bold. Machine-readable tables including p-values for targets that were not inspected (NI), as well as the tidal variation (TV), no tidal variation (NT), period problem (PP), and no secondary (NS) subclasses, are available online.}
\end{table}

\begin{figure}[h!]
\centering
\includegraphics[width=0.49\columnwidth]{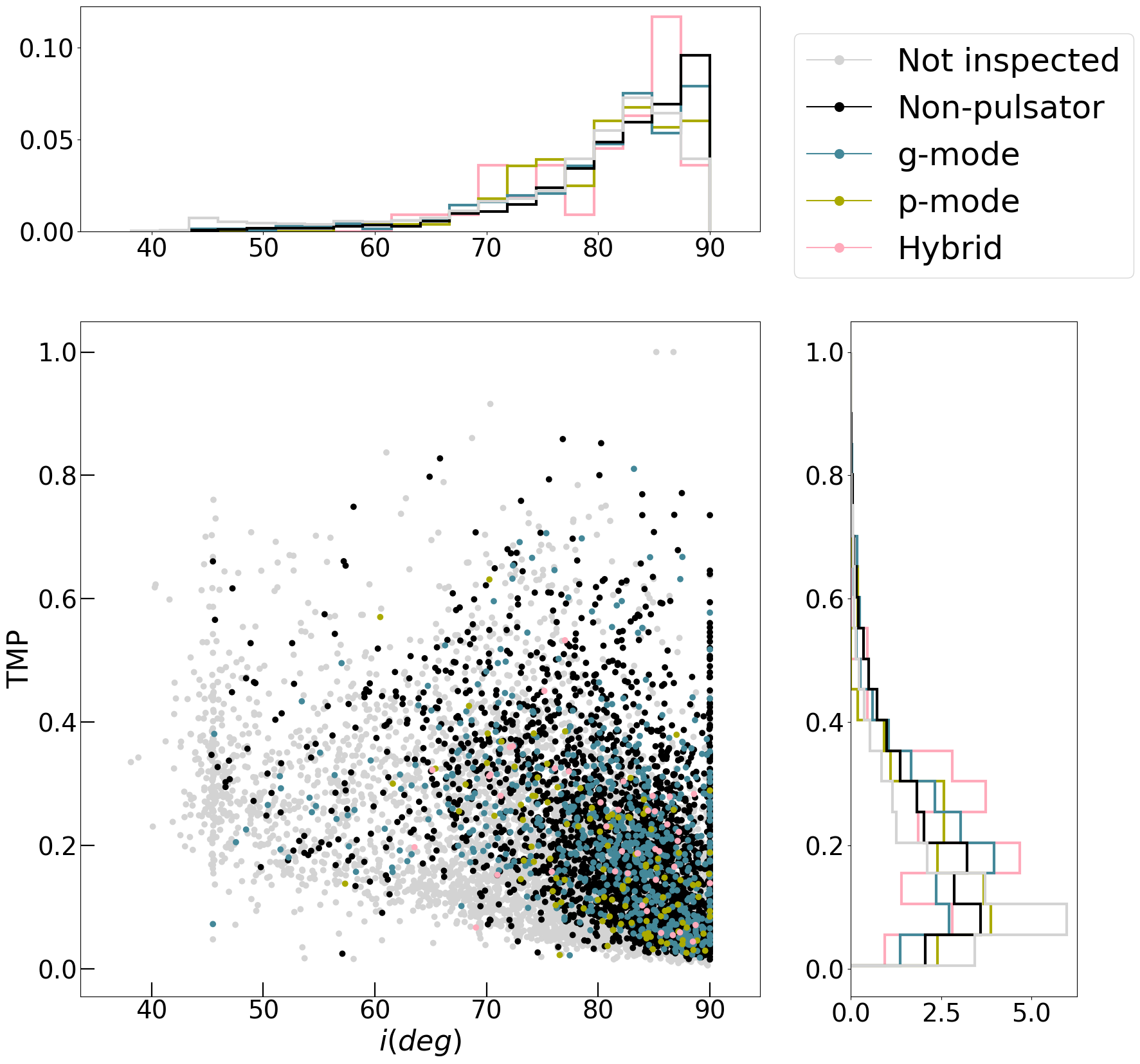}
\caption{{Orbital inclination $i$ vs Tidal morphology parameter (TMP).}}
\label{fig:i_morph}
\end{figure}

\begin{figure*}[h!]
\begin{subfigure}{0.49\columnwidth}
\centering
\includegraphics[width=0.99\columnwidth]{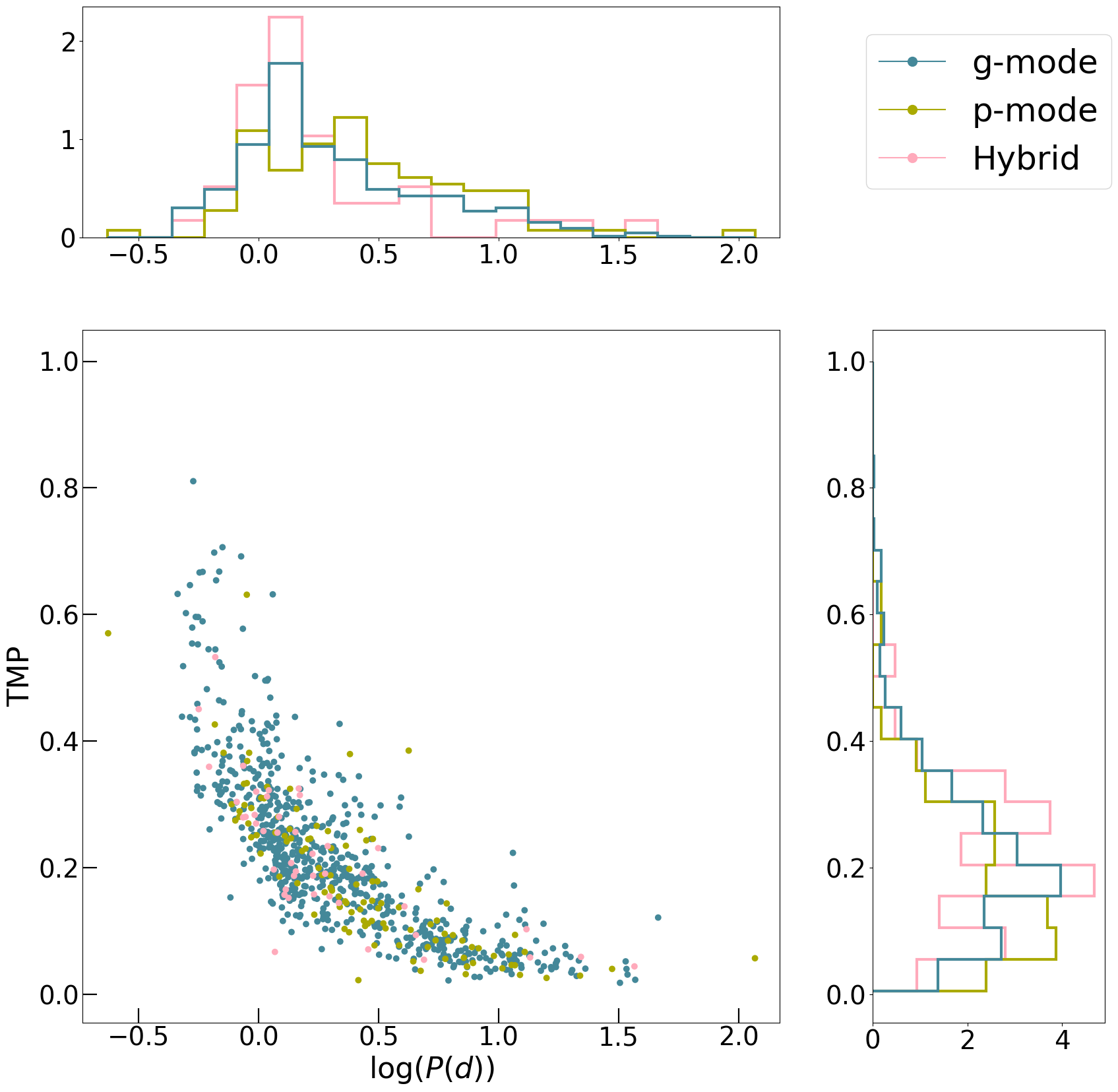}
\end{subfigure}%
\begin{subfigure}{0.49\columnwidth}
\centering
\includegraphics[width=0.99\columnwidth]{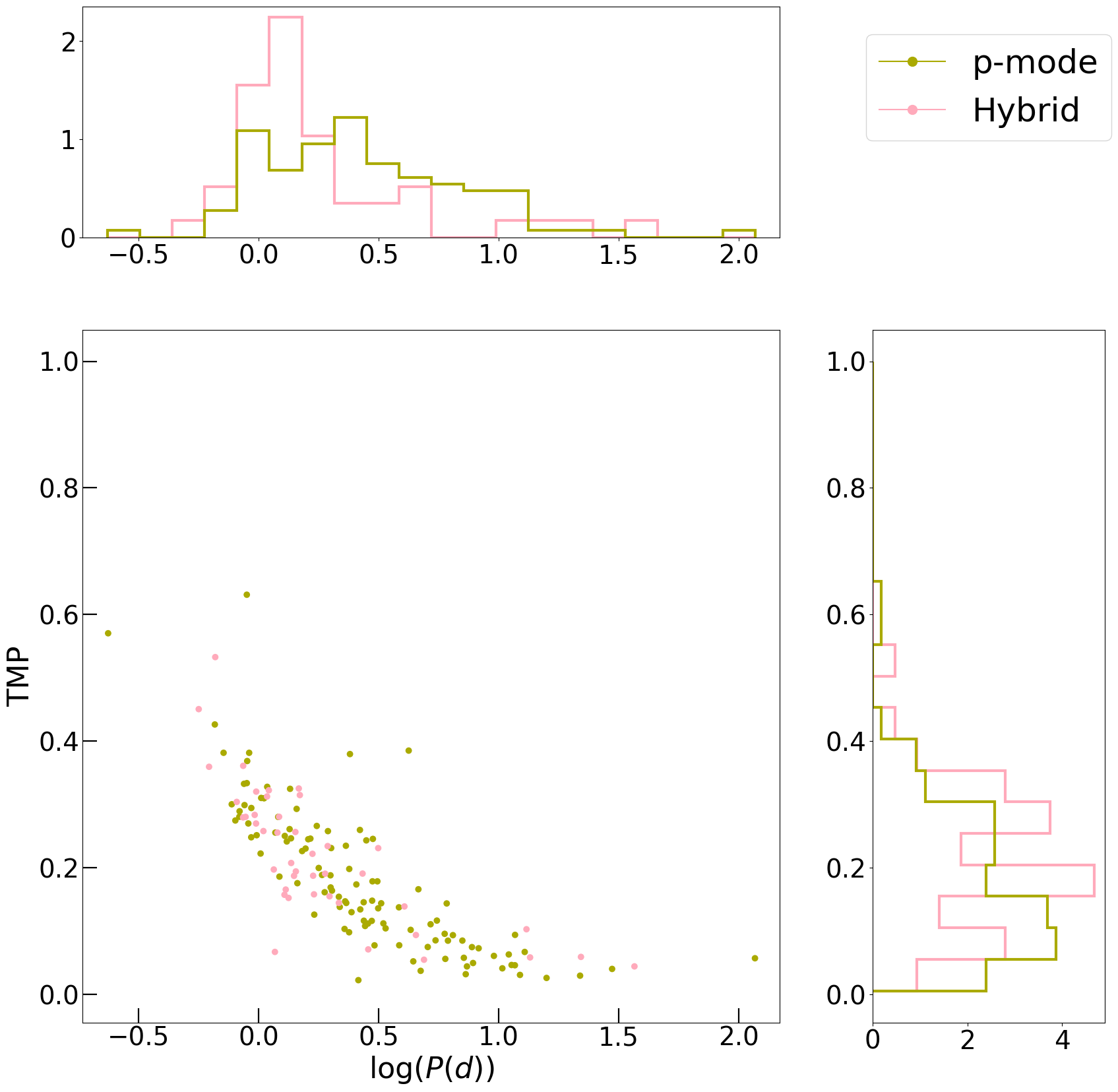}
\end{subfigure}
\caption{Orbital period vs tidal morphology parameter (TMP), highlighting the different pulsator classes.}
\label{fig:puls_morph}
\end{figure*}

\begin{figure*}[h!]
\centering
\begin{subfigure}{0.49\columnwidth}
\centering
\includegraphics[width=0.99\columnwidth]{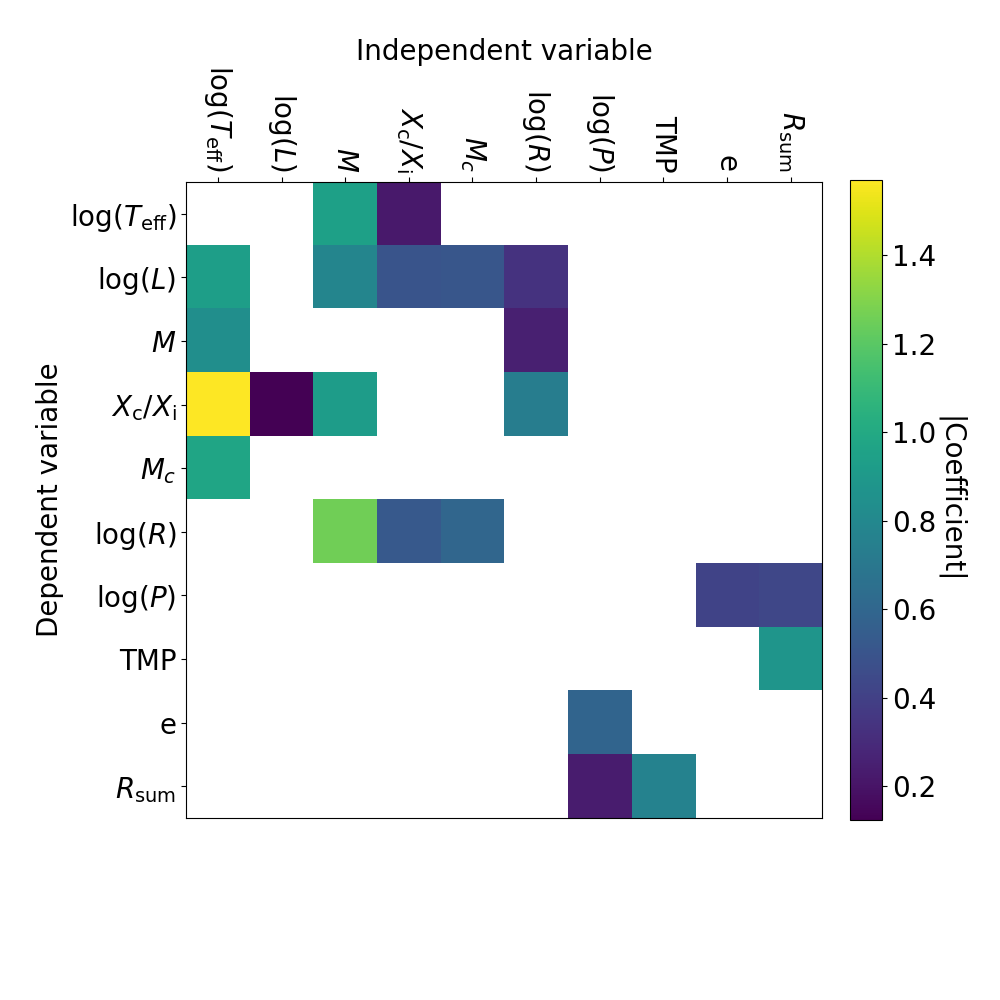}
\end{subfigure}%
\begin{subfigure}{0.49\columnwidth}
\centering
\includegraphics[width=0.99\columnwidth]{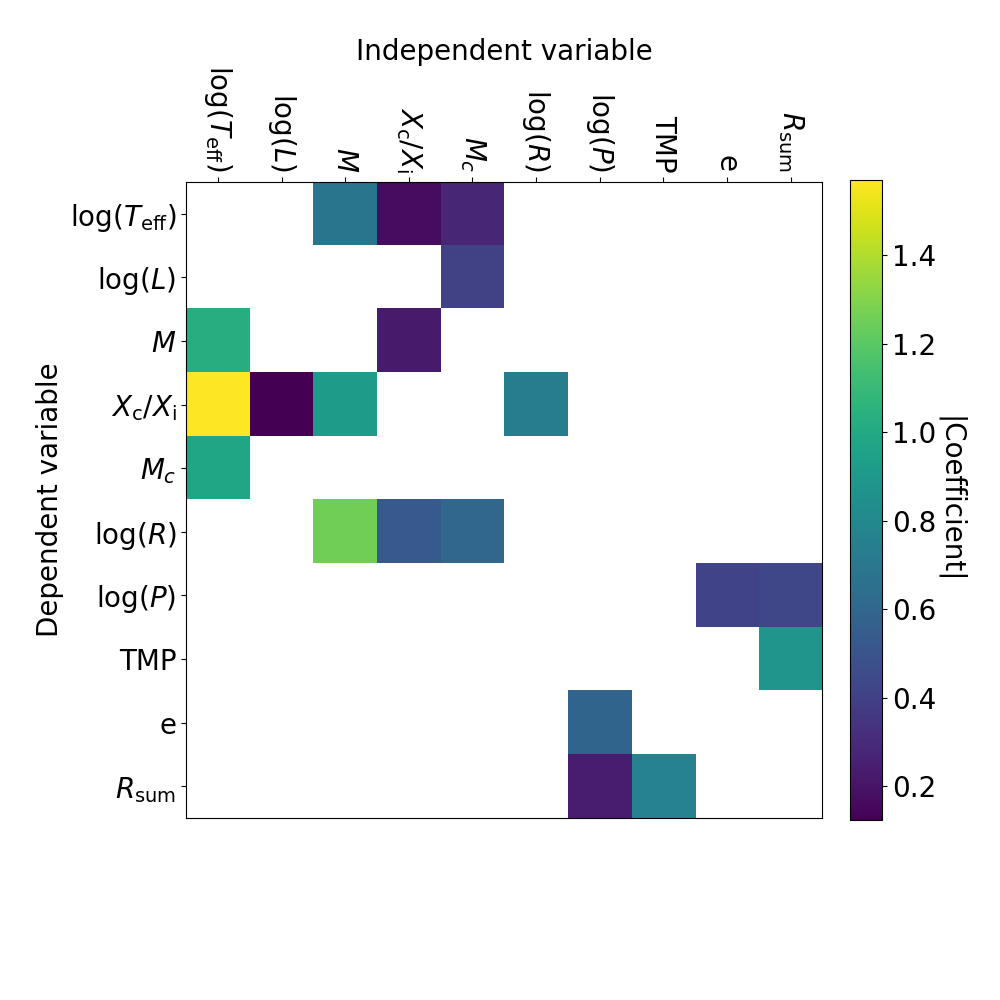}
\end{subfigure}

\begin{subfigure}{0.49\columnwidth}
\centering
\includegraphics[width=0.99\columnwidth]{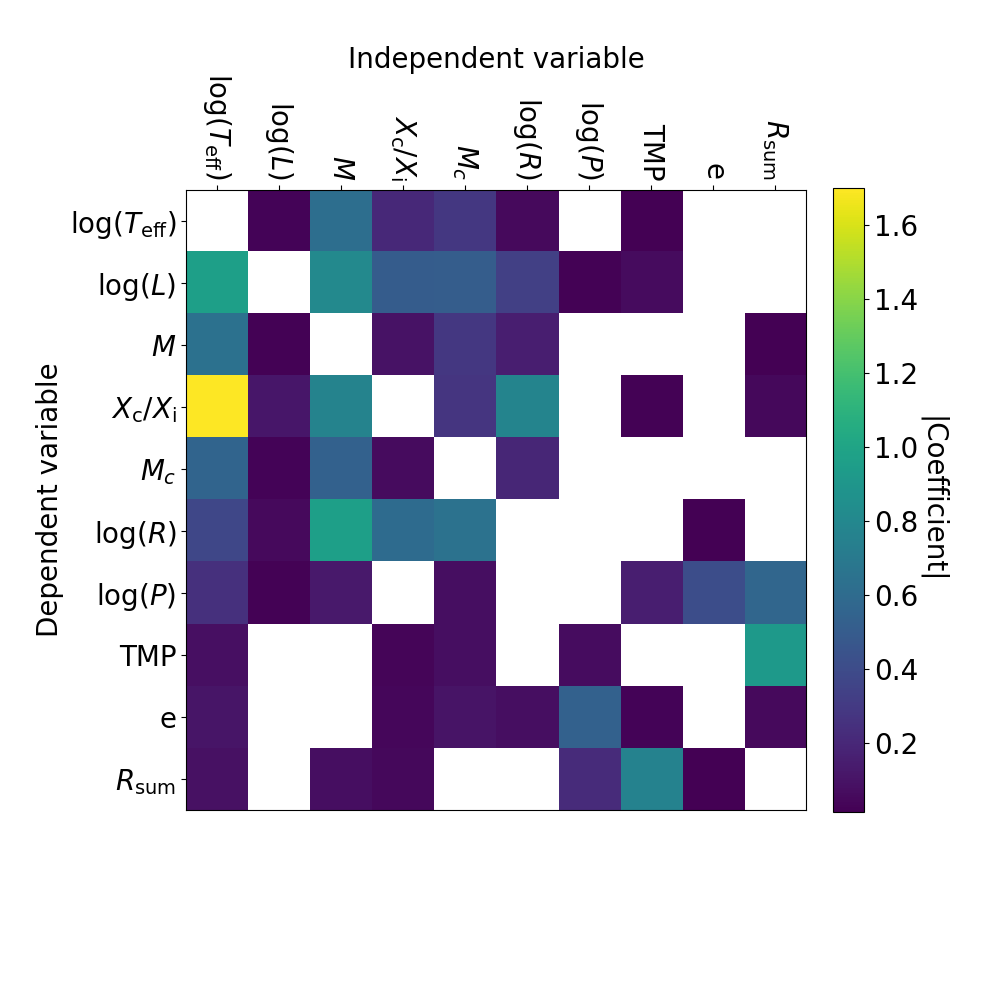}
\end{subfigure}%
\begin{subfigure}{0.49\columnwidth}
\centering
\includegraphics[width=0.99\columnwidth]{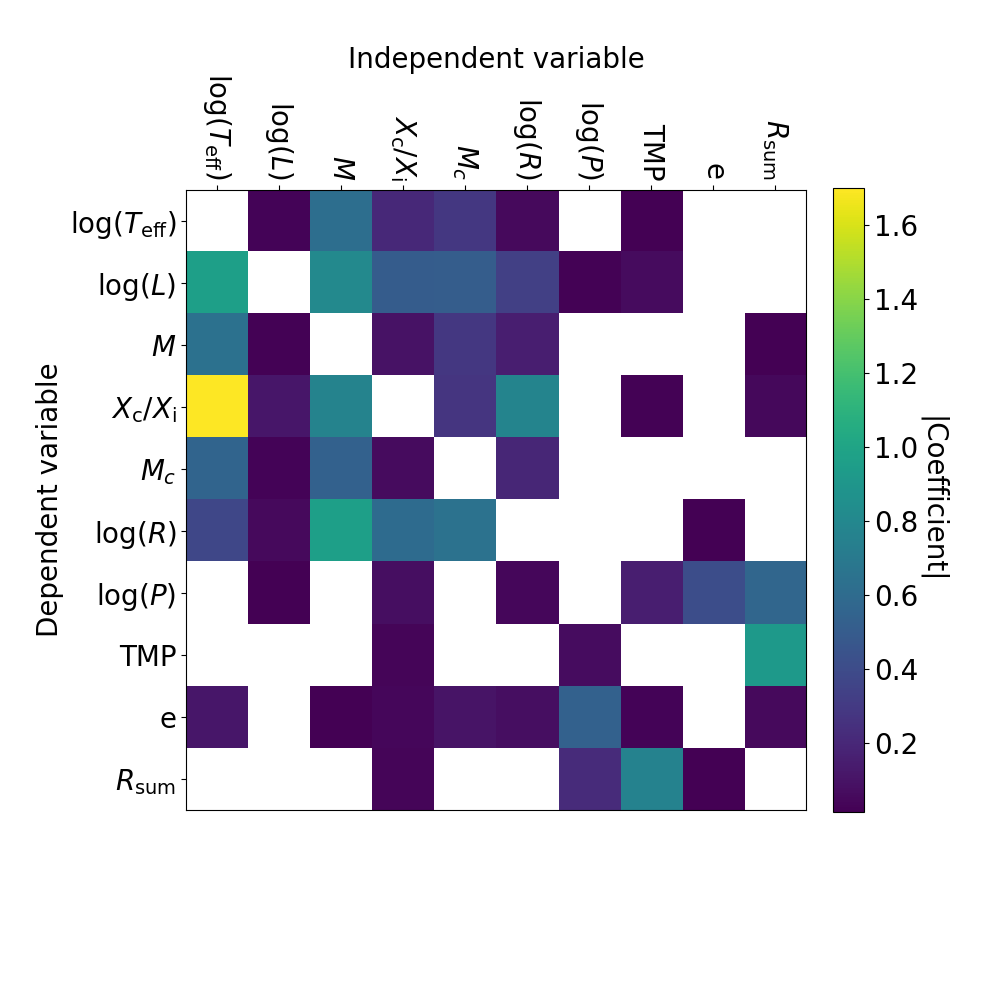}
\end{subfigure}

\caption{Results of our multivariate linear regression (MLR) based importance analysis. For each dependent variable, a predictive model relying on linear regression is built by iteratively adding (forward method) or removing (reverse method) independent variables according to their contribution to the $R^2$ of the linear model. This process is then repeated for each (dependent) variable. All variables are normalised to have a mean of 0 and a standard deviation of 1. Each pixel is coloured by the absolute value of the coefficient for each independent variable in the resulting linear model, where a higher absolute value of the coefficient implies higher importance of the independent variable in predicting the behaviour of the dependent variable. The left panels show the results of the reverse method (removing variables one at a time based on their $R^2$ contribution), while the right panels show the results of the forward method (adding variables one at a time based on their $R^2$ contribution). The two upper panels enforce a stopping criterion demanding at least a 0.01 improvement in $R^2$, while the two lower panels have a more relaxed stopping criterion that demands only a 0.0001 improvement in $R^2$.}
\label{fig:mlr}
\end{figure*}

\begin{figure*}[h!]
\centering
\begin{subfigure}{0.49\columnwidth}
\centering
\includegraphics[width=0.99\columnwidth]{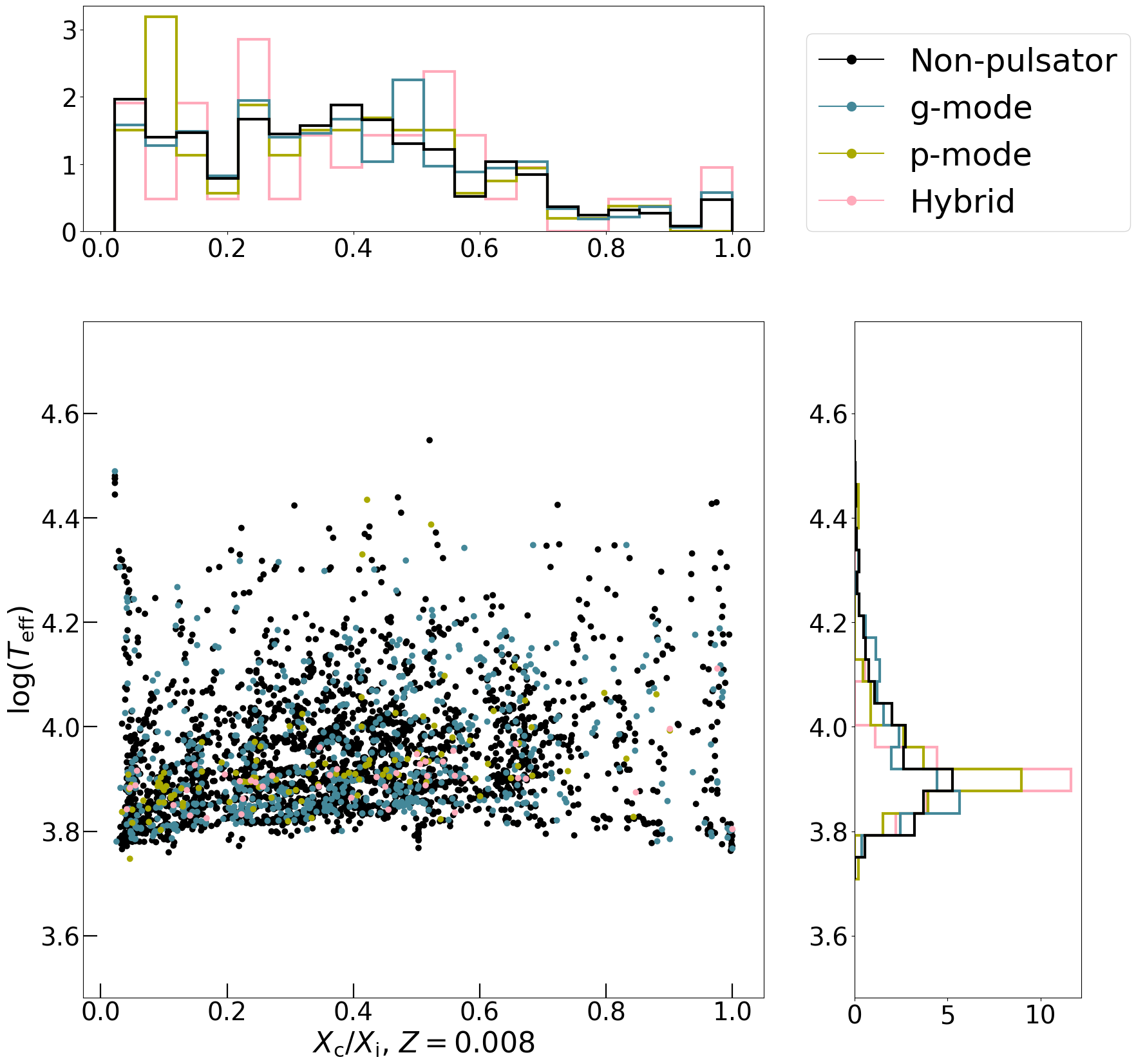}
\end{subfigure}%
\begin{subfigure}{0.49\columnwidth}
\centering
\includegraphics[width=0.99\columnwidth]{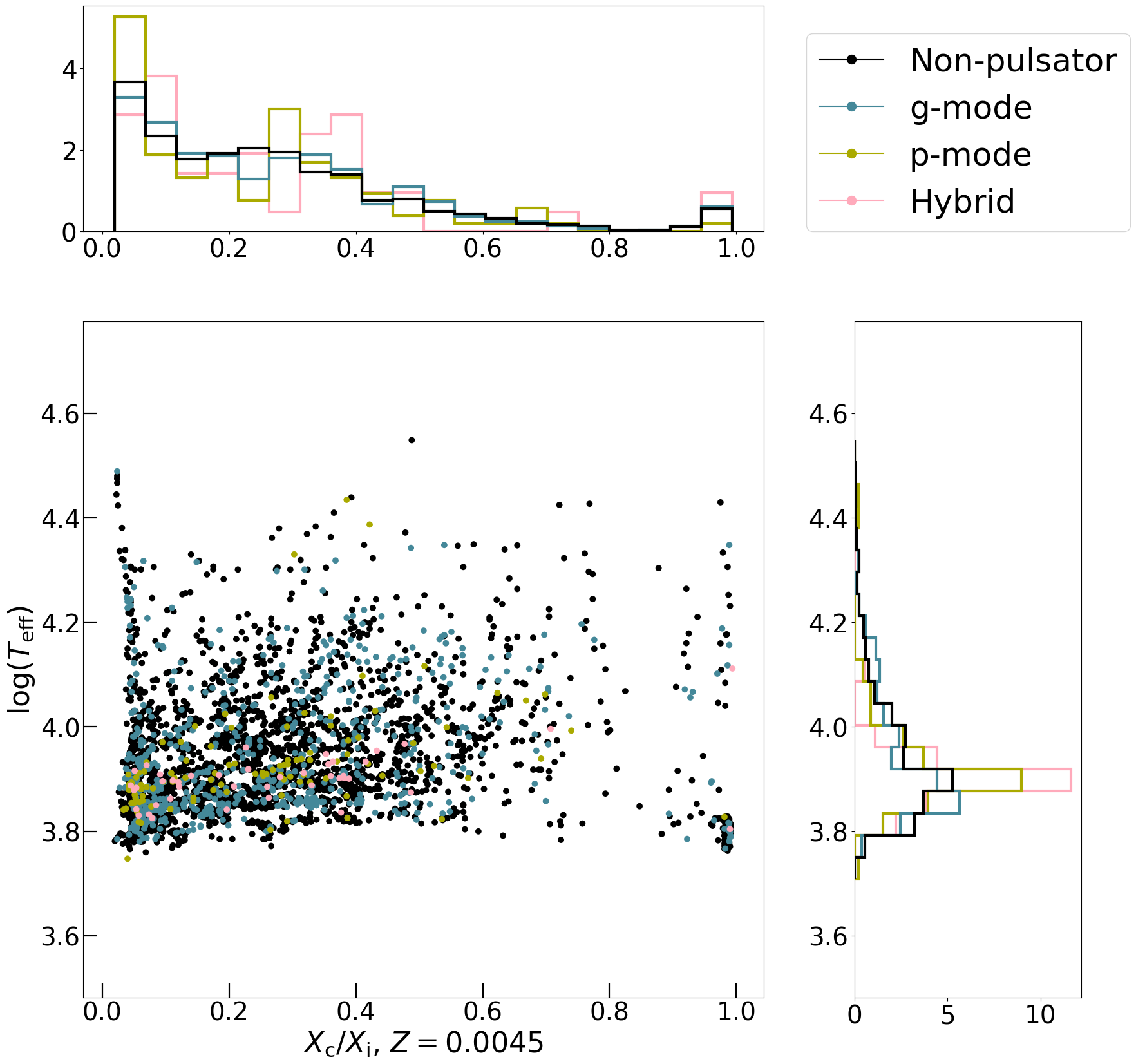}
\end{subfigure}

\begin{subfigure}{0.49\columnwidth}
\centering
\includegraphics[width=0.99\columnwidth]{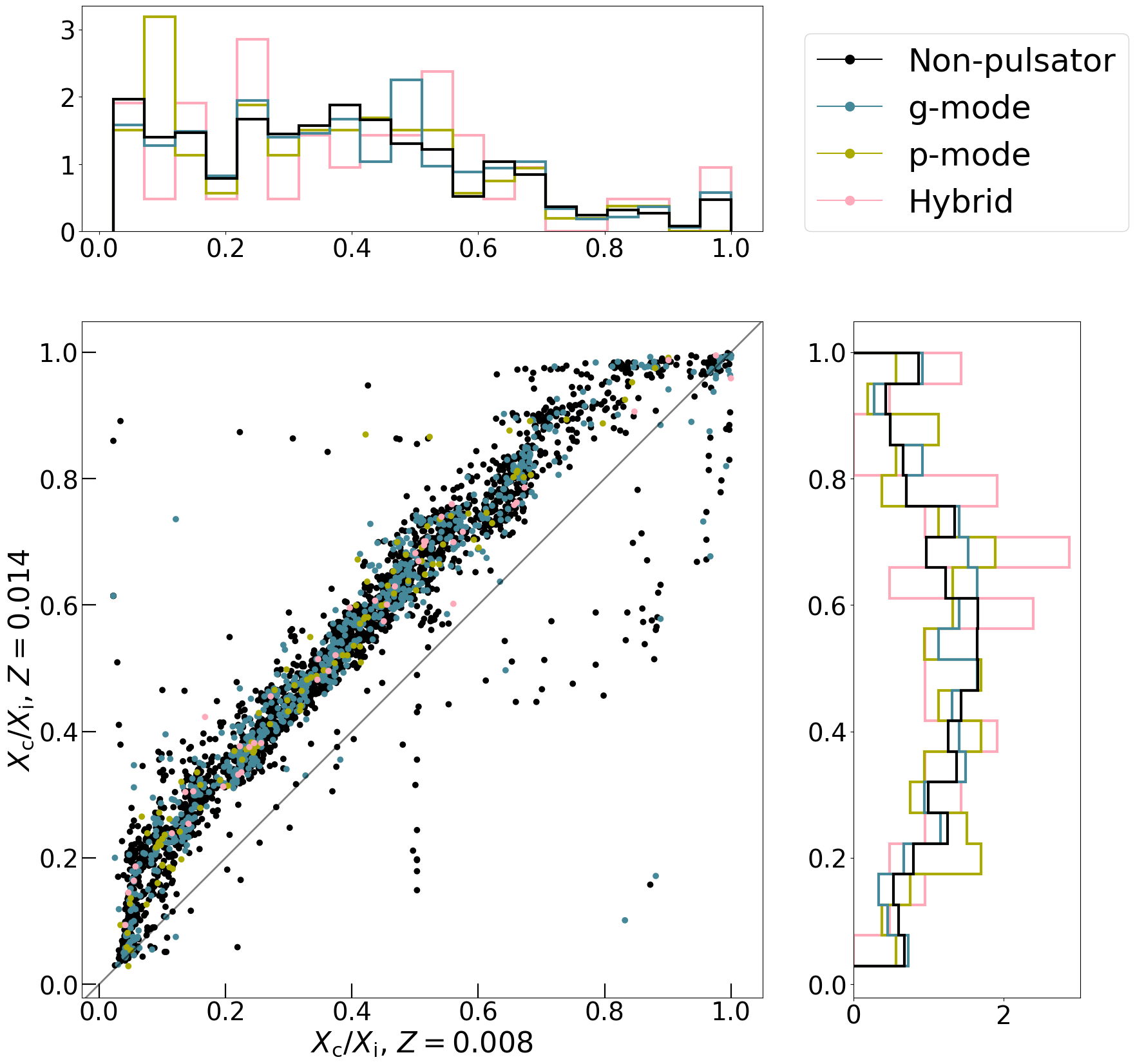}
\end{subfigure}%
\begin{subfigure}{0.49\columnwidth}
\centering
\includegraphics[width=0.99\columnwidth]{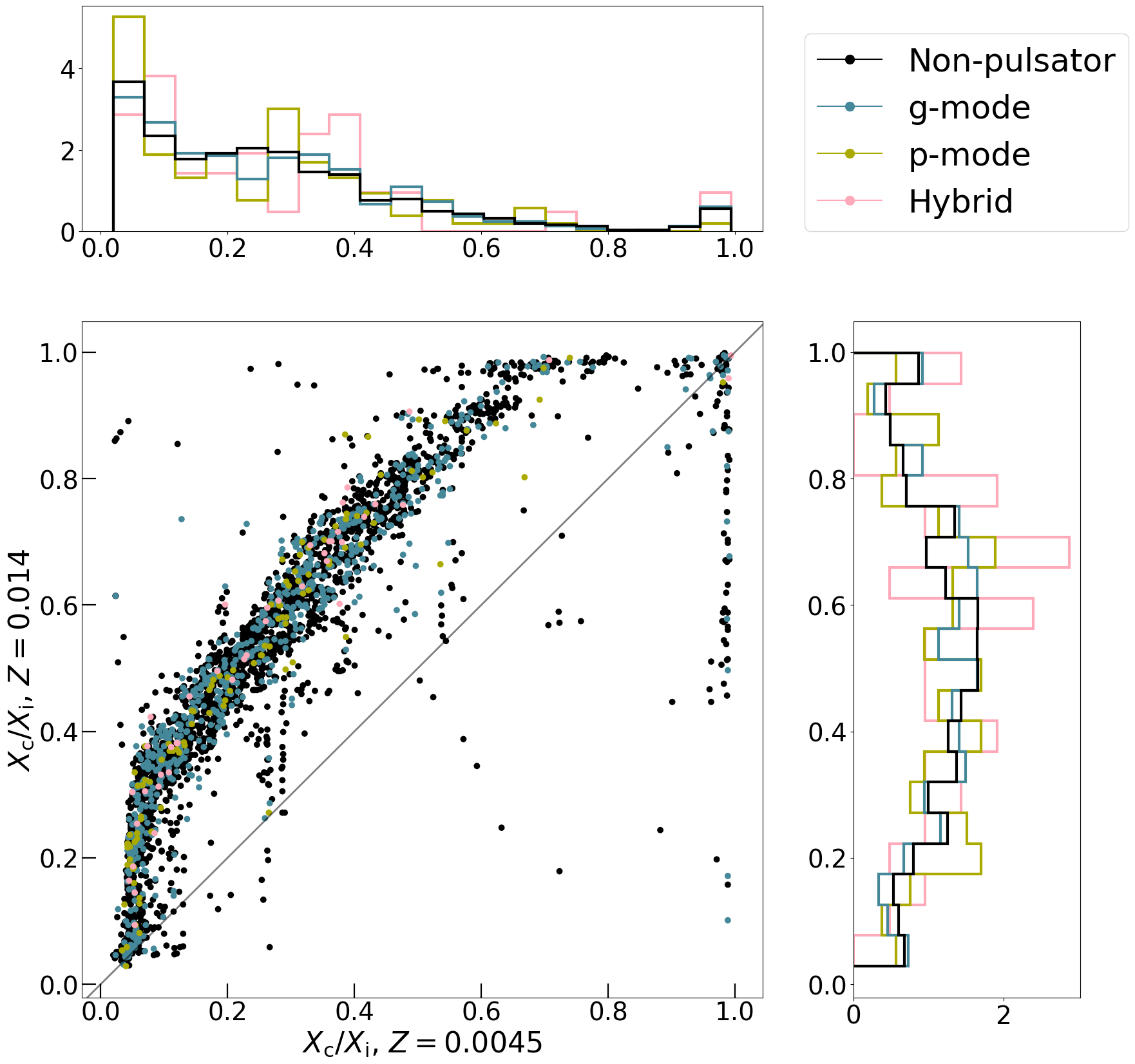}
\end{subfigure}

\caption{Metallicity variation of the \xcx\ distribution, comparing \Z~=~0.008 (left) and \Z~=~0.0045 (right) grids from M24 with \teff\ (top) and \xcx\ inferred from the \Z~=~0.014 grid (bottom).}
\label{fig:xcx_varz}
\end{figure*}

\begin{figure}
\centering
\includegraphics[width=0.7\columnwidth]{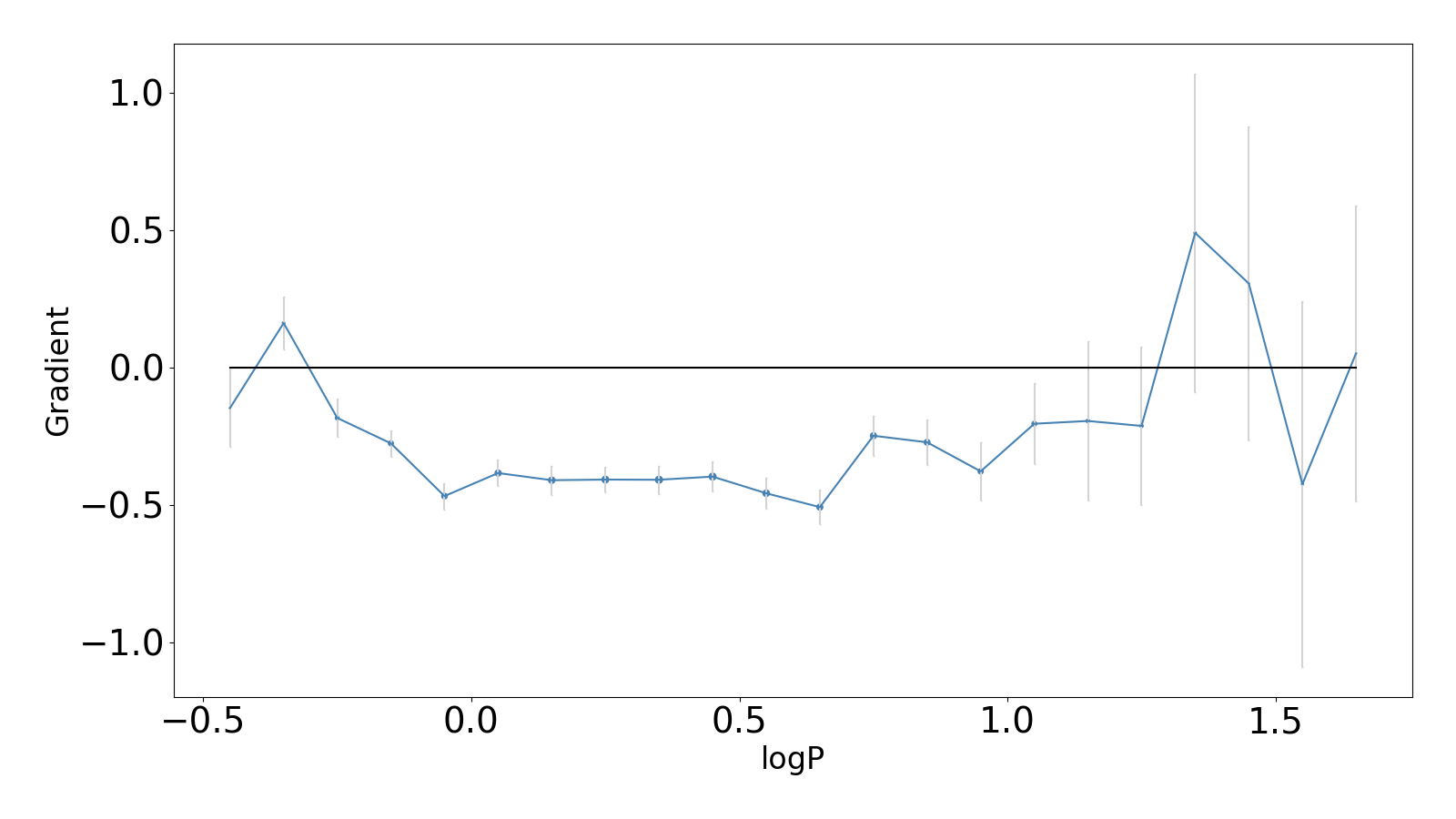}
\caption{{Gradient of the linear fit between TMP and \xcx\ vs the centre of each slice in $\log(P)$. Negative gradients imply that more evolved stars (lower values of \xcx) have higher values of TMP in that $\log(P)$ slice, as we expect from tidal theory. The gradient is normalised by the mean value of the TMP for each slice, preventing gradient bias due to the larger range of TMP values present at lower orbital periods. The equivalent figure without this orbital-period dependent TMP normalisation is available in the online supplementary material online, along with an animation illustrating the fit for each $\log(P)$ slice.}}
\label{fig:gradient}
\end{figure}

\begin{figure}
\centering
\begin{subfigure}{0.5\columnwidth}
\centering
\includegraphics[width=0.9\columnwidth]{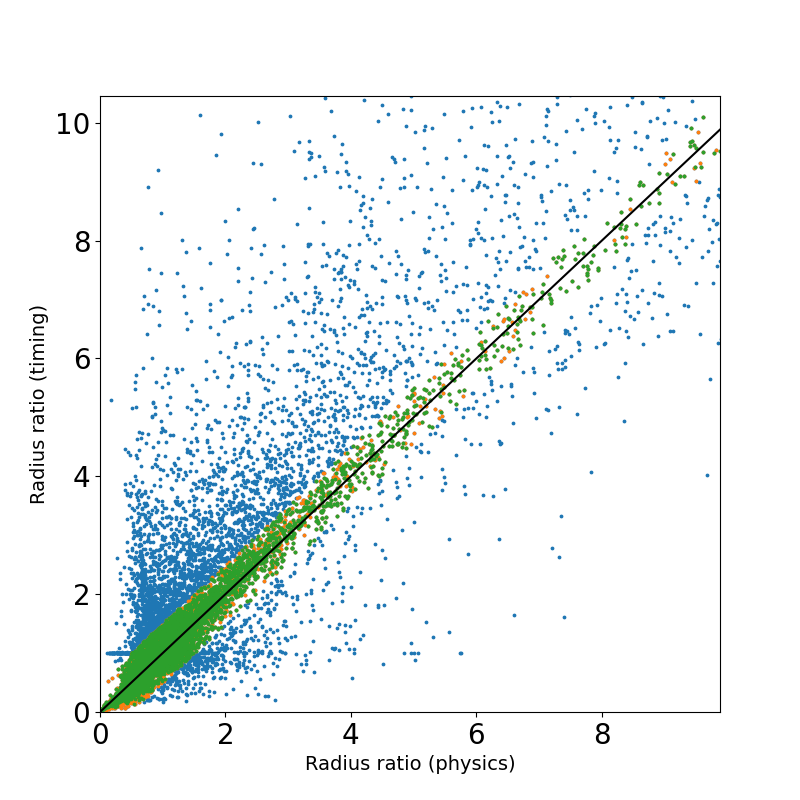}
\end{subfigure}%
\begin{subfigure}{0.5\columnwidth}
\centering
\includegraphics[width=0.9\columnwidth]{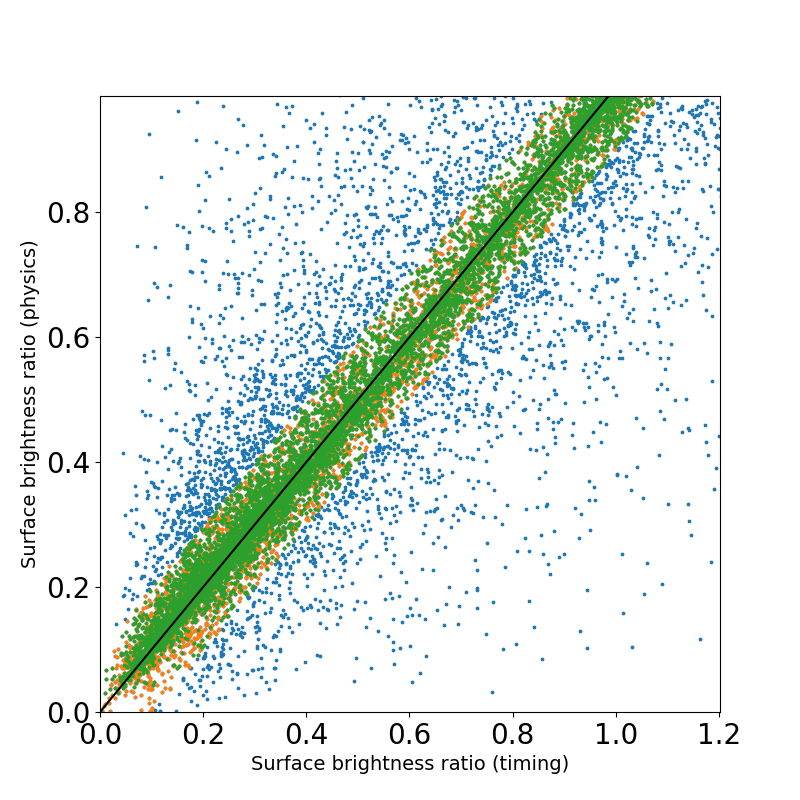}
\end{subfigure}
\caption{Parameter recovery for the radius ratio and surface brightness ratio from the timing and physics-based formulae \citep{ijspeert2024}. Orange targets satisfy the selection condition (see main text for details) for the relevant variable (radius ratio or surface brightness ratio), while green targets satisfy the condition for both the radius ratio and the surface brightness ratio. From the original 14377 sample, 10409 targets satisfy the surface brightness condition, 9728 satisfy the radius ratio condition, and 7823 satisfy both conditions.}
\label{fig:rrat_sbrat}
\end{figure}

\begin{figure*}[h!]
\centering
\begin{subfigure}{0.49\columnwidth}
\centering
\includegraphics[width=0.99\columnwidth]{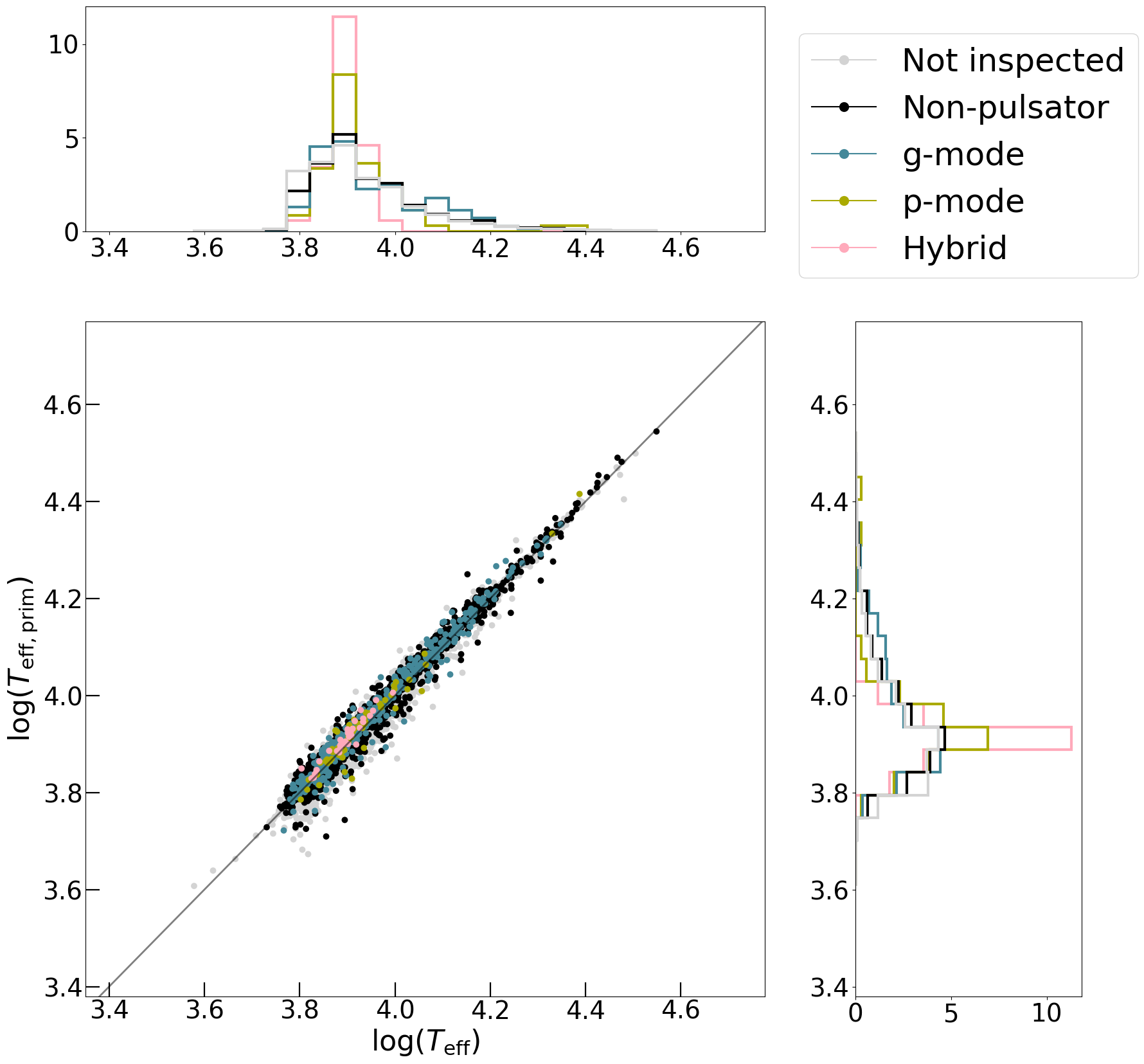}
\end{subfigure}%
\begin{subfigure}{0.49\columnwidth}
\centering
\includegraphics[width=0.99\columnwidth]{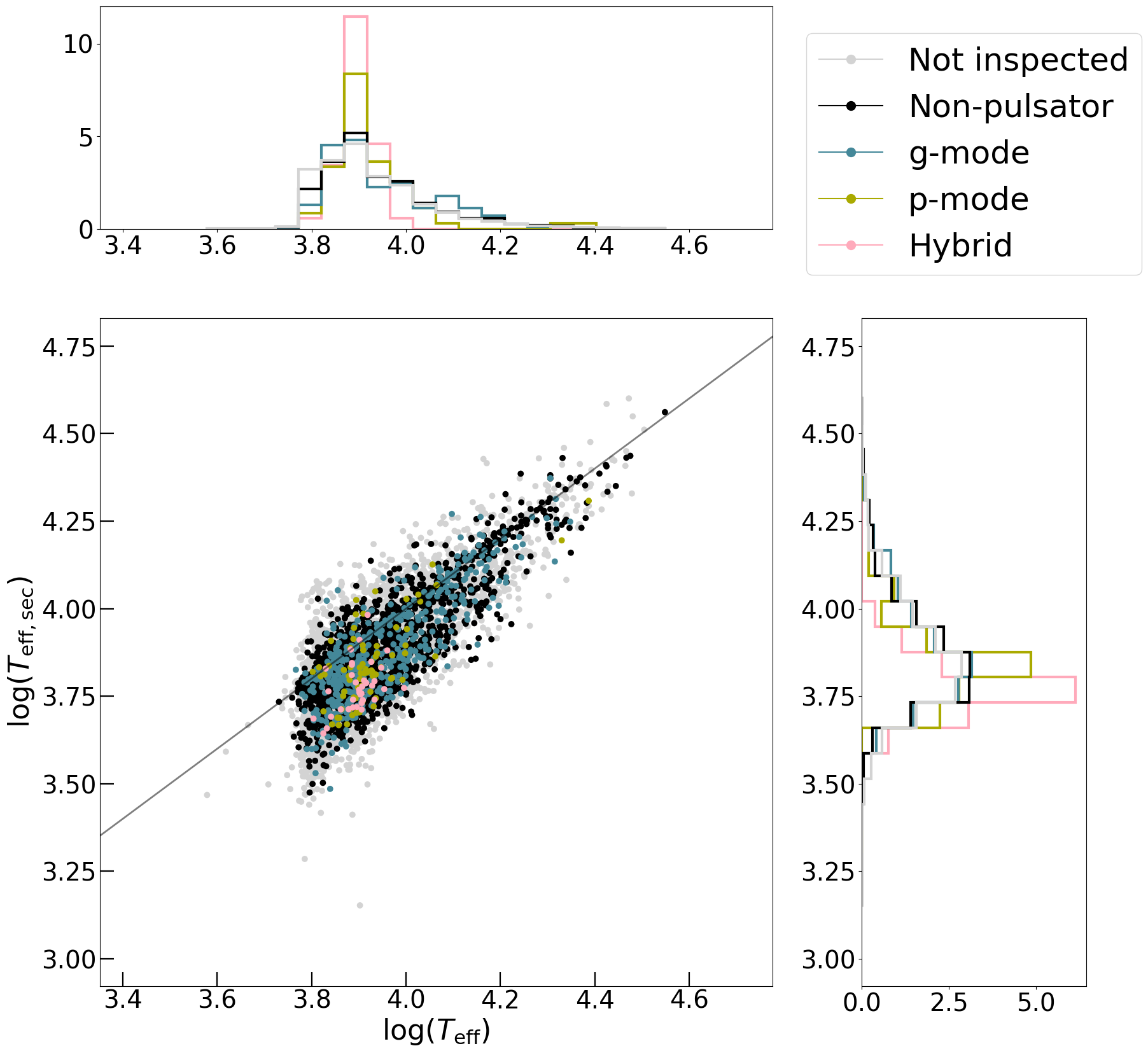}
\end{subfigure}

\begin{subfigure}{0.49\columnwidth}
\centering
\includegraphics[width=0.99\columnwidth]{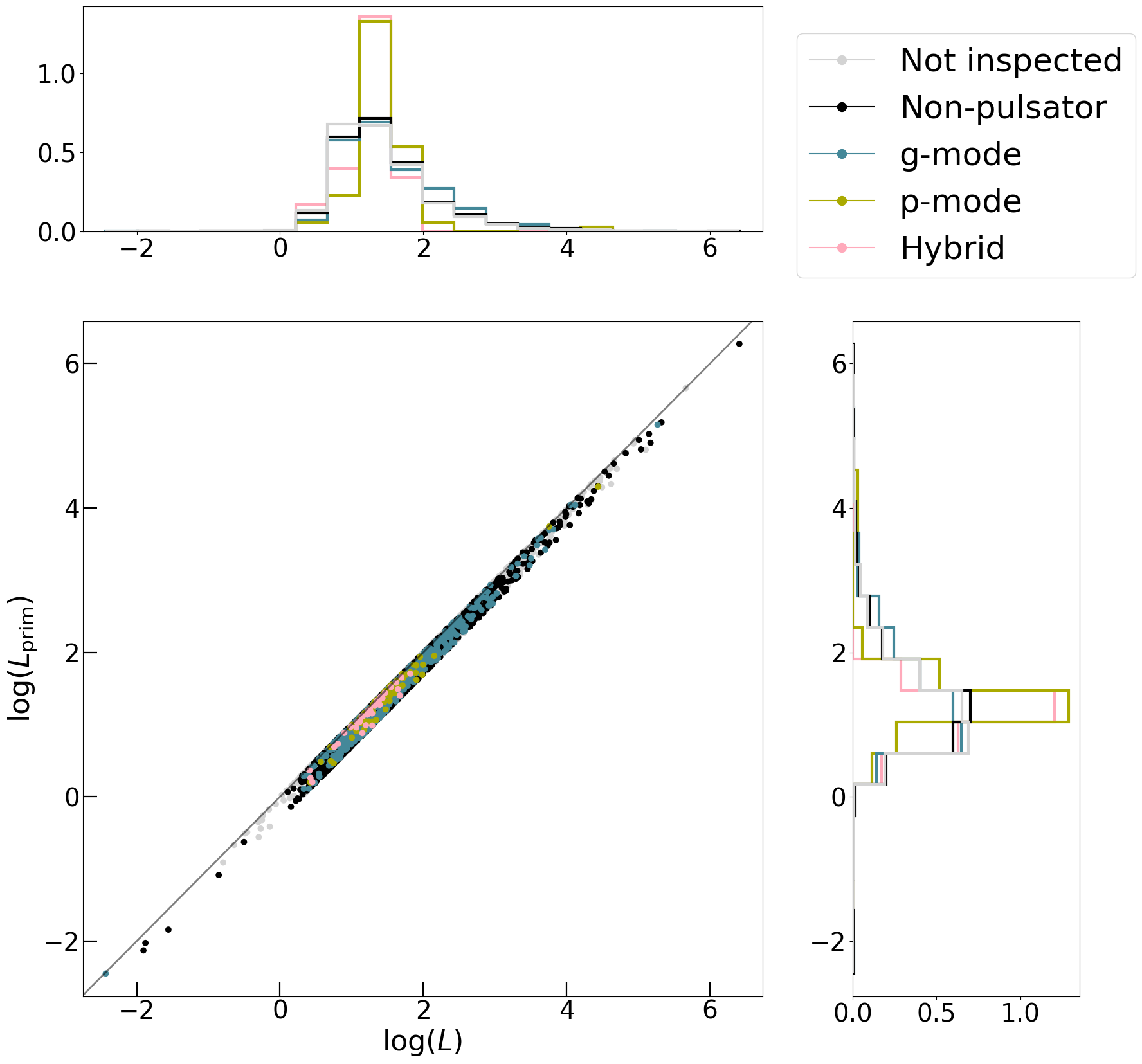}
\end{subfigure}%
\begin{subfigure}{0.49\columnwidth}
\centering
\includegraphics[width=0.99\columnwidth]{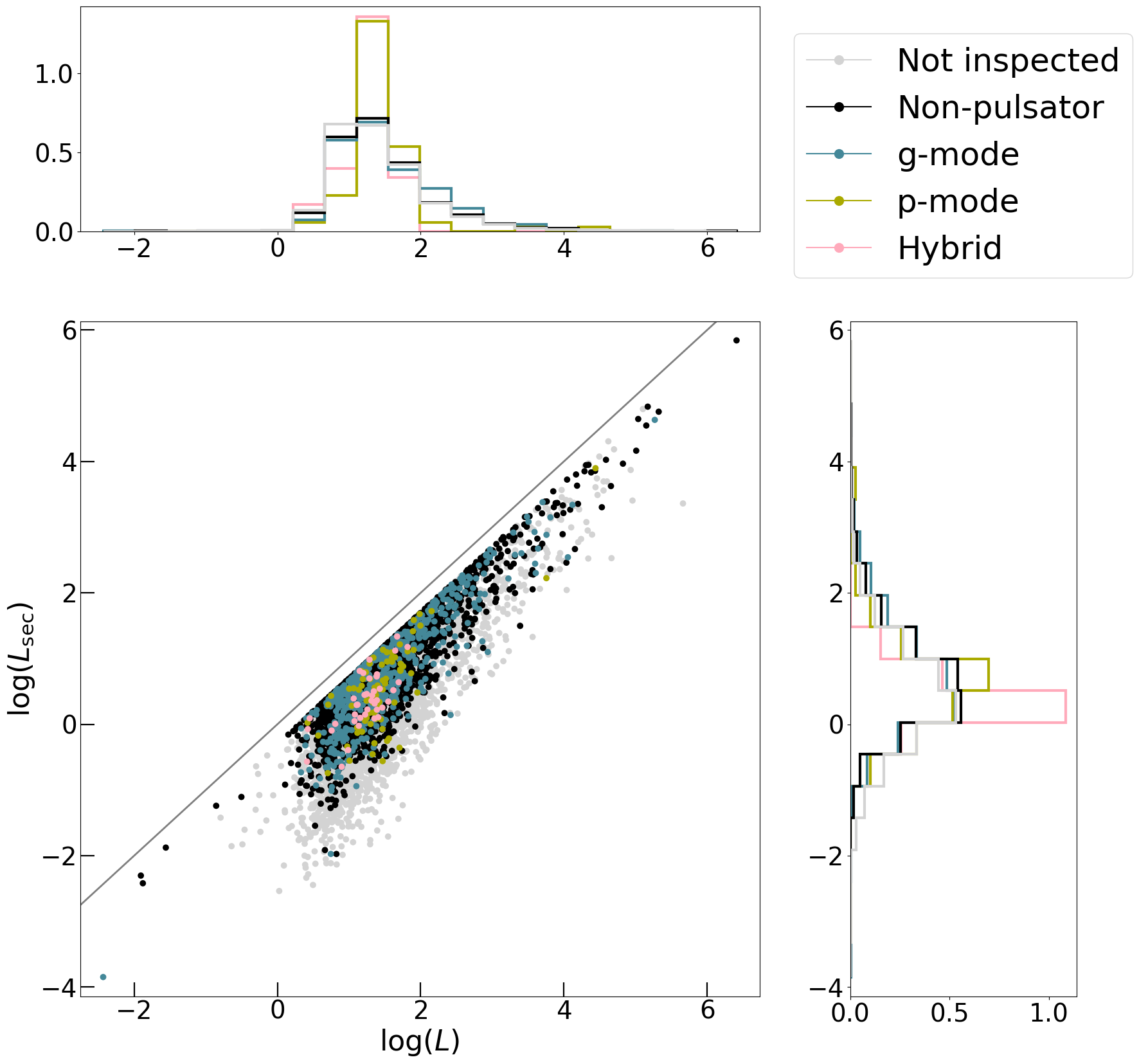}
\end{subfigure}

\caption{\logteff\ (upper panels) and \logl\ (lower panels) estimates for the primary based on the radius ratio and surface brightness ratios plotted against the \gaia\ values are shown in the left panels. Equivalent figures for the secondary are plotted in the right panels.}
\label{fig:teff_l_psb_recovery}
\end{figure*}

\begin{figure*}
\centering
\includegraphics[width=0.7\columnwidth]{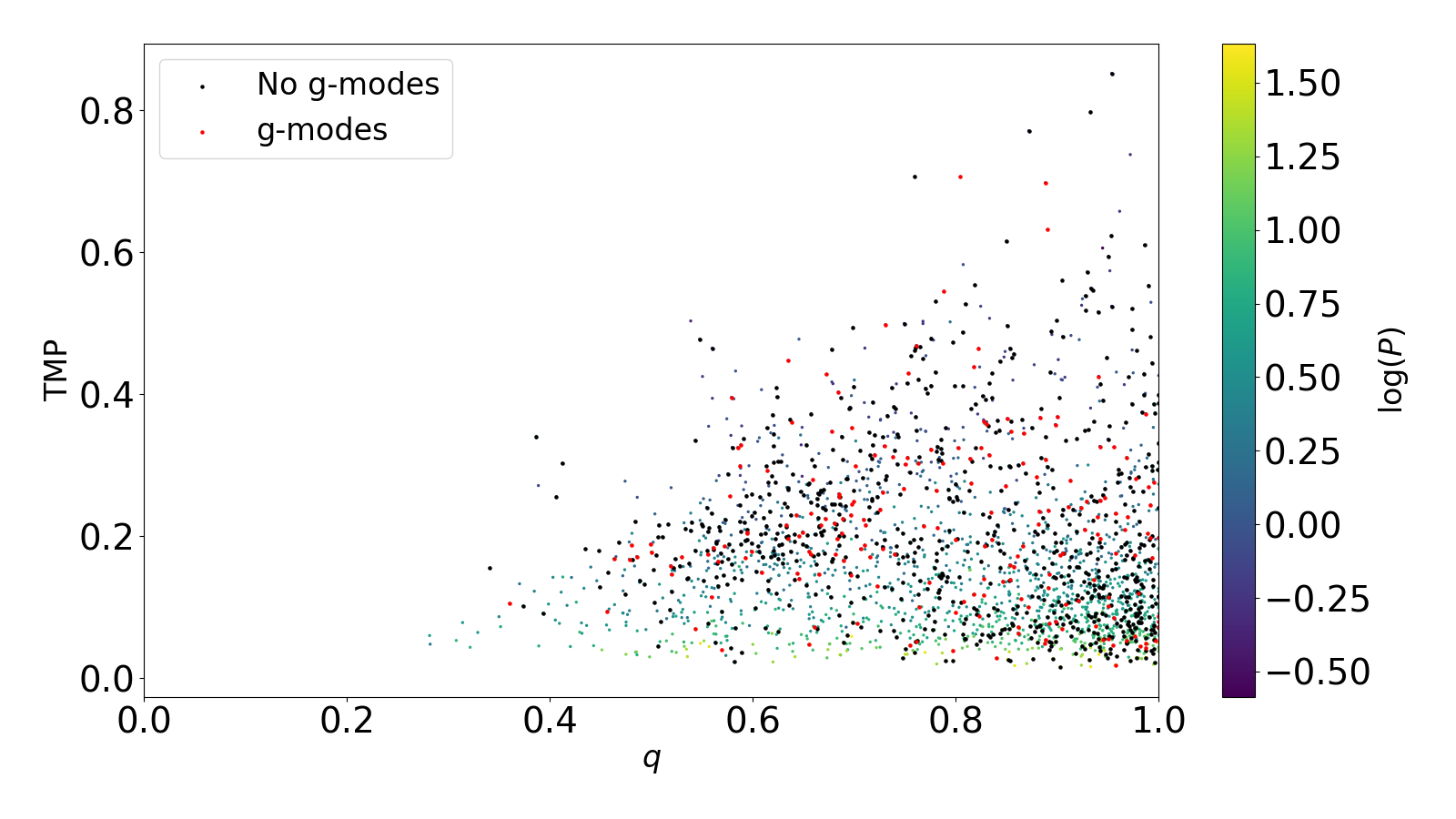}
\caption{{$q$ vs TMP, excluding points with obvious grid effects in the secondary mass. A correlation between q and TMP is present, particularly at higher values of q, although the orbital period dependence is clearly dominant. Considering targets exhibiting candidate g-mode (red) pulsations vs those found not to include g-mode pulsations, we find there is no evidence for g-mode pulsators having systematically higher values of TMP for when fixing $q$ (see, for example, Fig. \ref{fig:q_slice}.}}
\label{fig:q_tmp}
\end{figure*}

\begin{figure*}
\centering
\includegraphics[width=0.7\columnwidth]{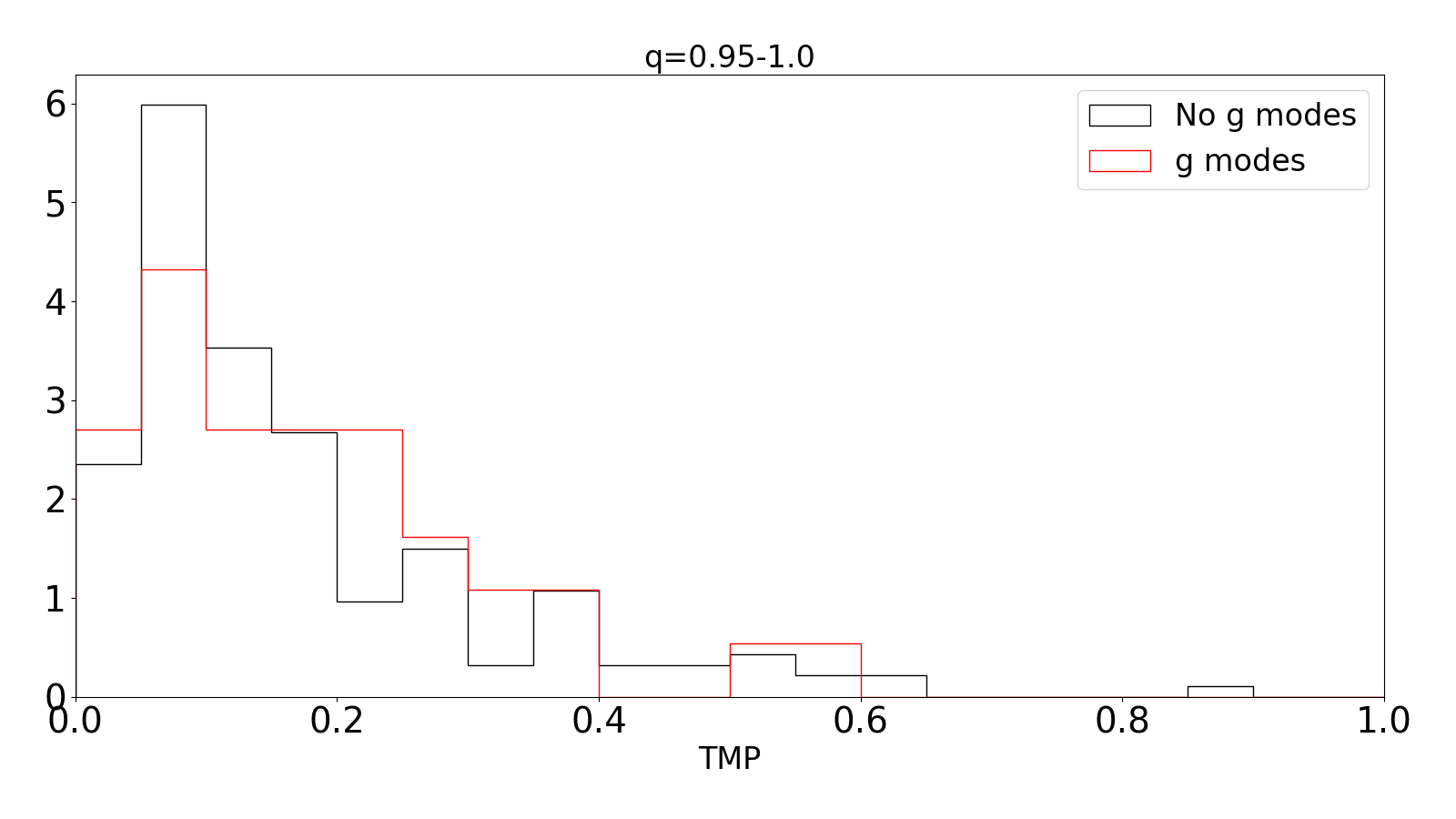}
\caption{{TMP distribution for pulsators with and without g~mode pulsations for targets with a $q$ between 0.95 and 1.0. Targets with obvious systematics affecting the secondary mass estimate are secondary mass are excluded. The presence of g~mode pulsations does not result in any significant bias towards higher values of TMP in this or any other q-slices (available online).}}
\label{fig:q_slice}
\end{figure*}

\begin{figure*}
\centering
\begin{subfigure}{0.49\columnwidth}
\centering
\includegraphics[width=0.90\columnwidth]{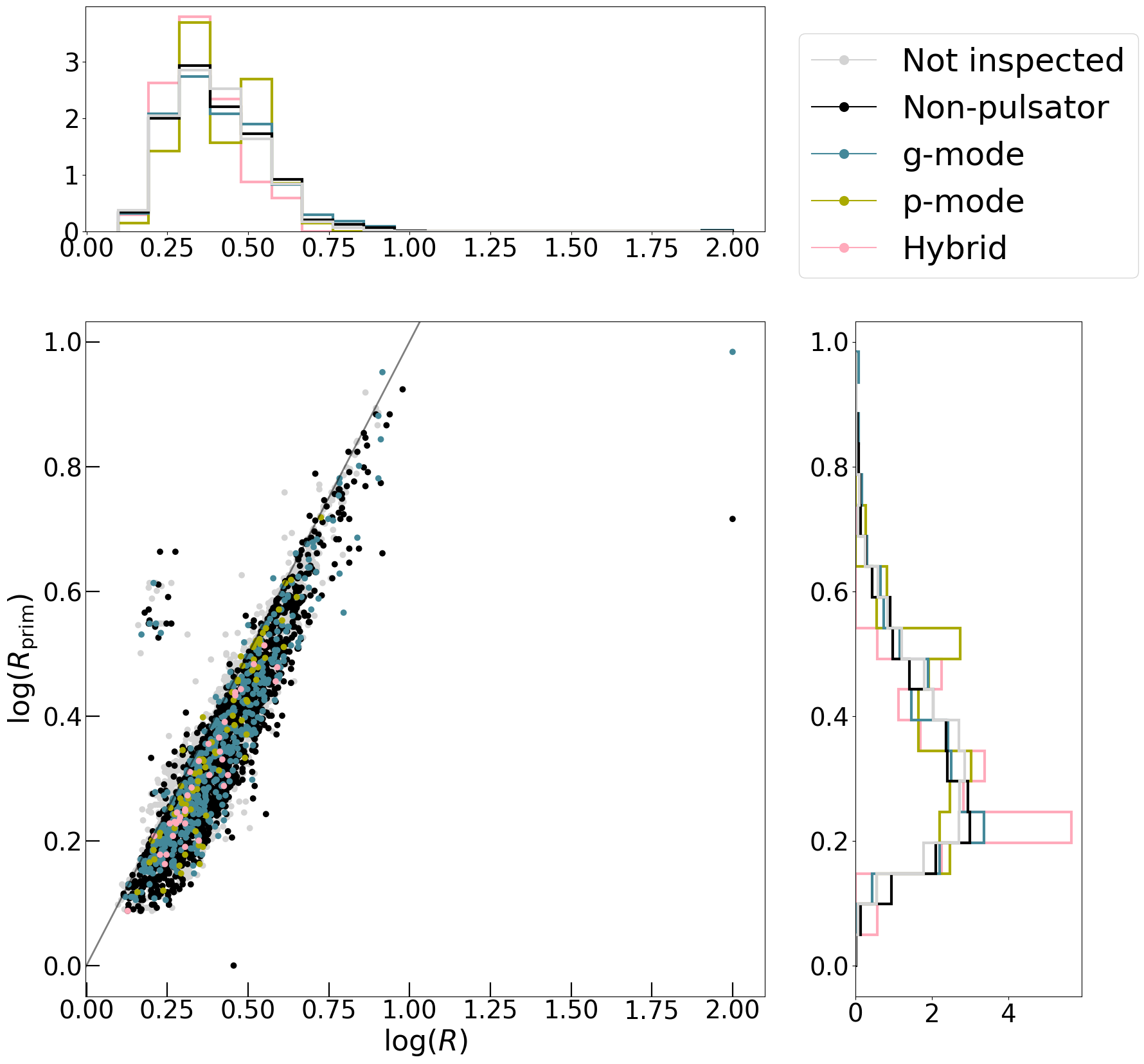}
\end{subfigure}%
\begin{subfigure}{0.49\columnwidth}
\centering
\includegraphics[width=0.90\columnwidth]{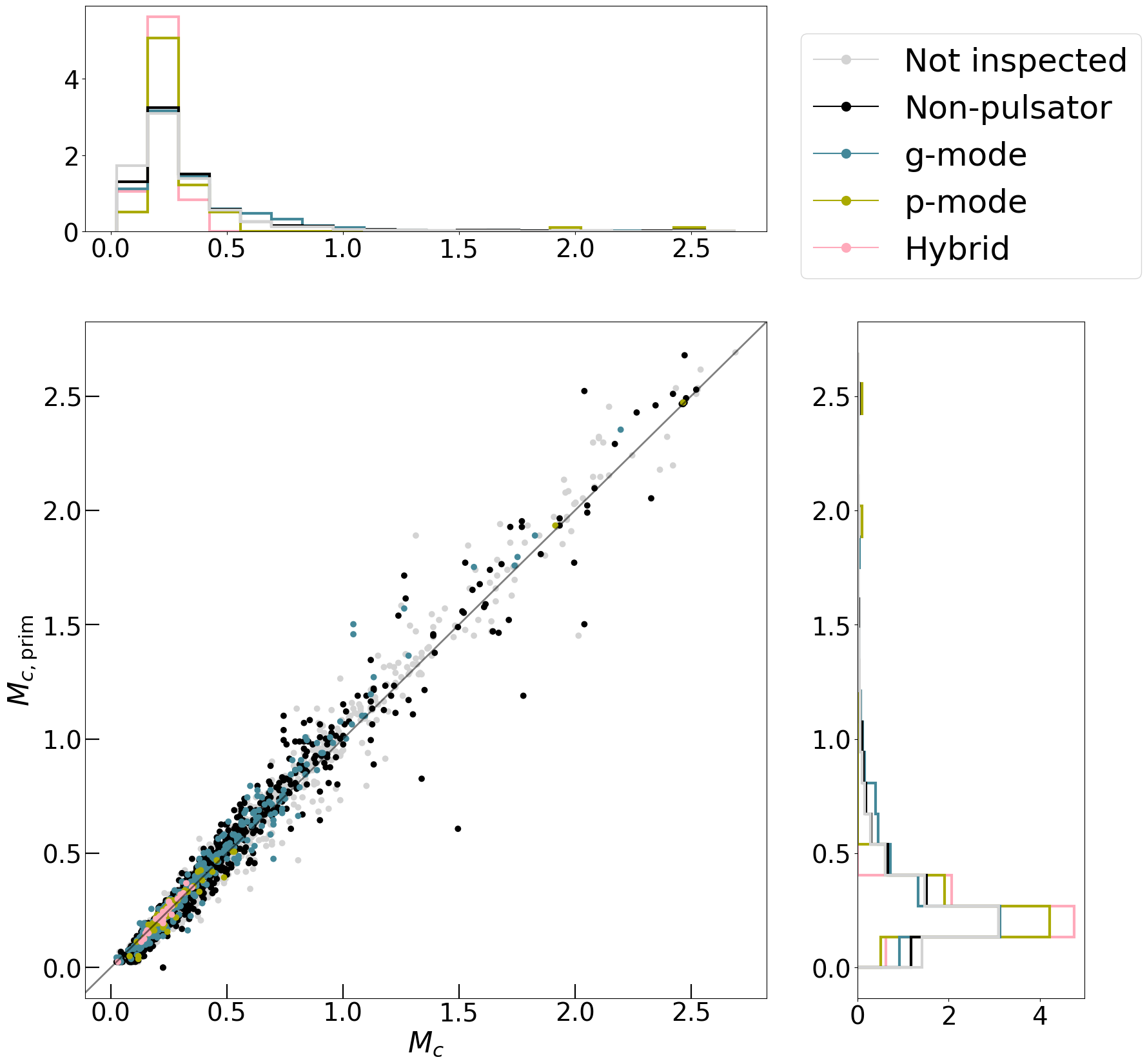}
\end{subfigure}

\begin{subfigure}{0.49\columnwidth}
\centering
\includegraphics[width=0.90\columnwidth]{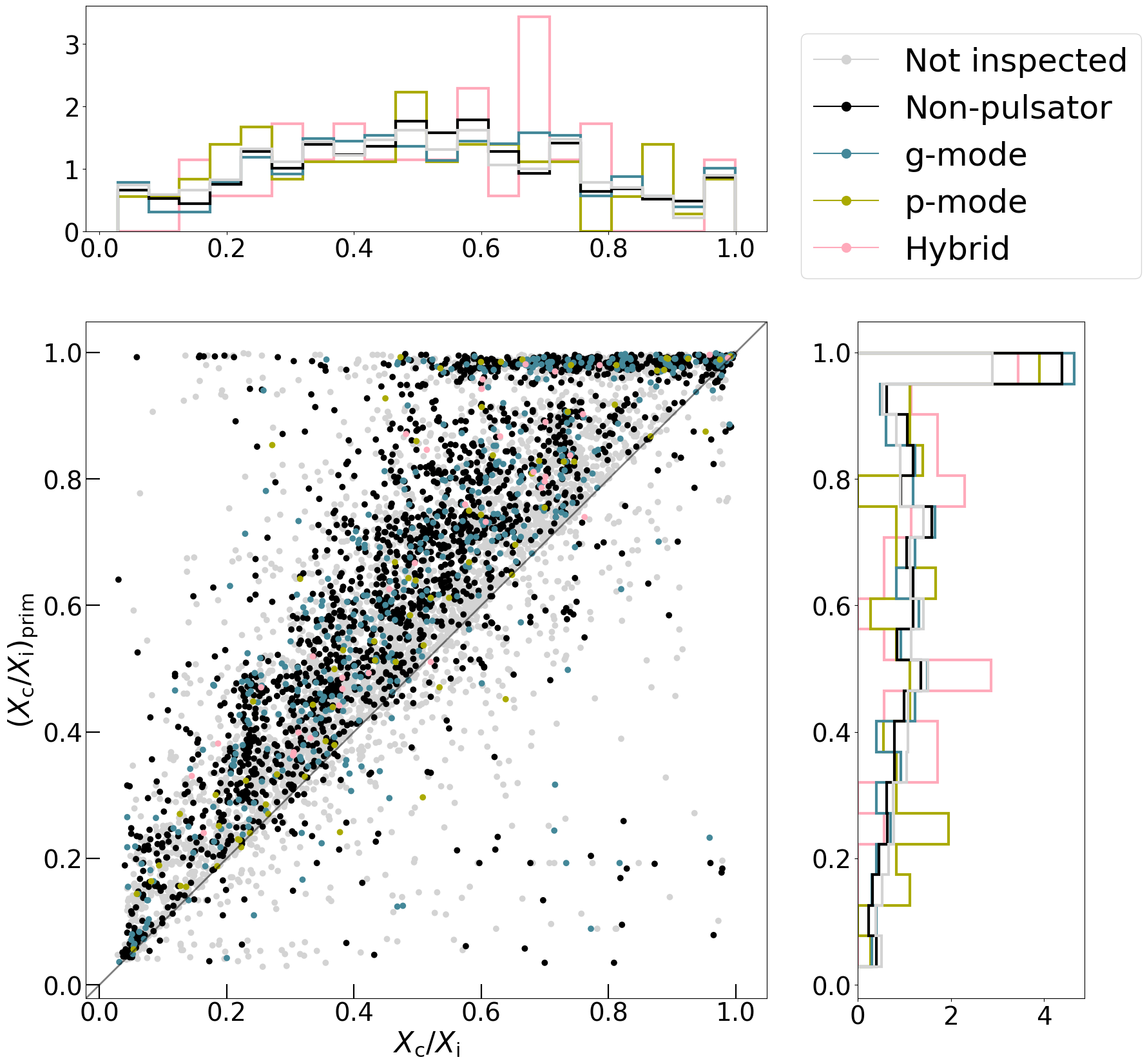}
\end{subfigure}%
\begin{subfigure}{0.49\columnwidth}
\centering
\includegraphics[width=0.90\columnwidth]{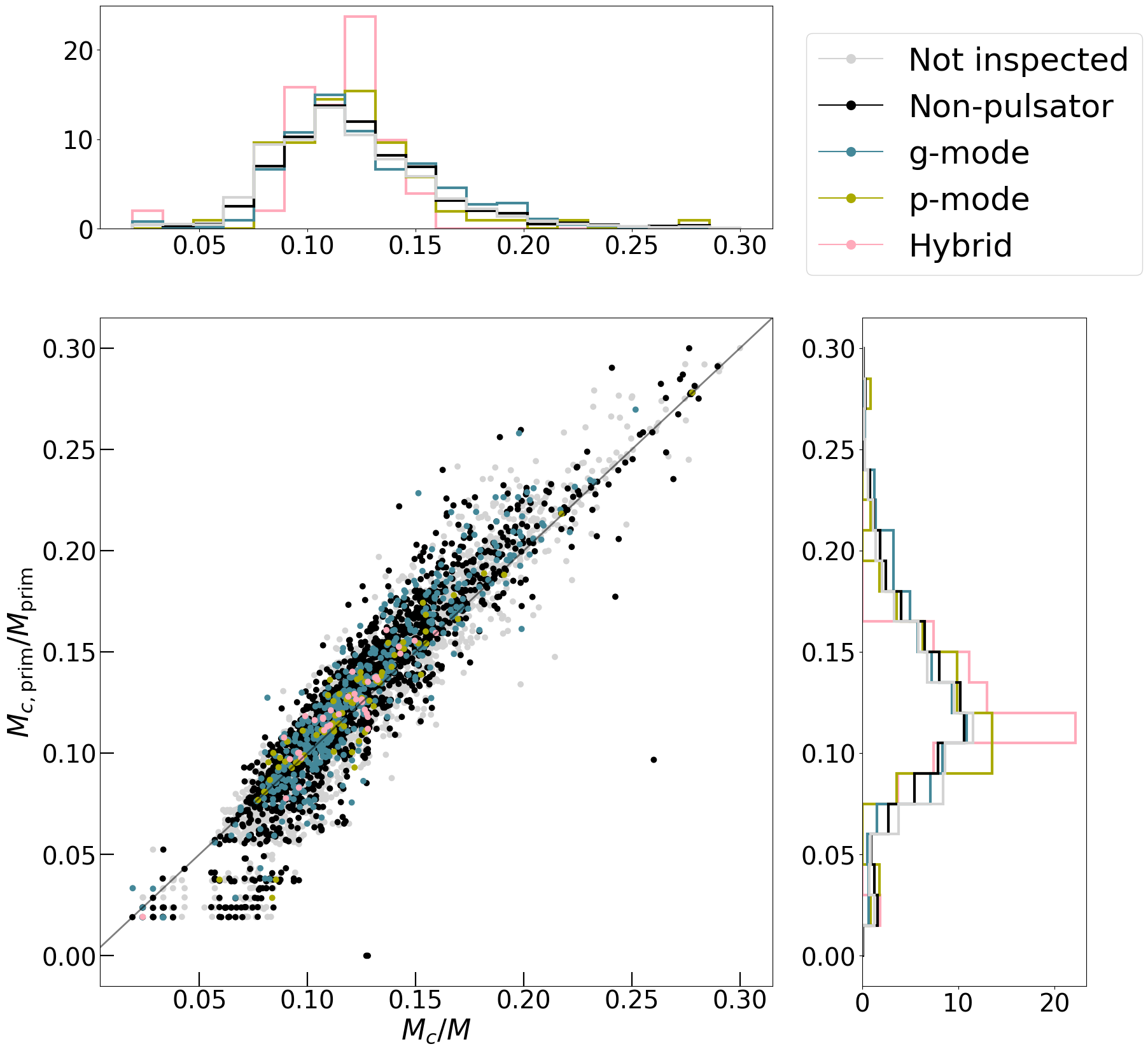}
\end{subfigure}

\caption{Distributions for the radius (upper left), core mass (upper right), central H fraction relative to the initial fraction \xcx\ (lower left) and core mass fraction (lower left) inferred from the primary's \logteff\ and \logl\ estimates, plotted against their respective \gaia\ inferences.}
\label{fig:partymixprim}
\end{figure*}

\begin{table*}[]
\caption{Numbers of targets with primary masses incompatible with the 2MASS \ks-magnitude measurement.}
\begin{tabular}{l|l|llllll}
\xc & Inference & M24 Valid & M24 Invalid & \% & MIST Valid & MIST Invalid & (\%) \\ \hline
0.69 & \gaia & 4359 & 884 & 16.86 & 4117 & 1126 & 21.48 \\
0.69 & $M_{\rm log(T)   + 0.1\ dex}$ & 3609 & 1634 & 31.17 & 3279 & 1964 & 37.46 \\
0.69 & $M_{\rm log(L)   - 0.2\ dex}$ & 4689 & 554 & 10.57 & 4577 & 666 & 12.70 \\
0.69 & $M_{\rm prim}$ & 4496 & 747 & 14.25 & 4303 & 940 & 17.93 \\ \hline
0.35 & \gaia & 1192 & 4051 & 77.26 & 1026 & 4217 & 80.43 \\
0.35 & $M_{\rm log(T)   + 0.1\ dex}$ & 743 & 4500 & 85.83 & 593 & 4650 & 88.69 \\
0.35 & $M_{\rm log(L)   - 0.2\ dex}$ & 2152 & 3091 & 58.95 & 1999 & 3244 & 61.87 \\
0.35 & $M_{\rm prim}$ & 1694 & 3549 & 67.69 & 1577 & 3666 & 69.92 \\ \hline
0.01 & \gaia & 107 & 5136 & 97.96 & 92 & 5151 & 98.25 \\
0.01 & $M_{\rm log(T)   + 0.1\ dex}$ & 76 & 5167 & 98.55 & 66 & 5177 & 98.74 \\
0.01 & $M_{\rm log(L)   - 0.2\ dex}$ & 211 & 5032 & 95.98 & 192 & 5051 & 96.34 \\
0.01 & $M_{\rm prim}$ & 175 & 5068 & 96.66 & 165 & 5078 & 96.85
\end{tabular}
\tablefoot{`\%' columns express the number of incompatible (invalid) targets as a fraction of the total targets (valid + invalid) for each case as a percentage.} 
\label{tab:jastab}
\end{table*}

\end{document}